\documentclass[12pt, a4paper]{article}

\pdfoutput=1

\usepackage{amsmath}
\usepackage{amsfonts}
\usepackage{amssymb}
\usepackage{bbm}
\usepackage{verbatim}
\usepackage{dsfont}
\usepackage{booktabs}
\usepackage{slashed}
\usepackage{nicefrac}

\usepackage{epsfig}
\usepackage{color}
\usepackage[table]{xcolor}
\usepackage{graphicx}
\usepackage[font=small]{caption}
\usepackage[listofformat=empty,subrefformat=empty]{subfig}

\usepackage{float}
\usepackage{overpic}
\usepackage{tikz}

\usepackage{enumerate}
\usepackage{hhline}
\usepackage{multirow}

\usepackage{cite}
\usepackage{xspace}
\usepackage{setspace}

\usepackage[pdftitle={},
  pdfauthor={},
  pdfsubject={},
  bookmarksopen, bookmarksnumbered, bookmarksopenlevel=2, colorlinks=false, linkcolor=blue, citecolor=blue, urlcolor=blue]{hyperref}

\renewcommand{\tfrac}{\genfrac{}{}{}1}
\newcommand{\ttfrac}{\genfrac{}{}{}2}

\newcommand{\Z}{\mathcal{Z}}
\newcommand{\Ss}{\mathcal{S}_7}
\newcommand{\Sn}{\mathcal{S}_9}
\newcommand{\Kbi}{\ensuremath{K_B^{-1}}}

\begin{document}

\thispagestyle{empty}

\vskip .8 cm
\begin{center}
{\Large {\bf Global tensor-matter transitions in F-theory}}\\[12pt]
\bigskip
\bigskip 
{
{\bf{Markus Dierigl$^{a,b,}$}\footnote{E-mail: m.j.dierigl@uu.nl}},
{\bf{Paul-Konstantin Oehlmann$^{c,}$}\footnote{E-mail: paul.oehlmann@desy.de}},  
{\bf{Fabian Ruehle$^{d,}$}\footnote{E-mail: fabian.ruehle@physics.ox.ac.uk}},
\bigskip}\\[0pt]
\vspace{0.23cm}
{\it 
$^a$ Institute for Theoretical Physics, Utrecht University, Princetonplein 5, 3584 CC Utrecht, The Netherlands \\[5mm]
$^b$ Institute of Physics, University of Amsterdam, Science Park 904, 1098 XH Amsterdam, The Netherlands \\[5mm]
$^c$ Physics Department, Robeson Hall, Virginia Tech, Blacksburg, VA 24061, USA \\[5mm]
$^d$ Rudolf Peierls Centre for Theoretical Physics, University of Oxford, Parks Road, Oxford OX1 3PU, UK
}\\[20pt] 
\bigskip
\end{center}

\begin{abstract}
\noindent
We use F-theory to study gauge algebra preserving transitions of 6d supergravity theories that are connected by superconformal points. While the vector multiplets remain unchanged, the hyper- and tensor multiplet sectors are modified. In 6d F-theory models, these transitions are realized by tuning the intersection points of two curves, one of them carrying a non-Abelian gauge algebra, to a $(4,6,12)$ singularity, followed by a resolution in the base. The six-dimensional supergravity anomaly constraints are strong enough to completely fix the possible non-Abelian representations and to restrict the Abelian charges in the hypermultiplet sector affected by the transition, as we demonstrate for all Lie algebras and their representations. Furthermore, we present several examples of such transitions in torically resolved fibrations. In these smooth models, superconformal points lead to non-flat fibers which correspond to non-toric K\"ahler deformations of the torus-fibered Calabi-Yau 3-fold geometry.
\end{abstract}

\newpage 
\setcounter{page}{2}
\setcounter{footnote}{0}
\tableofcontents
\newpage

{\onehalfspacing

\section{Introduction}
\label{sec:intro}

String theory has become an essential tool to understand the dynamics of strongly coupled gauge theories and superconformal field theories in various dimensions. Their physics is encoded in the geometric properties of the compactification manifold as well as D-brane data. F-theory \cite{Vafa:1996xn} extends this concept by geometrizing the dynamics of the type IIB axio-dilaton and allows for a deep geometric understanding of string theory vacua including superconformal theories. 

The maximal dimension compatible with non-trivial superconformal theories is six\cite{Seiberg:1996vs,Seiberg:1996qx}, which is also the number of dimensions in which the geometrization through F-theory leads to the strictest constraints. Those insights lead to major progress in superconformal theories with (2,0) and (1,0) supersymmetry, see e.g.\ \cite{Heckman:2013pva,Heckman:2015bfa,Bhardwaj:2015xxa} and references therein. In this way, a classification of six-dimensional superconformal field theories was proposed by gluing together various minimal geometries.

The possibly simplest ingredient in that classification is the so called E-string theory that can be understood as an M5 brane probing one Ho$\check{\text{r}}$ava-Witten E$_8$ wall \cite{Seiberg:1996vs} or in its heterotic dual picture as an heterotic E$_8$ instanton of vanishing size \cite{Ganor:1996mu,Morrison:1996pp}. For a recent study of such transitions in 6d and 4d F-theory models see \cite{Braun:2018ovc}. In the F-theory geometry such an object is a point in the base, where the Weierstrass model has a non-minimal singularity of vanishing order ord$(f,g,\Delta)=(4,6,12)$. Such a singularity cannot be resolved crepantly in the fiber, but requires instead a blow-up in the base with a curve of self intersection $-1$  \cite{Bershadsky:1996nu} leaving a smooth non-reduced fiber, which represents the tensor branch of the theory.

Superconformal theories are always obtained from local geometries where gravity is decoupled, since the Planck scale necessarily breaks conformal invariance. Hence, the aforementioned singularity within a compact geometry \cite{DelZotto:2014fia} does not lead to a superconformal theory, but defines a strongly coupled subsector of the supergravity (SUGRA) theory. Such a coupling to gravity is highly non-trivial and might lead to gaugings of the flavor symmetries or is even completely forbidden by the stringent gravitational anomalies. On the other hand, the compact geometry can also contain Abelian symmetries which can be gauged and coupled to the superconformal sector, as was recently proposed for $(1,0)$ \cite{Lee:2018ihr} and $(2,0)$ \cite{Anderson:2018heq} theories.

Furthermore, E-string theories can be used to connect two theories by superconformal matter transitions \cite{Anderson:2015cqy} with very different matter content, which can lead to exotic matter representations \cite{Cvetic:2015ioa, Klevers:2016jsz}. Turning to models of resolved codimension one fibers, $(4,6,12)$ singularities of the Weierstrass model in codimension two appear as non-flat fibrations \cite{Hu:2000pr,Braun:2011ux,Borchmann:2013hta, Braun:2013nqa, Lawrie:2012gg,Anderson:2016cdu} where the fiber dimension jumps. Furthermore, in the resolved models classified in \cite{Buchmuller:2017wpe}, it was observed that two models were related by transitions where several matter multiplicities in one geometry got exchanged by the additional presence of non-flat fibers in the other.

Since $(4,6,12)$ points admit a tensor branch, we can always obtain a well-defined SUGRA theory and make use of its stringent anomaly constraints in six dimensions. In the simplest type of those transitions, the gauge algebra is left unchanged and matter multiplets are exchanged with tensor multiplets in the global geometry, which is why we call them global tensor-matter transitions. Since these transitions relate F-theory vacua with different matter representations in a non-perturbative way, we want to study which SUGRA consistency conditions have to be satisfied by these global tensor-matter transitions using F-theory. In this way, this work generalizes the transitions of \cite{Anderson:2015cqy} to all semi simple Lie groups. However, we restrict ourselves to smooth divisors without double (or higher) point singularities.

The transitions discussed in this paper are constructed as follows. We start with a torus-fibered Calabi-Yau 3-fold $Y_3$ with gauge algebra 
\begin{align}
\overline{G} = G \times \prod_{i=1}^r \text{U(1)}_i \,,
\end{align}
where $G$ is an arbitrary semi-simple Lie algebra over a smooth base divisor $\Z$. This model has a well-defined supergravity description and therefore has to satisfy all six-dimensional anomaly constraints. By this we mean that the irreducible anomalies vanish identically and the reducible anomalies are accounted for by the Green-Schwarz mechanism involving the $T$ tensor multiplets in the theory \cite{Green:1984bx, Sagnotti:1992qw, Sadov:1996zm}. This leads to constraints for the matter spectrum $\mathcal{S}$ of the theory. In the F-theory formulation all anomaly coefficients can be determined explicitly from the geometric properties of $Y_3$, see e.g.\ \cite{Park:2011ji, Park:2011wv, Morrison:2012ei} and references therein.

Next, we perform a complex structure deformation in $Y_3$ which leads to non-flat fiber points, i.e.\ singularities of vanishing order $\text{ord}(f,g,\Delta) = (4,6,12)$ at the intersection of $\Z$ with another base divisor $D$. As discussed above, this can be understood as coupling a strongly coupled subsector to the supergravity theory. This strong coupling makes it difficult to extract sensible information about the theories and their anomalies directly\footnote{However, we suggest formulae for these transitions in Section~\ref{subsec:SCPtoric}.}, which is why we resolve the strongly coupled points (SCP) by a blow-up in the base. This leads to a modified Calabi-Yau manifold $\tilde{Y}_3$ as well as a change in the number of tensor multiplets $\tilde{T} = T + n_{\text{SCP}}$. Here, we focus on transitions which leave the gauge algebra $\overline{G}$ unchanged. Since the resulting theory has again a well-defined supergravity description, all anomaly constraints have to be satisfied.

Due to the unchanged gauge algebra and the changed number of tensor multiplets, we see that the matter spectrum, i.e.\ the number of hypermultiplets, has to change for the irreducible gravitational anomaly to vanish. Moreover, the specific form of the blow-up in the base via a curve of self-intersection $-1$ allows to uniquely determine the change in the non-Abelian matter spectrum, i.e.\ the matter states which transform non-trivially with respect to $G$. We denote the modified matter spectrum by
\begin{align}
\tilde{\mathcal{S}} = \mathcal{S} + \Delta \mathcal{S}
\label{spectrumchange}
\end{align}
Moreover, since we discuss compact geometries, there might be Abelian factors contained in the full gauge algebra $\overline{G}$.
\begin{figure}
\begin{center}
	\begin{tikzpicture}[scale=1.3]
	\node (A) at (-4,0) [draw,rounded corners,very thick,text width=3cm,align=center] {6d supergravity \\ $(\overline{G}, \mathcal{S}, T)_{Y_3}$
		};
	\draw[very thick, ->] (-2.5,0) -- (-1.2,0);
	\node at (-1.9,0.25) {{\small CS}}; .
	\node at (-1.9,-0.2) {{\small deform.}};
	\node (B) at (0,0) [draw,rounded corners,very thick,text width=2.5cm,align=center] {Theory with SCPs};
	\draw[very thick, ->] (1.2,0) -- (2.5,0);
	\node at (1.9,0.25) {{\small blow-up}};
	\node at (1.9,-0.2) {{\small in base}};
	\node (C) at (4,0) [draw,rounded corners,very thick,text width=3cm,align=center] {6d supergravity \\ $(\overline{G}, \tilde{\mathcal{S}}, \tilde{T})_{\tilde{Y}_3}$
		};
	\end{tikzpicture}
\end{center}
\caption{Theories connected by a global tensor-matter transition.}
\label{SCPtransition}
\end{figure}
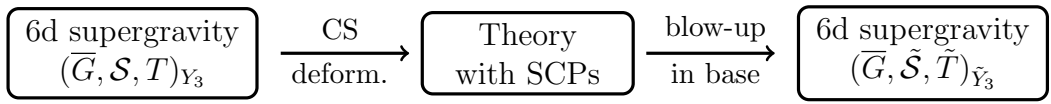
Even though the Abelian anomaly constraints are more difficult to constrain using only the modification of the base space, we can restrict the charges of matter states in $\Delta \mathcal{S}$ under an additional assumption. This assumption can be argued for within the framework of toric hypersurface models and is satisfied in all our models. Therefore, we can fix the non-Abelian part of $\Delta \mathcal{S}$ uniquely and constrain the Abelian charges of non-Abelian and singlet matter. The two steps of the transitions that we investigate in this work are summarized in Figure~\ref{SCPtransition}.

The rest of the paper is organized as follows. In Section~\ref{sec:anomblow} we review the six-dimensional anomaly constraints as well as their connection to the geometric properties of the torus-fibered Calabi-Yau 3-fold in F-theory compactifications. Moreover, we discuss the modification of the geometric properties by the blow-up procedure of the base manifold. Using these considerations, we can derive constraints on the matter spectrum in the tensor-matter transitions under discussion in Section~\ref{sec:transanom}. After two illustrative examples discussing the two interesting non-Abelian grand unified gauge algebras SU(5) and SO(10) in detail, we proceed with a classification of the transitions for all semi-simple Lie algebras. With an additional assumption, we derive restrictions on Abelian charges of states involved in the transition. In Section~\ref{sec:torichyper} we describe an explicit construction of tensor-matter transitions within the framework of toric hypersurface models. In these models we can also justify the additional assumptions for the Abelian charges. In Section~\ref{sec:Examples} we discuss five interesting examples that illustrate the general considerations of the earlier sections. We summarize and conclude this work in Section~\ref{sec:conclusions}. Appendix~\ref{App:speccomp} provides more details on the toric constructions.

\section{Anomalies and blow-up}
\label{sec:anomblow}

In this section we briefly recall the connection between the geometry of the base manifold $B$ and the 6d anomaly coefficients established in \cite{Park:2011ji, Morrison:2012ei} and references therein. We then describe the blow-up procedure in the base $B$ which resolves the SCPs arising by tuning complex structure moduli. This leads in turn to a change in the number of tensor multiplets and, correspondingly, to a modification of anomaly cancellation via the generalized Green-Schwarz mechanism \cite{Green:1984bx, Sagnotti:1992qw, Erler:1993zy, Sadov:1996zm}\footnote{See also \cite{Monnier:2017oqd} for constraints due to the global realization of the gauge algebra.}.

\subsection{Anomalies and base geometry}
\label{subsec:anom}

In six dimensions the anomaly constraints are especially stringent, since pure gravitational anomalies exist \cite{AlvarezGaume:1983ig}. The anomaly polynomial for a 6d supergravity with gauge algebra $\overline{G} = G \times \text{U(1)}^r$, where $G$ is a semi-simple Lie algebra, reads\footnote{We mainly follow the notational conventions of \cite{Park:2011ji, Park:2011wv}.}
\begin{align}
\begin{split}
\mathcal{I}_8 =& - \tfrac{1}{5760} (H - V + 29 T - 273) \left(\text{tr} R^4 + \tfrac{5}{4} (\text{tr} R^2)^2 \right) \\
& - \tfrac{1}{128} (9 - T) (\text{tr} R^2)^2 \\
& - \tfrac{1}{96} \text{tr} R^2 \Big(A_{\text{adj}} - \sum_{\mathbf{R}} n[\mathbf{R}] A_{\mathbf{R}} \Big) \text{tr} F^2 + \tfrac{1}{24} \Big( B_{\text{adj}} - \sum_{\mathbf{R}} n[\mathbf{R}] B_{\mathbf{R}} \Big) \text{tr} F^4 \\
& +\tfrac{1}{24} \Big( C_{\text{adj}} + \sum_{\mathbf{R}} n[\mathbf{R}] C_{\mathbf{R}} \Big) (\text{tr} F^2)^2+ \tfrac{1}{96} \text{tr} R^2 \sum_{i,j} n[q_i,q_j] q_i q_j \, F_i F_j \\
& - \tfrac{1}{6} \sum_{\mathbf{R}, i} n[\mathbf{R},q_i] q_i E_{\mathbf{R}} \, \text{tr} F^3 \, F_i - \tfrac{1}{4} \sum_{\mathbf{R}, i, j} n[\mathbf{R}, q_i, q_j] q_i q_j A_{\mathbf{R}} \, \text{tr} F^2 \, F_i F_j \\
& - \tfrac{1}{24} \sum_{i, j, k, l} n[q_i,q_j,q_k,q_l] q_i q_j q_k q_l \, F_i F_j F_k F_l \,.
\end{split}
\label{anomalypoly}
\end{align}
where $F$ denotes the non-Abelian field strength and $F_i$ the Abelian field strengths with $i \in \{1, \dots, r\}$. The parameters $n[\mathbf{R}]$, $n[q_i]$, and $n[\mathbf{R},q_i]$ take into account the multiplicity of fields transforming in representation $\mathbf{R}$ of $G$ and with charge $q_i$ under $U(1)_i$. The letters $H$, $V$, and $T$ denote the overall number of hyper-, vector-, and tensor multiplets, respectively. Furthermore, we decompose the traces in terms of a reference representation tr using techniques described in~\cite{vanRitbergen:1998pn},
\begin{align}
\begin{split}
\text{tr}_{\mathbf{R}} F^2 &= A_{\mathbf{R}} \, \text{tr} F^2 \,, \quad \text{tr}_{\mathbf{R}} F^3 = E_{\mathbf{R}} \, \text{tr} F^3 \,, \\
\text{tr}_{\mathbf{R}} F^4 &= B_{\mathbf{R}} \, \text{tr} F^4 + C_{\mathbf{R}} (\text{tr} F^2)^2 \,.
\end{split}
\end{align}
The irreducible contributions to the anomaly polynomial, corresponding to the terms proportional to $\text{tr}R^4$, $\text{tr}F^4$, and $\text{tr}F^3$, have to vanish, which leads to the constraints
\begin{align}
\begin{split}
H - V +29 T - 273 &= 0 \,, \\
B_{\text{adj}} - \sum_{\mathbf{R}} n[\mathbf{R}] B_{\mathbf{R}} &= 0 \,, \\
\sum_{\mathbf{R},i} n[\mathbf{R},q_i] q_i E_{\mathbf{R}} &= 0 \,.
\end{split}
\label{irredanomalies}
\end{align}
The remaining anomalies can be cancelled by a generalized version of the Green-Schwarz mechanism involving the $T + 1$ tensor fields. Following \cite{Park:2011ji, Morrison:2012ei}, the factorized anomaly has the form 
\begin{align}
\mathcal{I}_8 = - \tfrac{1}{32} \, \Omega_{\alpha \beta} X^{\alpha} X^{\beta} \,,
\label{facanomaly}
\end{align}
where the individual factors are of the form
\begin{align}
X^{\alpha} = \tfrac{1}{2} a^{\alpha} \, \text{tr} R^2 - \tfrac{2}{\lambda} b^{\alpha} \, \text{tr} F^2 - \sum_{i,j} 2 b^{\alpha}_{ij} \, F_i F_j \,.
\end{align}
The coefficients $\lambda$ depend on the non-Abelian gauge algebra $G$ and are given by
\begin{align*}
\begin{tabular}{| c || c | c | c | c | c | c | c | c | c |}
\hline
$G$ & SU$(N)$ & SO$(2N +1)$ & Sp$(2N)$ & SO$(2N)$ & E$_6$ & E$_7$ & E$_8$ & F$_4$ & G$_2$ \\ \hline
$\lambda$ & 1 & 2 & 1 & 2 & 6 & 12 & 60 & 6 & 2 \\ \hline
\end{tabular}
\end{align*}
The parameters $a^{\alpha}$, $b^{\alpha}$, and $b^{\alpha}_{ij}$ are anomaly coefficients, which are contracted by the SO$(1,T)$ metric $\Omega_{\alpha \beta}$. By matching the factorized form \eqref{facanomaly} to the reducible part of \eqref{anomalypoly}, one finds the defining equations for the anomaly coefficients, cf.\ \cite{Park:2011ji, Park:2011wv},
\begin{align}
\begin{split}
a \cdot a &= 9 - T \,, \\
a \cdot b &= - \tfrac{\lambda}{6} \Big( A_{\text{adj}} - \sum_{\mathbf{R}} n[\mathbf{R}] A_{\mathbf{R}} \Big) \,, \\
a \cdot b_{ij} &= \tfrac{1}{6} \sum_{i,j} n[q_i,q_j] q_i q_j \,, \\
b \cdot b &= - \tfrac{\lambda^2}{3} \Big( C_{\text{adj}} - \sum_{\mathbf{R}} n[\mathbf{R}] C_{\mathbf{R}} \Big) \,, \\
b \cdot b_{ij} &= \lambda \sum_{\mathbf{R},i,j} n[\mathbf{R},q_i,q_j] q_i q_j A_{\mathbf{R}} \,, \\
b_{ij} \cdot b_{kl} + b_{ik} \cdot b_{jl} + b_{il} \cdot b_{jk} &= \sum_{i,j,k,l} n[q_i,q_j,q_k,q_l] q_i q_j q_k q_l \,,
\end{split}
\label{redanomalies}
\end{align}
where the dot product indicates contraction with $\Omega_{\alpha \beta}$.

Part of the beauty of F-theory is that all these coefficients have an interpretation in terms of the geometry of the base manifold $B$ \cite{Park:2011ji, Morrison:2012ei}. First, we choose a basis $\{H_{\alpha}\}$ for the second homology $H_2 (B, \mathbbm{Z})$ of the base $B$. The SO$(1,T)$ metric in \eqref{facanomaly} can then be naturally associated with the intersection matrix on $B$,
\begin{align}
\Omega_{\alpha \beta} = H_{\alpha} \cdot H_{\beta} \,.
\label{intersection}
\end{align}
Moreover, the gravitational anomaly coefficients $a^{\alpha}$ are the coefficients in the expansion of the anti-canonical class $K^{-1}_B$ of the base $B$ in the basis $\{H_{\alpha}\}$ of $H_2 (B, \mathbbm{Z})$, i.e.\
\begin{align}
K^{-1}_B = \sum_{\alpha} a^{\alpha} H_{\alpha} \,.
\end{align}
Similarly, the non-Abelian coefficient $b^{\alpha}$ can be identified with the expansion coefficients of the base divisor $\mathcal{Z}$ carrying the non-Abelian gauge algebra $G$,
\begin{align}
\mathcal{Z} = \sum_{\alpha} b^{\alpha} H_{\alpha} \,.
\end{align}
The interpretation of the Abelian anomaly coefficients $b_{ij}^{\alpha}$ is more complicated due to the global nature of the Abelian gauge algebra factors. In F-theory, these are generated by the free part of the Mordell-Weil group\footnote{Without the zero section, which does not lead to a Abelian gauge algebra.}, i.e.\ the rational sections $s_i$, see \cite{Morrison:2012ei}. In order to evaluate the U(1) charge of a matter field, we need to orthogonalize the U(1) generator with respect to the Cartan subalgebra of $G$, which is done via the Shioda map $\sigma$. The map can be written as
\begin{align}
\sigma(s_i) = S_i - S_0 - (S_i \cdot S_0 \cdot B_{\alpha} + a^{\alpha}) B_{\alpha} + \sum_{I,J} (S_i \cdot \alpha_{I}) (C^{-1}_G)_{IJ} T_{J} \, ,
\label{Shioda}
\end{align}
where $S_0$ and $S_i$ denote the homology class of the zero section $s_0$ and rational sections $s_i$ in $H_4 (Y_3, \mathbbm{Z})$, respectively. The vertical divisors $B_{\alpha}$ are defined via
\begin{align}
B_{\alpha} = \pi^{-1} (H_{\alpha}) \,,
\end{align}
with $\pi: Y_3 \rightarrow B$ the projection to the base. The fibral divisors $T_{J}$ can be obtained by fibering the complex fiber curve $\alpha_J$, corresponding to a simple root of the non-Abelian gauge algebra $G$, over the base divisor $\mathcal{Z}$. The inverse Cartan matrix of $G$ is denoted by $C^{-1}_G$. The Shioda map is constructed in such a way that
\begin{align}
- \pi (\sigma(s_i) \cdot C) = 0 \,, \quad \text{for} \quad C \in \text{span}( \{ S_0, B_{\alpha}, T_{J}\}) \,,
\label{orthoshioda}
\end{align}
where the map
\begin{align}
- \pi (\, \cdot \,): H_4(Y_3) \times H_4(Y_3) \rightarrow H_2(B) \,,
\end{align}
is often called height-pairing, see e.g.\ \cite{Wazir:2001dec, shioda1990}. In terms of these quantities, the Abelian anomaly coefficients can be written as the height-pairing of the Shioda maps,
\begin{align}
- \pi (\sigma(s_i) \cdot \sigma(s_j)) = \sum_{\alpha} b^{\alpha}_{ij} H_{\alpha} \,.
\end{align}
For the Abelian anomaly coefficients we can make use of the orthogonality condition \eqref{orthoshioda} of the Shioda map \eqref{Shioda} to write the height-pairing as
\begin{align}
b^{\alpha}_{ij} H_{\alpha} = - \pi \big(S_i \cdot \sigma(s_j)\big) = - \pi \big( \sigma(s_i) \cdot S_j \big) \,.
\end{align}
Moreover, for an arbitrary rational section $s_i$, we have \cite{Morrison:2012ei}
\begin{align}
- \pi (S_i \cdot S_i) = K_B^{-1} \,, \quad - \pi (S_i \cdot T_J) = - (S_i \cdot \alpha_J) \mathcal{Z} \,, \quad - \pi (S_i \cdot B_{\alpha}) = - H_{\alpha} \,.
\end{align}
For the anomaly coefficients involving the same U(1) factor we hence find
\begin{align}
b_{ii}^{\alpha} H_{\alpha} = 2 a^{\alpha} H_{\alpha} + 2 \,  \pi (S_i \cdot S_0) - c_{ii} b^{\alpha} H_{\alpha} \,,
\label{samehp}
\end{align}
with
\begin{align}
c_{ij} \equiv \sum_{I,J} (S_i \cdot \alpha_{I}) (C^{-1}_G)_{IJ} (S_j \cdot \alpha_J) \,.
\end{align}
Similarly, for distinct U(1) factors we have
\begin{align}
\begin{split}
b^{\alpha}_{ij} H_{\alpha} &= a^{\alpha} H_{\alpha} - \pi (S_i \cdot S_j) + \pi (S_i \cdot S_0) + \pi (S_j \cdot S_0) - c_{ij} b^{\alpha} H_{\alpha} \\
&\equiv \big( a^{\alpha} - \sigma^{\alpha}_{ij} + \sigma^{\alpha}_{i0} + \sigma^{\alpha}_{j0} - c_{ij} b^{\alpha} \big) H_{\alpha} \,.
\end{split}
\label{distincthp}
\end{align}
This concludes our revision of the connection between anomaly coefficients and the geometry of the base.

\subsection{Base blow-up}
\label{subsec:blow}

In the previous section, we described how the anomaly coefficients can be understood in terms of the second homology of the base $B$. However, in the transitions under investigation, we perform a blow-up in the base in order to resolve the SCPs. This leads to a modification of the geometry due to the blown-up base $\tilde{B}$. Here, we discuss the explicit blow-up procedure and the consequences for the anomaly coefficients, whose change can be constrained by the blow-up \cite{Bershadsky:1996nu}.

Recall that the SCPs we focus on arise after a complex structure deformation at the intersections of the discriminant locus $\mathcal{Z}$ with another base divisor $D$, the latter of which carries no non-Abelian gauge algebra. Consequently, the blow-up has to resolve these intersection points and introduces additional divisors $E_a$ with $a \in \{1,\dots, k\}$, schematically depicted in Figure~\ref{blowup}.
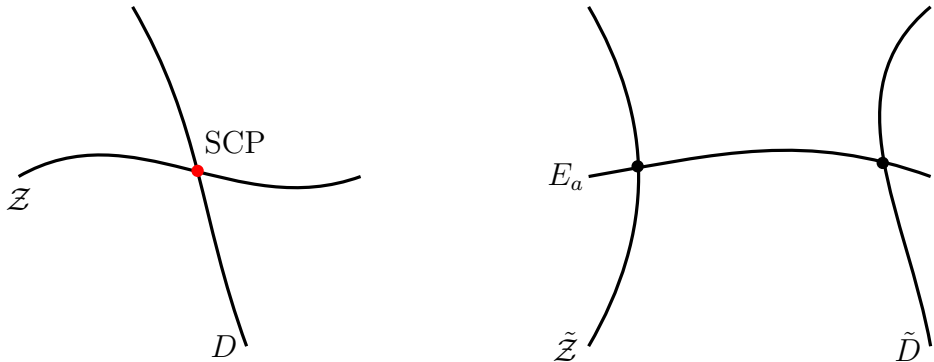
\begin{figure}
\begin{center}
\begin{tikzpicture}[scale=1.5]
	\draw[very thick] (-4,0) to[out = 30, in = 200] (-1,0);
	\draw[very thick] (-3,1.5) to[out = -60, in = 110] (-2,-1.5);
	\draw[very thick, red, fill=red] (-2.43,0.05) circle (0.04);
	\node at (-4,-0.2) {$\mathcal{Z}$};
	\node at (-2.2, -1.5) {$D$};
	\node at (-2.1,0.3) {SCP};
	\draw[very thick] (1,1.5) to[out = -60, in = 60] (1,-1.5);
	\draw[very thick] (1,0) to[out = 10, in = 160] (4,0);
	\draw[very thick] (4,1.5) to[out = 220, in = 100] (4,-1.5);
	\draw[very thick, fill = black] (1.43,0.09) circle (0.04);
	\draw[very thick, fill = black] (3.58, 0.12) circle (0.04);
	\node at (0.8,-1.5) {$\tilde{\mathcal{Z}}$};
	\node at (0.8,0) {$E_a$};
	\node at (3.8,-1.5) {$\tilde{D}$};
\end{tikzpicture}
\end{center}
\caption{Resolution of SCPs in codimension-two.}
\label{blowup}
\end{figure}
This resolution reduces the vanishing order of the singularity by $\text{ord}(4,6,12)$, which, since we restrict to $(4,6,12)$ singularities, leads to a smooth fiber over all exceptional divisors $E_a$ \cite{Aspinwall:1997ye}. We indicate the fact that the base divisors get modified by a tilde. Moreover, this procedure affects our choice of basis for the base homology $H_2 (\tilde{B},\mathbbm{Z})$. To account for that, we define the blow-down map $\beta: \tilde{B} \rightarrow B$, which, via push-forward, leads to a map of the second homology,
\begin{align}
\beta_{\ast}: H_2 (\tilde{B}, \mathbbm{Z}) \rightarrow H_2 (B, \mathbbm{Z}) \,.
\end{align}
For the basis of $H_2 (B, \mathbbm{Z})$ defined above, we define the full pre-image as
\begin{align}
\beta_{\ast}^{-1} (H_{\alpha}) = \tilde{H}_{\alpha} + \sum_a h^a_{\alpha} E_a \,,
\label{homdecomp}
\end{align}
where the coefficients $h^{\alpha}_a$ are non-negative and the new $\tilde{H}_{\alpha}$ are irreducible. In this way, we find a basis for $H_2 (\tilde{B},\mathbbm{Z})$ in terms of $\{\tilde{H}_{\alpha}\}$ and $\{E_a\}$, which we collectively denote by $\{\tilde{H}_A\}$ with
\begin{align}
\tilde{H}_{A} = \tilde{H}_{\alpha} \quad \text{for} \quad A = \alpha \,, \quad \tilde{H}_A = E_a \quad \text{for} \quad A = a \,.
\end{align}
Importantly, the map $\beta$ respects the intersection product \cite{Bershadsky:1996nu}, which means that for two divisors $D$ and $D'$ in $H_2 (B, \mathbbm{Z})$, we have
\begin{align}
\beta_{\ast}^{-1} (D) \cdot \beta^{-1}_{\ast} (D') |_{\tilde{B}} = D \cdot D' |_B \,,
\end{align}
where we specified the manifold on which the intersection is evaluated, which will be omitted in the following. More specifically, this also implies that the exceptional divisors $\{E_a\}$ generate the kernel of $\beta_{\ast}$, i.e.\ for all $D \in H_2 (B, \mathbbm{Z})$ and $E_a$ we have
\begin{align}
\beta_{\ast}^{-1} (D) \cdot E_a = 0 \,.
\end{align}
Moreover, the exceptional divisors considered in the transitions have the intersection form
\begin{align}
E_a \cdot E_b = - \delta_{ab} \,.
\label{excintersection}
\end{align}
This changes if one resolves higher order singularities \cite{Aspinwall:1997ye}. Moreover, since we restrict to SCPs on single transverse intersections of $D$ and $\mathcal{Z}$, we need to introduce exactly one exceptional divisor for each SCP, so that $n_{\text{SCP}} = k$.

With these properties we can investigate the change in the anomaly constraints induced by the blow-up in the base.

\subsubsection*{Intersection form}

First, we derive the new intersection form $\tilde{\Omega}_{AB}$ on the base $\tilde{B}$ using the definition of the basis for $H_2 (\tilde{B}, \mathbbm{Z})$ and the properties of the blow-down map $\beta$ discussed above. In analogy to \eqref{intersection}, the new intersection form is defined as
\begin{align}
\tilde{\Omega}_{AB} = \tilde{H}_A \cdot \tilde{H}_B \,.
\end{align}
With the intersection of the exceptional divisors given by \eqref{excintersection}, we readily see that
\begin{align}
\tilde{\Omega}_{ab} = E_a \cdot E_b = - \delta_{ab}\,.
\end{align}
Additionally, since the exceptional divisors $E_a$ are in the kernel of $\beta_{\ast}$, we find\footnote{Summation over repeated indices is implied from now on.}
\begin{align}
0 = \beta^{-1}_{\ast} (H_{\alpha}) \cdot E_a = (\tilde{H}_{\alpha} + h^{a}_{\alpha} E_a) \cdot E_b = \tilde{\Omega}_{\alpha b} - h^a_{\alpha}  \delta_{ab} \,,
\end{align}
which leads to the identification
\begin{align}
\tilde{\Omega}_{\alpha b} = h_{\alpha}^a \delta_{ab} \,.
\end{align}
Finally, we can deduce the elements $\tilde{\Omega}_{\alpha \beta}$ from
\begin{align}
\begin{split}
\Omega_{\alpha \beta} &= H_{\alpha} \cdot H_{\beta} = \beta^{-1}_{\ast} (H_{\alpha}) \cdot \beta^{-1}_{\ast} (H_{\beta}) = (\tilde{H}_{\alpha} + h^a_{\alpha} E_a) \cdot (\tilde{H}_{\beta} + h^b_{\beta} E_b) \\
&= \tilde{\Omega}_{\alpha \beta} + h^a_{\alpha} h^a_{\beta} \,.
\end{split}
\end{align}
Hence, we can summarize the new intersection matrix on $\tilde{B}$ as
\begin{align}
\tilde{\Omega}_{AB} = \begin{pmatrix} \Omega_{\alpha \beta} - h^a_{\alpha} h^a_{\beta} & h^a_{\alpha} \\ h^b_{\beta} & - \delta_{ab} \end{pmatrix} \,.
\end{align}
This further defines the new SO$(1, \tilde{T})$ metric appearing in the anomaly cancellation on $\tilde{B}$. We can already anticipate that $\tilde{T} = T + k$, where $k$ is the number of exceptional divisors, from the rank of the intersection matrix $\tilde{\Omega}_{AB}$.

\subsubsection*{Gravitational anomaly coefficient}

As discussed in Section \ref{subsec:anom}, the gravitational anomaly coefficient is related to the anti-canonical class of the base space. Since we performed a blow-up from $B$ to $\tilde{B}$, we expect the coefficients to change accordingly. The very specific type of blow-up allows us to determine the gravitational coefficients $\tilde{a}^A$ on $\tilde{B}$ in terms of $a^{\alpha}$ on $B$ and the coefficients in the decomposition \eqref{homdecomp}. Using the relation \cite{Bershadsky:1996nu}
\begin{align}
\begin{split}
K^{-1}_{\tilde{B}} &= \beta^{-1}_{\ast} (K^{-1}_B) - \sum_a E_a = a^{\alpha} \tilde{H}_{\alpha} + a^{\alpha} h^{a}_{\alpha} E_a - \sum_a E_a \\
&= a^{\alpha} \tilde{H}_{\alpha} + (a^{\alpha} h^a_{\alpha} - 1) E_a = \tilde{a}^A \tilde{H}_A \,,
\end{split}
\end{align}
we read off
\begin{align}
\tilde{a}^A = \begin{pmatrix} \tilde{a}^{\alpha} \\ \tilde{a}^A \end{pmatrix} = \begin{pmatrix} a^{\alpha} \\ a^{\alpha} h^a_{\alpha} - 1 \end{pmatrix} \,.
\end{align}
From the anomaly constraints \eqref{redanomalies} we can now calculate the number of tensor multiplets $\tilde{T}$ on the blown-up base $\tilde{B}$,
\begin{align}
\begin{split}
\tilde{T} &= 9 - \tilde{a} \cdot \tilde{a} \\
&= 9 - (\Omega_{\alpha \beta} - h^a_{\alpha} h^a_{\beta}) a^{\alpha} a^{\beta} - 2 h^a_{\alpha} a^{\alpha} (h^a_{\beta} a^{\beta} - 1) + \delta_{ab} (h^a_{\alpha} a^{\alpha} - 1)(h^b_{\beta} a^{\beta} - 1) \\
&= 9 - a \cdot a + \delta_{aa} = T + k = T + n_{\text{SCP}} \,,
\end{split}
\label{tensorchange}
\end{align}
which verifies our expectation.

\subsubsection*{Non-Abelian anomaly coefficient}

The non-Abelian anomaly coefficients are defined via the base divisor $\mathcal{Z}$ over which the fiber degenerates in a way determined by the non-Abelian gauge algebra $G$. Since we restrict to singularities of vanishing order $(4,6,12)$, we know that the fiber over all blow-up divisors $E_a$ is smooth and in particular does not lead to any non-Abelian gauge algebras. Consequently, the modified base divisor $\tilde{\mathcal{Z}} = \tilde{b}^A \tilde{H}_A$ that carries the non-Abelian gauge algebra $G$ does not contain any exceptional divisors and, due to the decomposition \eqref{homdecomp}, we can identify the new non-Abelian anomaly coefficients as
\begin{align}
\tilde{b}^A = \begin{pmatrix} \tilde{b}^{\alpha} \\ \tilde{b}^a \end{pmatrix} = \begin{pmatrix} b^{\alpha} \\ 0 \end{pmatrix} \,.
\end{align}
With this identification, we are able to calculate the change of the genus of $\tilde{\mathcal{Z}}$ from the genus of $\Z = b^{\alpha} H_{\alpha}$, which is given by
\begin{align}
g_{\mathcal{Z}} = 1 - \tfrac{1}{2} \, a \cdot b + \tfrac{1}{2} \, b \cdot b \,,
\label{genuseq}
\end{align}
and the blow-up data. The genus of the base divisor carrying a non-Abelian gauge algebra counts the number of hypermultiplets in the adjoint representation. The genus of $\tilde{\Z}$ is consequently given by
\begin{align}
g_{\tilde{\mathcal{Z}}} &= 1 - \tfrac{1}{2} \, \tilde{a} \cdot \tilde{b} + \tfrac{1}{2} \, \tilde{b} \cdot \tilde{b} = 1 - \tfrac{1}{2} \, a \cdot b + \tfrac{1}{2} \, b \cdot b + \sum_a h^a_{\alpha} b^{\alpha} - h^a_{\alpha} h^a_{\beta} b^{\alpha} b^{\beta} \\
&= g_{\mathcal{Z}} +  \sum_a h^a_{\alpha} b^{\alpha} - h^a_{\alpha} h^a_{\beta} b^{\alpha} b^{\beta} \,.
\label{newgenus}
\end{align}
As we shall see now, we can further evaluate and simplify this expression by considering the intersection with the divisors that lead to SCPs.

\subsubsection*{Superconformal points}

For the transitions we discuss here we have $n_{\text{SCP}} = k$, i.e.\ we have to introduce one exceptional divisor for each SCP in the theory. The number of SCPs after complex structure deformation is obtained via the intersection number 
\begin{align}
n_{\text{SCP}} = \mathcal{Z} \cdot D = b \cdot d \,,
\end{align}
where we have decomposed $D$ as $D = d^{\alpha} H_{\alpha}$. After a resolution in the base, we find the coefficients for $\tilde{D}$ (remember that the fiber over $E_a$ is smooth) similar to $\tilde{\mathcal{Z}}$ above,
\begin{align}
\tilde{d}^A = \begin{pmatrix} \tilde{d}^{\alpha} \\ \tilde{d}^a \end{pmatrix} = \begin{pmatrix} d^{\alpha} \\ 0 \end{pmatrix} \,.
\label{SCPcoef}
\end{align}
Demanding that all SCPs are resolved, we find
\begin{align}
\tilde{\mathcal{Z}} \cdot \tilde{D} = \tilde{b} \cdot \tilde{d} = b \cdot d - h^a_{\alpha} h^a_{\beta} b^{\alpha} d^{\beta} = n_{\text{SCP}} - h^a_{\alpha} h^a_{\beta} b^{\alpha} d^{\beta} = 0 \,. 
\label{SCPintersec}
\end{align}
Moreover, since each SCP is resolved by an individual blow-up divisor in the transitions under discussion, and since the intersection product is preserved under the blow-down map, it follows that
\begin{align}
\tilde{\mathcal{Z}} \cdot E_a = h^a_{\alpha} b^{\alpha} = 1 \,, \quad \tilde{D} \cdot E_a = h^a_{\alpha} d^{\alpha} = 1 \,,
\end{align}
for all $a \in \{1, \dots, k\}$. Together with \eqref{SCPintersec}, this shows that $n_{\text{SCP}} = k$.

Furthermore, using that $h^a_{\alpha} b^{\alpha} = 1$, we can evaluate the genus of $\tilde{\mathcal{Z}}$, given in \eqref{newgenus}, to find
\begin{align}
g_{\tilde{\mathcal{Z}}} = g_{\mathcal{Z}} + \sum_a 1 - k = g_{\mathcal{Z}} \,.
\end{align}
This means that the genus of the base divisor carrying the non-Abelian part of the gauge algebra remains unchanged, as does the multiplicity of the matter states in the adjoint representation.

\subsubsection*{Matter multiplicities}

Similar to the SCPs, the multiplicities of matter transforming non-trivially with respect to the non-Abelian gauge algebra $G$ is obtained by intersecting $\mathcal{Z}$ with another divisor $D_{\mathbf{R}_i}$, where the index $i$ accounts for the possibility of matter transforming in the same representation $\mathbf{R}$ but with different U(1) charges,
\begin{align}
n[\mathbf{R}_i] = \mathcal{Z} \cdot D_{\mathbf{R}_i} = b \cdot d_{\mathbf{R}_i} \,.
\end{align}
For the transitions under investigation there is no non-Abelian matter that arises at the intersection of $\mathcal{Z}$ with the exceptional divisors $E_a$. Then, similarly to \eqref{SCPcoef}, the modified divisors $\tilde{D}_{\mathbf{R}_i}$ have coefficients
\begin{align}
\tilde{d}^{A}_{\mathbf{R}_i} = \begin{pmatrix} \tilde{d}^{\alpha}_{\mathbf{R}_i} \\ \tilde{d}^{a}_{\mathbf{R}_i} \end{pmatrix} = \begin{pmatrix} d^{\alpha}_{\mathbf{R}_i} \\ 0 \end{pmatrix} \,.
\end{align}
Accordingly, the new matter multiplicities are given by
\begin{align}
\tilde{n} [\mathbf{R}_i] = \tilde{\mathcal{Z}} \cdot \tilde{D} = \tilde{b} \cdot \tilde{d}_{\mathbf{R}_i} = b \cdot d_{\mathbf{R}_i} - h^a_{\alpha} h^{a}_{\beta} b^{\alpha} d^{\beta}_{\mathbf{R}_i} = n [\mathbf{R}_i] - \sum_a h^a_{\alpha} d^{\alpha}_{\mathbf{R}_i} \,.
\end{align}
However, since the intersections
\begin{align}
\tilde{D}_{\mathbf{R}_i} \cdot E_a = h^a_{\alpha} d^{\alpha}_{\mathbf{R}_i} \geq 0 \,,
\end{align}
are non-negative, we conclude that
\begin{align}
\Delta n [\mathbf{R}_i] = \tilde{n} [\mathbf{R}_i] - n [\mathbf{R}_i] = - \sum_a h^a_{\alpha} d^{\alpha}_{\mathbf{R}_i} \leq 0 \,.
\label{mattermultchange}
\end{align}
Hence, in the process of the tensor-matter transitions we are considering, the multiplicity of non-Abelian matter is not increased.

\subsubsection*{Abelian anomaly coefficients}

After the complex structure deformation leading to the SCPs and their subsequent resolution in the base, the 4-cycle $S_i$ corresponding to the rational section $s_i$ might in general change;  see Figure~\ref{Abelblow} for a schematic depiction. We denote the element in $H_4 (\tilde{Y}_3)$ corresponding to $s_i$ by $\tilde{S}_i$. Accordingly, the Shioda map gets modified to
\begin{align}
\tilde{\sigma}(s_i) = \tilde{S}_i - \tilde{S}_0 - (\tilde{S}_i \cdot \tilde{S}_0 \cdot \tilde{B}_{A} + \tilde{a}^A) \tilde{B}_A + \sum_{I,J} (\tilde{S}_i \cdot \alpha_I) (C^{-1}_G)_{IJ} \tilde{T}_J \,,
\end{align}
with $\tilde{B}_{A} = \pi^{-1} (\tilde{H}_A)$ the new vertical divisors on $\tilde{Y}_3$. Nevertheless, the new Shioda map has to satisfy the analogous orthonormality condition \eqref{orthoshioda} on the resolved geometry. Again, the height-pairings of identical \eqref{samehp} and distinct \eqref{distincthp} U(1) factors read
\begin{align}
\begin{split}
\tilde{b}^A_{ii} \tilde{H}_A &= 2 \tilde{a}^A \tilde{H}_A + 2 \, \pi (\tilde{S}_i \cdot \tilde{S}_0) - \tilde{c}_{ii} \tilde{b}^A \tilde{H}_A \,, \\
\tilde{b}^{A}_{ij} \tilde{H}_A &= \tilde{a}^{A} \tilde{H}_A - \pi(\tilde{S}_i \cdot \tilde{S}_j) + \pi (\tilde{S}_i \cdot \tilde{S}_0) + \pi (\tilde{S}_j \cdot \tilde{S}_0) - \tilde{c}_{ij} \tilde{b}^A \tilde{H}_A \,,
\end{split}
\end{align}
with
\begin{align}
\tilde{c}_{ij} = \sum_{I,J} (\tilde{S}_i \cdot \alpha_I) (C^{-1}_G)_{IJ} (\tilde{S}_j \cdot \alpha_J) \,.
\end{align}
In general, all contributions of the form $\pi (\tilde{S}_i \cdot \tilde{S}_j)$, $\pi (\tilde{S}_i \cdot \tilde{S}_0)$, and $\tilde{c}_{ij}$ might change due to the complex structure deformations and the base blow-up.
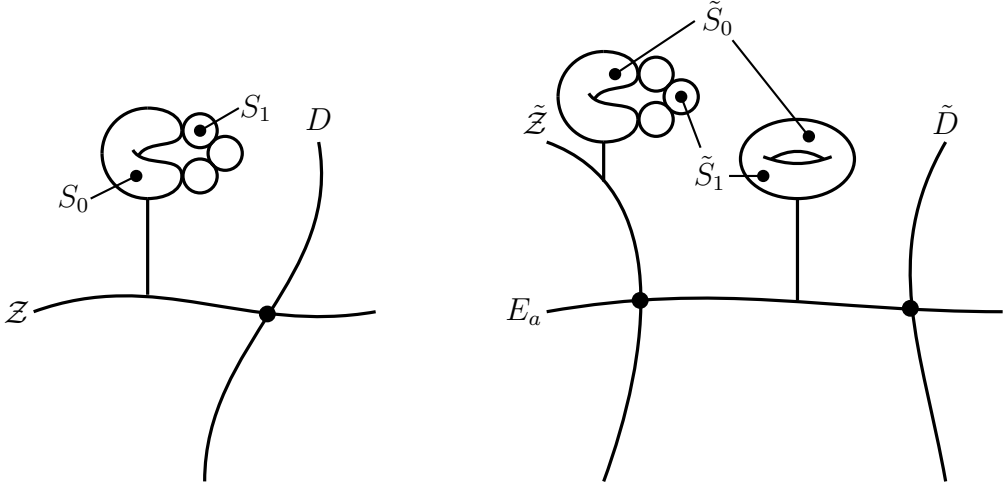
\begin{figure}
\begin{center}
\begin{tikzpicture}[scale=1.5]
	\draw[very thick] (-4,0) to[out = 20, in = 190] (-1,0);
	\node at (-4.15,0) {$\mathcal{Z}$};
	\draw[very thick] (-2.5,-1.5) to[out = 90, in = -80] (-1.5,1.5);
	\node at (-1.5,1.7) {$D$};
	\draw[very thick, fill = black] (-1.95,-0.02) circle (0.06);
	\draw[very thick] (-3,0.15) -- (-3,1);
	\draw[very thick] (-3.4,1.4) to[out = -90, in = 180] (-3,1);
	\draw[very thick] (-3.4,1.4) to[out = 90, in = 180] (-3,1.8);
	\draw[very thick] (-3,1.8) to[out = 0, in = 90] (-2.7,1.6);
	\draw[very thick] (-3,1) to[out = 0, in = -90] (-2.7,1.2);
	\draw[very thick] (-2.7,1.6) to[out = -90, in = 60] (-3.08,1.38);
	\draw[very thick] (-2.7, 1.2) to[out = 90, in = -60] (-3.13,1.43);
	\draw[very thick] (-2.54, 1.6) circle (0.15);
	\draw[very thick] (-2.54, 1.2) circle (0.15);
	\draw[very thick] (-2.32,1.4) circle (0.15);
	\draw[very thick, fill = black] (-3.1,1.2) circle (0.04);
	\draw[very thick, fill = black] (-2.54,1.6) circle (0.04);
	\draw[thick] (-3.1,1.2) -- (-3.5,1.0);
	\node at (-3.65,1) {$S_0$};
	\draw[thick] (-2.54,1.6) -- (-2.2,1.8);
	\node at (-2.05,1.8) {$S_1$};
	\draw[very thick] (0.5,1.5) to[out = -20, in = 70] (1,-1.5);
	\node at (0.4,1.7) {$\tilde{\mathcal{Z}}$};
	\draw[very thick] (4,1.5) to[out = 240, in = 100] (4,-1.5);
	\node at (4,1.7) {$\tilde{D}$};
	\draw[very thick] (0.5,0) to[out = 10, in = 180] (4.5,0);
	\draw[very thick, fill = black] (1.32,0.1) circle (0.06);
	\draw[very thick, fill = black] (3.69,0.03) circle (0.06);
	\node at (0.3,0) {$E_a$};
	\draw[very thick] (1,1.15) -- (1,1.5);
	\draw[very thick] (0.6,1.9) to[out = -90, in = 180] (1,1.5);
	\draw[very thick] (0.6,1.9) to[out = 90, in = 180] (1,2.3);
	\draw[very thick] (1,2.3) to[out = 0, in = 90] (1.3,2.1);
	\draw[very thick] (1,1.5) to[out = 0, in = -90] (1.3,1.7);
	\draw[very thick] (1.3,2.1) to[out = -90, in = 60] (0.92,1.88);
	\draw[very thick] (1.3, 1.7) to[out = 90, in = -60] (0.87,1.93);
	\draw[very thick] (1.46, 2.1) circle (0.15);
	\draw[very thick] (1.46, 1.7) circle (0.15);
	\draw[very thick] (1.68, 1.9) circle (0.15);
	\draw[very thick, fill = black] (1.1,2.1) circle (0.04);
	\draw[thick] (1.1,2.1) -- (1.8,2.57);
	\node at (2,2.6) {$\tilde{S}_0$};
	\draw[very thick, fill = black] (1.68, 1.9) circle (0.04);
	\draw[thick] (1.68,1.9) -- (1.85,1.45);
	\node at (1.93,1.23) {$\tilde{S}_1$};
	\draw[very thick] (2.7,0.1) -- (2.7,1);
	\draw[very thick] (2.7,1.35) ellipse (0.5 and 0.35);
	\draw[very thick] (2.4,1.37) to[out = -20, in = 200] (3,1.37); 
	\draw[very thick] (2.47,1.35) to[out = 30, in = 150] (2.93,1.35);
	\draw[very thick, fill = black] (2.4,1.2) circle (0.04);
	\draw[very thick, fill = black] (2.8,1.55) circle (0.04);
	\draw[thick] (2.4,1.2) -- (2.1,1.2);
	\draw[thick] (2.8,1.55) -- (2.13,2.4);
\end{tikzpicture}
\end{center}
\caption{Schematic picture of the change in rational sections over blow-up divisors with gauge group $\overline{G} = \text{SU}(4) \times \text{U}(1)$.}
\label{Abelblow}
\end{figure}
Hence, the modified Abelian anomaly coefficients read
\begin{align}
\tilde{b}^{A}_{ij} = \begin{pmatrix} a^{\alpha} - \tilde{\sigma}^{\alpha}_{ij} + \tilde{\sigma}^{\alpha}_{i0} + \tilde{\sigma}^{\alpha}_{j0} - \tilde{c}_{ij} b^{\alpha} \\ (a^{\alpha} h^a_{\alpha} - 1) + \Delta b^a_{ij} \end{pmatrix} \,, \quad \tilde{b}^{A}_{ii} = \begin{pmatrix} 2 a^{\alpha} + 2 \tilde{\sigma}^{\alpha}_{i0} - \tilde{c}_{ii} b^{\alpha} \\ 2 (a^{\alpha} h^a_{\alpha} - 1) + \Delta b^a_{ii} \end{pmatrix} \,.
\end{align}
We write the changes in the components $\tilde{b}^{\alpha}_{ij}$ and $\tilde{b}^{\alpha}_{ii}$, which are not fixed by the blow-up procedure, as
\begin{align}
\tilde{b}^A_{ij} = \begin{pmatrix} b^{\alpha}_{ij} + \Delta b^{\alpha}_{ij} \\ (a^{\alpha} h^a_{\alpha} -1) + \Delta b^a_{ij}  \end{pmatrix} \,, \quad \tilde{b}^A_{ii} = \begin{pmatrix} b^{\alpha}_{ii} + \Delta b^{\alpha}_{ii} \\ 2 (a^{\alpha} h^a_{\alpha} - 1) + \Delta b^a_{ii} \end{pmatrix} \,,
\label{Abeliananomcoef}
\end{align}
which will lead to constraints on the U(1) charges of matter involved in the tensor-matter transitions under an additional assumption.

\section{Tensor-matter transitions and anomalies}
\label{sec:transanom}

In this section we work out the constraints imposed by the absence of anomalies in transitions of the type described above. After working out the general formulae and elucidating our procedure in two specific examples, we proceed with a classification of tensor-matter transitions for an arbitrary semi-simple non-Abelian gauge algebra located on $\mathcal{Z}$.

We explicitly determine the allowed changes in the non-Abelian matter spectrum in all cases for a single SCP. They are fixed uniquely for the type of transitions discussed. Moreover, under an additional assumption we can restrict the Abelian charges for the matter states involved in the transition. Even though the Abelian constraints are not strong enough to fix the change in the matter spectrum uniquely, it is often enough to know the charges of the non-Abelian states in order to predict the full modification of the hypermultiplet sector.

\subsection{General constraints}
\label{subsec:genconst}

Resolving $n_{\text{SCP}} = k$ SCPs leads to the appearance of $k$ additional tensor multiplets \eqref{tensorchange},
\begin{align}
\tilde{T} = T + k \,.
\end{align}
Since the gauge algebra $\overline{G}$, and consequently the number of vector multiplets, does not change during the transitions, we see that the irreducible gravitational anomaly in \eqref{irredanomalies} dictates that
\begin{align}
\tilde{H} = H - 29 k \,.
\end{align}
So the matter spectrum has to change and 29 degrees of freedom per SCP in the hypermultiplet sector have to disappear. Since the change in the matter multiplicities \eqref{mattermultchange} is negative, we conclude that only representations with dimension $\text{dim} (\mathbf{R}) < 29$ can be involved\footnote{Note that there is always at least one degree of freedom in the uncharged singlet sector due to the tuning in the complex structure. Moreover, for (pseudo)-real representations, one can have half-hypermultiplets which leads to $\text{dim}(\mathbf{R}) < 58$. We will include these cases in the following discussion. These states are then necessarily uncharged with respect to Abelian gauge factors.}. Moreover, we know that the remaining irreducible anomalies vanish for the spectrum $\mathcal{S}$ as well as for the spectrum $\tilde{\mathcal{S}}$ after the resolution in the base manifold. With the definition in \eqref{spectrumchange} and $\tilde{n} = n + \Delta n$ this implies
\begin{align}
\begin{split}
\sum_{\mathbf{R}} \Delta n[\mathbf{R}] B_{\mathbf{R}} &= 0 \,, \\
\sum_{\mathbf{R},i} \Delta n[\mathbf{R}, q_i] q_i E_{\mathbf{R}} &= 0 \,.
\end{split}
\label{specirred}
\end{align}
For the reducible anomalies, the consistency constraints are more interesting, since the presence of an additional tensor multiplet also extends the possibilities of anomaly cancellation via the Green-Schwarz mechanism. Nevertheless, the specific form of the blow-up in connection with a restriction to $(4,6,12)$ singularities, discussed in Section~\ref{sec:anomblow}, allows to constrain possible transitions significantly.

The reducible gravitational anomaly is indeed canceled by the modified anomaly coefficients, since we have seen in \eqref{tensorchange} that
\begin{align}
\tilde{a} \cdot \tilde{a} = 9 - \tilde{T} = 9 - T - k\,.
\end{align}
The reducible non-Abelian anomaly on the blown-up base $\tilde{B}$ demands
\begin{align}
\tilde{b} \cdot \tilde{b} = b \cdot b - h^a_{\alpha} h^a_{\beta} b^{\alpha} b^{\beta} = b \cdot b - k \,,
\end{align}
leading to the following constraint on $\Delta \mathcal{S}$
\begin{align}
\tilde{b} \cdot \tilde{b} - b \cdot b = \tfrac{\lambda^2}{3} \sum_{\mathbf{R}} \Delta n [\mathbf{R}] C_{\mathbf{R}} = - k \,.
\label{changenonA}
\end{align}
In the same way, the cancellation of mixed gravitational anomalies leads to
\begin{align}
\tilde{a} \cdot \tilde{b} - a \cdot b = \tfrac{\lambda}{6} \sum_{\mathbf{R}} \Delta n[\mathbf{R}] A_{\mathbf{R}} = - k \,.
\label{changemixed}
\end{align}
These two equations, together with the irreducible non-Abelian anomaly above, fix the change in the non-Abelian matter spectrum uniquely, as we will see below.

For the Abelian part of the gauge algebra, the restrictions are less severe because the change in the Shioda map is in general not determined by the blow-up procedure in the base alone. Using the expressions \eqref{Abeliananomcoef} for the modified Abelian anomaly coefficients, we can calculate the new mixed anomalies
\begin{align}
\begin{split}
\tilde{b}_{ij} \cdot \tilde{b} &= b_{ij} \cdot b + \Omega_{\alpha \beta} \Delta b^{\alpha}_{ij} b^{\beta} - \sum_a \big( h^a_{\alpha} (b^{\alpha}_{ij} + \Delta b^{\alpha}_{ij}) - a^{\alpha} h^a_{\alpha} + 1 - \Delta b^a_{ij} \big) \,, \\
\tilde{b}_{ii} \cdot \tilde{b} &= b_{ii} \cdot b + \Omega_{\alpha \beta} \Delta b^{\alpha}_{ii} b^{\beta} - \sum_a \big( h^a_{\alpha} (b^{\alpha}_{ii} + \Delta b^{\alpha}_{ii}) - 2 a^{\alpha} h^a_{\alpha} + 2 - \Delta b^a_{ii} \big) \,, \\
\tilde{b}_{ij} \cdot \tilde{a} &= b_{ij} \cdot a + \Omega_{\alpha \beta} \Delta b^{\alpha}_{ij} a^{\beta} - \sum_a \big( h^a_{\alpha} (b^{\alpha}_{ij} + \Delta b^{\alpha}_{ij}) - a^{\alpha} h^a_{\alpha} + 1 - \Delta b^a_{ij} \big) \,, \\
\tilde{b}_{ii} \cdot \tilde{a} &= b_{ii} \cdot a + \Omega_{\alpha \beta} \Delta b^{\alpha}_{ii} a^{\beta} - \sum_a \big( h^a_{\alpha} (b^{\alpha}_{ii} + \Delta b^{\alpha}_{ii}) - 2 a^{\alpha} h^a_{\alpha} + 2 - \Delta b^a_{ii} \big) \,.
\end{split}
\end{align}
We see that the anomaly coefficients depend in a complicated way on the change in the height-pairings. However, we can form a combination in which almost all of the unknown contributions drop out
\begin{align}
\begin{split}
\tilde{b}_{ij} \cdot (\tilde{b} - \tilde{a}) &= b_{ij} \cdot (b - a) + \Omega_{\alpha \beta} \Delta b^{\alpha}_{ij} (b^{\beta} - a^{\beta}) \,, \\
\tilde{b}_{ii} \cdot (\tilde{b} - \tilde{a}) &= b_{ii} \cdot (b - a) + \Omega_{\alpha \beta} \Delta b^{\alpha}_{ii} (b^{\beta} - a^{\beta}) \,.
\end{split}
\label{Abeliananomconstr}
\end{align}
Hence, if we assume that the second term on the right hand side vanishes, i.e.\
\begin{align}
\Omega_{\alpha \beta} \Delta b^{\alpha}_{ij} (b^{\beta} - a^{\beta}) = 0 \,, \quad \Omega_{\alpha \beta} \Delta b^{\alpha}_{ii} (b^{\beta} - a^{\beta}) = 0  \,,
\label{assumption}
\end{align}
we can constrain the Abelian charges of the states modified in the transition. Moreover, we will show in Section~\ref{subsec:toricrational} that for toric hypersurface models $\Delta b^{\alpha}_{ii}$ and $\Delta b^{\alpha}_{ij}$ are proportional to $b^{\alpha}$ with real proportionality constants $\kappa_{ii}$ and $\kappa_{ij}$, see \eqref{abelianprop}. Plugging this back into the equation \eqref{assumption} and using the expression for the genus of the base divisor $\mathcal{Z}$, see \eqref{genuseq}, we find the modified assumption
\begin{align}
\kappa_{ij} (2 g_{\mathcal{Z}} - 2) = 0 \,, \quad \kappa_{ii} (2 g_{\mathcal{Z}} - 2) = 0 \,,
\end{align}
where $\kappa_{ii}$ an $\kappa_{ij}$ depend on the details of the model. This has two possible solutions: Either $\kappa_{ij}$ and $\kappa_{ii}$ have to vanish or the divisor $\mathcal{Z}$ has to be a genus-one curve in the base. In particular, this means that for $g_{\mathcal{Z}} = 1$ the intersections of the rational sections in the height pairing can change without affecting the Abelian anomaly constraints. The vanishing of the coefficients $\kappa_{ii}$ and $\kappa_{ij}$ is a generic feature of all toric models we studied, see Section~\ref{subsec:toricrational}.

Hence, using \eqref{assumption} in \eqref{Abeliananomconstr} we have
\begin{align}
\tilde{b}_{ij} \cdot (\tilde{b} - \tilde{a}) - b_{ij} \cdot (b - a) = 0 \,, \quad \tilde{b}_{ii} \cdot (\tilde{b} - \tilde{a}) - b_{ii} \cdot (b - a) = 0 \,,
\end{align}
which constrains the Abelian matter spectrum affected by the tensor-matter transitions. With \eqref{redanomalies} we then find
\begin{align}
\begin{split}
\lambda \sum_{\mathbf{R},i,j} \Delta n [\mathbf{R}, q_i, q_j] q_i q_j A_{\mathbf{R}} - \tfrac{1}{6} \sum_{i,j} \Delta n [q_i, q_j] q_i q_j &= 0 \,, \\
\lambda \sum_{\mathbf{R},i} \Delta n [\mathbf{R}, q_i] q_i^2 A_\mathbf{R} - \tfrac{1}{6} \sum_{i} \Delta n [q_i] q_i^2 &= 0 \,,
\end{split}
\label{specAbelian}
\end{align}
which is indeed satisfied for all examples discussed in Section~\ref{sec:Examples}. Moreover, using that the U(1) charges of singlets have to be integer and the U(1) charges of non-Abelian matter representations are fractional \cite{Grimm:2015wda, Cvetic:2017epq}, we can restrict the singlet charges in the transition using knowledge of the charges of the non-Abelian representations. Even though this does not allow for a unique determination of the complete change in the matter spectrum $\Delta \mathcal{S}$, it turns out to be very useful in the investigation of specific models.

\subsection{Warm-up examples}
\label{subsec:warmup}

Before we begin the classification of tensor-matter transition arising at the intersection of a divisor carrying no gauge algebra with a divisor carrying a non-Abelian gauge algebra in Section~\ref{subsec:class}, we discuss two interesting examples, i.e.\ the grand unified gauge algebras SU(5) and SO(10). These two gauge algebras have been intensively investigated in the F-theory literature, see e.g.\ \cite{Chen:2010ts, Esole:2011sm, Marsano:2011hv, Tatar:2012tm, Mayrhofer:2012zy, Braun:2013nqa, Braun:2013yti, Borchmann:2013hta, Borchmann:2013jwa, Cvetic:2013uta, Hayashi:2013lra, Anderson:2013xka, Hayashi:2014kca, Lawrie:2015hia, Baume:2015wia, Buchmuller:2017wpe} and references therein, and serve to illustrate the general procedure.

\subsubsection*{SU(5) transitions}

The relevant representations for SU(5) with dim$(\mathbf{R}) < 29$ are of dimension 5, 10, and 15, with group theory coefficients specified by
\begin{align*}
\begin{tabular}{| c | c | c | c | c |}
\hline
dim$(\mathbf{R})$ & $A_{\mathbf{R}}$ & $B_{\mathbf{R}}$ & $C_{\mathbf{R}}$ & $E_{\mathbf{R}}$ \\ \hline \hline
5 & 1 & 1 & 0 & 1 \\ \hline
10 & 3 & -3 & 3 & 1 \\ \hline
15 & 8 & 13 & 3 & 9 \\ \hline
\end{tabular}
\end{align*}
Vanishing of the irreducible $\text{tr}F^4$ SU(5) anomaly demands
\begin{align}
\Delta n[\mathbf{5}] -3 \, \Delta n[\mathbf{10}] + 13 \, \Delta n[\mathbf{15}] = 0 \,,
\label{su5irred}
\end{align}
For the resolution of a single SCP, the modifications in the reducible anomaly coefficients in the non-Abelian sector given by \eqref{changenonA} and \eqref{changemixed} yield the following equations
\begin{align}
\begin{split}
 3 \, \Delta n[\mathbf{10}] + 3 \, \Delta n[\mathbf{15}]  &= - 3 \,, \\
\Delta n[\mathbf{5}] + 3 \, \Delta n[\mathbf{10}] + 8 \, \Delta n[\mathbf{15}] &= - 6 \,.
\end{split}
\end{align}
Together with \eqref{su5irred}, these equations have a unique solution given by
\begin{align}
\Delta n[\mathbf{5}] = -3 \,, \quad \Delta n[\mathbf{10}] = -1 \,, \quad \Delta n[\mathbf{15}] = 0 \,,
\end{align}
making up 25 degrees of freedom lost in the non-Abelian hypermultiplet sector. Hence, the change in the matter spectrum can be summarized as
\begin{align}
\text{SU(5):} \quad \Delta \mathcal{S} = - (\mathbf{10} \oplus 3 \times \mathbf{5} \oplus 3 \times \mathbf{1} \oplus \mathbf{1}) \,,
\end{align}
where the last singlet is neutral and corresponds to the complex structure deformation.

Next, we want to analyze possible constraints for the Abelian charges for a single U(1) factor under the assumption \eqref{assumption}. We parametrize the U(1) charges of the fundamental and antisymmetric representations as $\tfrac{1}{5} q_{\mathbf{5}_i}$ and $\tfrac{1}{5} q_{\mathbf{10}}$, respectively, with $q_{\mathbf{5}_i}, q_{\mathbf{10}} \in \mathbbm{Z}$, see e.g.\ \cite{Grimm:2015wda, Cvetic:2017epq}. Since the anomaly coefficients $E_{\mathbf{R}}$ do not vanish, the corresponding irreducible anomaly \eqref{specirred} demands
\begin{align}
\sum_{i=1}^3 q_{\mathbf{5}_i} + q_{\mathbf{10}} = 0 \,.
\end{align}
Employing the restriction \eqref{specAbelian}, keeping in mind that only three of the four singlets can be charged with charges $q_{\mathbf{1}_a} \in \mathbbm{Z}$, we deduce
\begin{align}
25 \sum_{a = 1}^3 q_{\mathbf{1}_a}^2 - \sum_{i=1}^3 q_{\mathbf{5}_i}^2 - 8 q_{\mathbf{10}}^2= 0 \,.
\end{align}
In order to demonstrate the predictive power of this constraint, we assume that the three charges of the fundamental representation are equal, $q_{\mathbf{5}_i} = q_{\mathbf{5}}$. Hence, we find that $q_{\mathbf{10}} = -3 q_{\mathbf{5}}$ and the equation above reads
\begin{align}
\sum_{a=1}^3 q_{\mathbf{1}_a}^2 = 3 q_{\mathbf{5}}^2 \,,
\end{align}
which is solved e.g.\ by $q_{\mathbf{1}_a} = q_{\mathbf{5}}$. As it turns out this is satisfied by the toric SU(5) example discussed in Section \ref{su5exa}.

\subsubsection*{SO(10) transitions}

The representations of SO(10) with dimension smaller than 29 are the vector and the spinor representation with
\begin{align*}
\begin{tabular}{| c | c | c | c | c |}
\hline
dim$(\mathbf{R})$ & $A_{\mathbf{R}}$ & $B_{\mathbf{R}}$ & $C_{\mathbf{R}}$ & $E_{\mathbf{R}}$ \\ \hline \hline
10 & 1 & 1 & 0 & 0 \\ \hline
16 & 2 & -1 & $\tfrac{3}{4}$ & 0 \\ \hline
\end{tabular}
\end{align*}
Vanishing of the irreducible $\text{tr}F^4$ SO(10) anomaly leads to the constraint
\begin{align}
\Delta n[\mathbf{10}] - \Delta n[\mathbf{16}] = 0 \,,
\end{align}
i.e., the transition involves the same number of fields in the vector and spinor representations. The constraints from reducible \eqref{changenonA} and mixed non-Abelian anomalies \eqref{changemixed} read
\begin{align}
\Delta n[\mathbf{16}] = -1 \,, \quad \Delta n[\mathbf{10}] + 2 \Delta n[\mathbf{16}] = -3 \,.
\end{align}
The solution to these equations is unique,
\begin{align}
\Delta n[\mathbf{10}] = -1 \,, \quad \Delta n[\mathbf{16}] = -1 \,,
\end{align}
leading to a change in the matter spectrum of the form
\begin{align}
\text{SO(10):} \quad \Delta \mathcal{S} = - (\mathbf{16} \oplus \mathbf{10} \oplus 2 \times \mathbf{1} \oplus \mathbf{1}) \,.
\end{align}
If the model has a single Abelian U(1) gauge factor and we employ the assumption \eqref{assumption}, we can further constrain the Abelian charges, keeping in mind that the singlet corresponding to the complex structure deformation is uncharged. Parametrizing the Abelian charges of the non-Abelian matter as $\tfrac{1}{2} q_{\mathbf{10}}$ and $\tfrac{1}{4} q_{\mathbf{16}}$, respectively, where $q_{\mathbf{10}},q_{\mathbf{16}} \in \mathbbm{Z}$ and the singlet charges as $q_{\mathbf{1}_a} \in \mathbbm{Z}$, we find the constraint
\begin{align}
\tfrac{1}{2} (q_{\mathbf{10}}^2 + q_{\mathbf{16}}^2) = \sum_{a=1}^2 q_{\mathbf{1}_a}^2 \,.
\end{align}
For the three toric models and their transitions considered in \cite{Buchmuller:2017wpe}, we find the possibilities
\begin{align*}
\begin{tabular}{| c || c | c | c |}
\hline
 & Model 1 & Model 2 & Model 3 \\
 & ($F_3$ top 2) & ($F_3$ top 3) & ($F_3$ top 5) \\ \hline \hline
$q_{\mathbf{10}}$ & $-1$ & $-1$ & $2$ \\ \hline
$q_{\mathbf{16}}$ & $-1$ & $3$ & $0$ \\ \hline
$\tfrac{1}{2} (q_{\mathbf{10}}^2 + q_{\mathbf{16}}^2)$ & 1 & 5 & 2 \\ \hline
\end{tabular}
\end{align*}
Due to the fact that the singlet charges are integers, this uniquely\footnote{Up to an obvious symmetry involving permutation of $q_{\mathbf{1}_1}$ and $q_{\mathbf{1}_2}$ and complex conjugation.} fixes them to
\begin{align*}
\begin{tabular}{| c || c | c | c |}
\hline
 & Model 1 & Model 2 & Model 3 \\
 & ($F_3$ top 2) & ($F_3$ top 3) & ($F_3$ top 5) \\ \hline \hline
$q_{\mathbf{1}_1}$ & $1$ & $1$ & $1$ \\ \hline
$q_{\mathbf{1}_2}$ & $0$ & $2$ & $1$ \\ \hline
\end{tabular}
\end{align*}
These are indeed the charges appearing in the transitions in \cite{Buchmuller:2017wpe}. For more than one Abelian gauge algebra factor, the corresponding constraints are similarly satisfied and fixed by the charges of the non-Abelian representations for all models discussed in \cite{Buchmuller:2017wpe}. Since the singlet charges are usually challenging to determine, the constraints \eqref{specAbelian} present a great simplification in the investigation of the change in the matter spectrum~$\Delta \mathcal{S}$.

\subsection{Classification of transitions}
\label{subsec:class}

In this section we classify the modification of the matter spectrum in the tensor-matter transitions described above. The anomaly constraints are enough to fix the change in the non-Abelian matter spectrum uniquely. The restrictions of the Abelian charges are derived assuming \eqref{assumption}, except for the cases SU(8), SU(7), and E$_{7}$, for which they are valid in general.

\subsubsection*{SU(N) transitions}

For the special unitary algebras we can restrict to $N < 29$, since otherwise there is no representation whose dimension is small enough to compensate for a single tensor multiplet, keeping in mind that one degree of freedom is accounted for by the neutral complex structure deformation. The case $N = 5$ was discussed above. For $N > 5$ the only relevant representations\footnote{We specify the representations in terms of their Dynkin labels. For ordering of the simple roots we follow the conventions of~\cite{Slansky:1981yr}.} are
\begin{align*}
\begin{tabular}{| c | c || c | c | c | c |}
\hline
$\mathbf{R}$ &dim$(\mathbf{R})$ & $A_{\mathbf{R}}$ & $B_{\mathbf{R}}$ & $C_{\mathbf{R}}$ & $E_{\mathbf{R}}$ \\ \hline \hline
\!$(1,0,0,0,\ldots,0)\!$&$N$ & 1 & 1 & 0 & 1 \\ \hline
\!$(2,0,0,0,\ldots,0)\!$&$\tfrac{N (N+1)}{2}$ & $N+2$ & $N+8$ & 3 & $(N+4)$ \\ \hline
\!$(0,1,0,0,\ldots,0)\!$&$\tfrac{N (N-1)}{2}$ & $N-2$ & $N-8$ & 3 & $(N-4)$ \\ \hline
\!$(0,0,1,0,\ldots,0)\!$&$\tfrac{N (N-1) (N-2)}{6}$ & $\tfrac{N^2 - 5N + 6}{2}$ & $\tfrac{N^2 - 17 N + 54}{2}$ & $3N -12$& $\tfrac{1}{2} (N^2 - 9N + 18)$ \\ \hline
\end{tabular}
\end{align*}
The smaller gauge algebras SU(4), SU(3), and SU(2) have a large number of additional representations and are discussed separately below. First, we discuss the case with $N > 6$ in which also the two-fold anti-symmetric representation is too large to contribute. Cancellation of the irreducible SU$(N)$ anomaly demands
\begin{align}
\Delta n [\mathbf{N}] + (N+8) \Delta n [\mathbf{\tfrac{N(N+1)}{2}}] + (N-8) \Delta n [\mathbf{\tfrac{N(N-1)}{2}}] = 0 \,.
\end{align}
Consistency with the formulas \eqref{changenonA} and \eqref{changemixed} further implies
\begin{align}
\begin{split}
\Delta n [\mathbf{N}] + (N+2) \Delta n [\mathbf{\tfrac{N(N+1)}{2}}] + (N-2) \Delta n [\mathbf{\tfrac{N(N-1)}{2}}] &= - 6 \,, \\
3 \Delta n [\mathbf{\tfrac{N(N+1)}{2}}] + 3  \Delta n [\mathbf{\tfrac{N(N-1)}{2}}] &= - 3 \,.
\end{split}
\end{align}
This has the unique solution
\begin{align}
\Delta n [\mathbf{N}] = -8 + N \,, \quad \Delta n [\mathbf{\tfrac{N(N+1)}{2}}] = 0 \,, \quad \Delta n [\mathbf{\tfrac{N(N-1)}{2}}] = -1 \,.
\end{align}
Since $\Delta n [\mathbf{R}] < 0$, the only two possible transitions are for $N = 7,8$. For both of them the change in the non-Abelian matter spectrum contains $28$ hypermultiplets, which, together with the uncharged singlet from the complex structure sector, make up for the 29 degrees of freedom contained in the new tensor multiplet. Moreover, since $E_{\mathbf{R}}$ does not vanish in the anti-symmetric representation, we find the full change in the matter spectrum including U(1) charges,
\begin{align}
\begin{split}
\text{SU}(8):& \quad \Delta \mathcal{S} = - ( \mathbf{28}_0 \oplus \mathbf{1}_0 ) \,, \\
\text{SU}(7):& \quad \Delta \mathcal{S} = - ( \mathbf{21}_{-\ttfrac{1}{3}q} \oplus \mathbf{7}_{q} \oplus \mathbf{1}_0 ) \,.
\end{split}
\end{align}
For SU(6) the restriction of $\Delta n [\mathbf{R}]$ to be negative fixes the change in the non-Abelian matter spectrum to 
\begin{align}
\Delta n [\mathbf{6}] = -2 \,, \quad \Delta n [\mathbf{15}] = -1 \,, \quad \Delta n [\mathbf{20}] = 0 \,, \quad \Delta n [\mathbf{21}] = 0 \,,
\end{align}
and we find
\begin{align}
\text{SU(6):} \quad \Delta \mathcal{S} = - (\mathbf{15} \oplus 2 \times \mathbf{6} \oplus \mathbf{1} \oplus \mathbf{1}) \,.
\end{align}
Equation \eqref{specAbelian} and the irreducible anomaly \eqref{specirred} further lead to the restrictions
\begin{align}
\sum_{i=1}^2 q_{\mathbf{6}_i} + 2 q_{\mathbf{15}} = 0 \,, \quad q_{\mathbf{15}}^2 = q_{\mathbf{1}}^2 \,,
\end{align}
where we defined the charges $\tfrac{1}{6} q_{\mathbf{6}_i}$ and $\tfrac{1}{3} q_{\mathbf{15}}$, with $q_{\mathbf{6}_i}, q_{\mathbf{15}} \in \mathbbm{Z}$. In general these equations are not strong enough to fix the full charge-dependence uniquely, but they simplify the analysis of specific models.

The gauge algebra SU(5) was discussed above, see Section~\ref{subsec:warmup}, so we go directly on to SU(4). For SU(4), three different 20-dimensional matter representations can in principle be involved in the transition. Their group theory factors read
\begin{align*}
\begin{tabular}{| c | c || c | c | c | c |}
\hline
$\mathbf{R}$&dim$(\mathbf{R})$ & $A_{\mathbf{R}}$ & $B_{\mathbf{R}}$ & $C_{\mathbf{R}}$ & $E_{\mathbf{R}}$ \\ \hline \hline
(1,1,0)&20 & 13 & $-11$ & 24 & 7 \\ \hline
(3,0,0)&$20'$ & 21 & 69 & 24 & 35 \\ \hline
(0,2,0)&$20''$ & 16 & $-54$ & 54 & 0 \\ \hline
\end{tabular}
\end{align*}
The irreducible and reducible non-Abelian anomaly constraints derived from \eqref{specirred}, \eqref{changenonA} and \eqref{changemixed} read
\begin{align}
\begin{split}
\Delta n[\mathbf{4}] - 4 \Delta n[\mathbf{6}] + 12 \Delta n[\mathbf{10}] - 11 \Delta n[\mathbf{20}] + 69 \Delta n[\mathbf{20'}] - 54 \Delta n[\mathbf{20''}] &= 0 \,, \\
\Delta n[\mathbf{6}] + \Delta n[\mathbf{10}] + 8 \Delta n[\mathbf{20}] + 8 \Delta n[\mathbf{20'}] + 18 \Delta n[\mathbf{20''}] &= -1 \,, \\
\Delta n[\mathbf{4}] + 2 \Delta n[\mathbf{6}] + 6 \Delta n[\mathbf{10}] + 13 \Delta n[\mathbf{20}] + 21 \Delta n[\mathbf{20'}] + 16 \Delta n[\mathbf{20''}] &= - 6 \,.
\end{split}
\end{align}
The transition is unique and given by
\begin{align}
\Delta n[\mathbf{4}] = -4 \,, \quad \Delta n[\mathbf{6}] = -1 \,,
\end{align}
with all other matter multiplicities unchanged. Therefore, we can summarize the change in the matter spectrum as
\begin{align}
\text{SU(4):} \quad \Delta \mathcal{S} = - (\mathbf{6} \oplus 4 \times \mathbf{4} \oplus 6 \times \mathbf{1} \oplus \mathbf{1}) \,.
\end{align}
Using \eqref{specAbelian} in combination with the irreducible Abelian mixed anomaly in \eqref{specirred}, we further obtain
\begin{align}
\sum_{i=1}^4 q_{\mathbf{4}_i} = 0 \,, \quad \sum_{i=1}^4 q_{\mathbf{4}_i}^2 + 12 q_{\mathbf{6}}^2 = \sum_{a=1}^6 q_{\mathbf{1}_a}^2 \,,
\end{align}
where we parametrized the charges for the fundamental and anti-symmetric representation as $\tfrac{1}{4} q_{\mathbf{4}_i}$ and $\tfrac{1}{2} q_{\mathbf{6}}$ with $q_{\mathbf{4}_i}, q_{\mathbf{6}} \in \mathbbm{Z}$, respectively.

Next we discuss the algebra SU(3). We have seen that we can restrict to representation with dimension smaller than $29$. Moreover, since the transition under discussion induce $\Delta n[\mathbf{R}] < 0$ and all algebra theory coefficients are positive, we can focus on representations with $A_{\mathbf{R}} < 6$ and $C_{\mathbf{R}} < 3$. The only remaining representation is the fundamental with
\begin{align*}
\begin{tabular}{| c | c || c | c | c | c |}
\hline
$\mathbf{R}$&dim$(\mathbf{R})$ & $A_{\mathbf{R}}$ & $B_{\mathbf{R}}$ & $C_{\mathbf{R}}$ & $E_{\mathbf{R}}$ \\ \hline \hline
(1,0) &3 & 1 & 0 & $\tfrac{1}{2}$ & 1 \\ \hline
\end{tabular}
\end{align*}
Solving the anomaly constraints we find
\begin{align}
\text{SU(3):} \quad \Delta \mathcal{S} = - (6 \times \mathbf{3} \oplus 10 \times \mathbf{1} \oplus \mathbf{1}) \,.
\end{align}
With an additional Abelian gauge algebra we further derive
\begin{align}
\sum_{i=1}^6 q_{\mathbf{3}_i} = 0 \,, \quad \sum_{i=1}^{6} q_{\mathbf{3}_i}^2 = 3 \sum_{a=1}^{10} q_{\mathbf{1}_a}^2 \,,
\end{align}
where we parametrized the U(1) charges of the triplets as $\frac{1}{3} q_{\mathbf{3}_i}$, with $q_{\mathbf{3}_i} \in \mathbbm{Z}$.

Similar considerations for the algebra SU(2) lead to the possible change in the matter spectrum involving only the fundamental representation with
\begin{align*}
\begin{tabular}{| c | c || c | c | c | c |}
\hline
$\mathbf{R}$&dim$(\mathbf{R})$ & $A_{\mathbf{R}}$ & $B_{\mathbf{R}}$ & $C_{\mathbf{R}}$ & $E_{\mathbf{R}}$ \\ \hline \hline
(1)&2 & 1 & 0 & $\tfrac{1}{2}$ & 0 \\ \hline
\end{tabular}
\end{align*}
The change is given by
\begin{align}
\text{SU(2):} \quad \Delta \mathcal{S} = - (6 \times \mathbf{2} \oplus 16 \times \mathbf{1} \oplus \mathbf{1}) \,.
\end{align}
The additional constraints on the matter spectrum for the Abelian charges read
\begin{align}
\sum_{i=1}^6 q_{\mathbf{2}_i}^2 = \sum_{a=1}^{16} q_{\mathbf{1}_a}^2 \,,
\end{align}
where we have written the charges of the doublets as $\tfrac{1}{2} q_{\mathbf{2}_i}$ with $q_{\mathbf{2}_i} \in \mathbbm{Z}$.

\subsubsection*{SO(N) transitions}

In the case of the gauge algebra $G$ being SO$(N)$, we can restrict the discussion to $N > 6$. For the lower-dimensional cases isomorphisms with other gauge algebras can be used,
\begin{align}
\begin{split}
& \text{SO}(3) \sim \text{SU}(2) \sim \text{Sp}(2) \, \quad \text{SO}(4) \sim \text{SU}(2) \oplus \text{SU}(2) \,, \\
& \text{SO}(5) \sim \text{Sp}(4) \,,\quad \text{SO}(6) \sim \text{SU}(4) \,.
\end{split}
\end{align}
The representations that appear in possible transitions for $N > 8$ are the $N$-dimensional vector representation and the spinor representation. Together with the adjoint representation, their group theory coefficients are given by \begin{align*}
\begin{tabular}{| c| c || c | c | c | c |}
\hline
$\mathbf{R}$&dim$(\mathbf{R})$ & $A_{\mathbf{R}}$ & $B_{\mathbf{R}}$ & $C_{\mathbf{R}}$ & $E_{\mathbf{R}}$ \\ \hline \hline
(1,0,\ldots,0,0)&$N$ & 1 & 1 & 0 & 0 \\ \hline
(0,0,\ldots,0,1)&$2^{\lceil N/2 \rceil - 1}$ & $2^{\lceil N/2 \rceil - 4}$ & $- 2^{\lceil N/2 \rceil - 5}$ & $3 \times 2^{\lceil N/2 \rceil - 7}$ & 0 \\ \hline
(0,1,\ldots,0,0)&$\tfrac{N(N-1)}{2}$ & $N-2$ & $N-8$ & 3 & 0 \\ \hline
\end{tabular}
\end{align*}
The irreducible SO$(N)$ anomaly \eqref{specirred} hence translates to
\begin{align}
\Delta n[\mathbf{N}] - 2^{\lceil N/2 \rceil - 5} \Delta n[ \mathbf{S} ] = 0 \,,
\end{align} 
where we denote the spinor representation by $\mathbf{S}$. The constraints \eqref{changenonA} and \eqref{changemixed} further demand
\begin{align}
\begin{split}
2^{\lceil N/2 \rceil - 5} \Delta n[\mathbf{S}]  &= -1 \,, \\
\Delta n[\mathbf{N}] + 2^{\lceil N/2 \rceil - 4} \Delta n[\mathbf{S}] &= -3 \,.
\end{split}
\end{align}
From the first of these equations we see that the number of spinor representations has to change. For $N \geq 11$ the spinor representation is already 32-dimensional and too large to appear in the transition. However, since it is a pseudo-real representation for SO(11) we can have a half-hypermultiplet in the spinor representation. The change in the spectrum is
\begin{align}
\text{SO}(11): \quad \Delta \mathcal{S} = - (\tfrac{1}{2} \times \mathbf{32} \oplus \mathbf{11} \oplus \mathbf{1} \oplus \mathbf{1}) \,.
\end{align}
The half-hypermultiplet is necessarily uncharged. Denoting the Abelian charge of the vector representation by $q_{\mathbf{11}}$, we derive
\begin{align}
q_{\mathbf{11}}^2 = q_{\mathbf{1}}^2 \,.
\end{align}
The case of SO(10) was discussed in full detail in Section~\ref{subsec:warmup} above, so we continue with SO(9). Again, the non-Abelian representations are fixed uniquely and the change in the matter spectrum is
\begin{align}
\text{SO}(9): \quad \Delta \mathcal{S} = - (\mathbf{16} \oplus \mathbf{9} \oplus 3 \times \mathbf{1} \oplus \mathbf{1}) \,.
\end{align}
With an additional Abelian gauge algebra, we further find
\begin{align}
2 q_{\mathbf{16}}^2 + 3 q_{\mathbf{9}}^2 = \sum_{a=1}^3 q_{\mathbf{1}_a}^2 \,,
\end{align}
where we parametrized the charges of the spinor and vector representation as $\tfrac{1}{2} q_{\mathbf{16}}$ and $q_{\mathbf{9}}$, respectively.

The two remaining cases are the algebras SO(7) and SO(8). For SO(8) the only relevant representations are the spinor, co-spinor, and vector representation that are all of dimension 8 with the following group theory coefficients:
\begin{align*}
\begin{tabular}{| c | c || c | c | c | c |}
\hline
$\mathbf{R}$&dim$(\mathbf{R})$ & $A_{\mathbf{R}}$ & $B_{\mathbf{R}}$ & $C_{\mathbf{R}}$ & $E_{\mathbf{R}}$ \\ \hline \hline
(1,0,0,0)&$8_v$ & 1 & 1 & 0 & 0 \\ \hline
(0,0,1,0)&$8_{cs}$ & 1 & $-\tfrac{1}{2}$ & $\tfrac{3}{8}$ & 0 \\ \hline
(0,0,0,1)&$8_{s}$ & 1 & $-\tfrac{1}{2}$ & $\tfrac{3}{8}$ & 0 \\ \hline
\end{tabular}
\end{align*}
The anomaly equations are solved uniquely by the following change in the non-Abelian matter spectrum
\begin{align}
\text{SO(8):} \quad \Delta \mathcal{S} = - (2 \times \mathbf{8}_{s/cs} \oplus \mathbf{8}_v \oplus 4 \times \mathbf{1} \oplus \mathbf{1}) \,.
\end{align}
Parameterizing the charges of the (co)-spinor and vector representations by $\tfrac{1}{2} q_{\mathbf{8}_{s/cs,i}}$ and $\tfrac{1}{2} q_{\mathbf{8}_v}$, respectively, we obtain
\begin{align}
\sum_{i=1}^2 q_{\mathbf{8}_{s/cs,i}}^2 + q_{\mathbf{8}_v}^2 = \sum_{a=1}^4 q_{\mathbf{1}_a}^2 \,.
\end{align}
For SO(7) the constraints are very similar to above with the same group coefficients for the 8-dimensional spinor representation as for SO(8)
\begin{align*}
\begin{tabular}{| c | c || c | c | c | c |}
\hline
$\mathbf{R}$&dim$(\mathbf{R})$ & $A_{\mathbf{R}}$ & $B_{\mathbf{R}}$ & $C_{\mathbf{R}}$ & $E_{\mathbf{R}}$ \\ \hline \hline
$(1,0,0)$& $7$ & 1 & 1 & 0 & 0 \\ \hline
$(0,0,1)$&$8$ & 1 & $-\tfrac{1}{2}$ & $\tfrac{3}{8}$ & 0 \\ \hline
$(2,0,0)$&$27$ & 9 & 60 & $\tfrac{33}{4}$ & 0 \\ \hline
\end{tabular}
\end{align*}
The 27-dimensional representation does not participate in the transition and we obtain for the change in the spectrum
\begin{align}
\text{SO(7):} \quad \Delta \mathcal{S} = - (2 \times \mathbf{8} \oplus \mathbf{7} \oplus 5 \times \mathbf{1} \oplus \mathbf{1}) \,.
\end{align}
With $\tfrac{1}{2} q_{\mathbf{8}_i}$ and $q_{\mathbf{7}}$ parameterizing the Abelian charges for non-Abelian representations involved in the transitions, the additional constraint reads
\begin{align}
\sum_{i=1}^2 q_{\mathbf{8}_i}^2 + 5 q_{\mathbf{7}}^2 = \sum_{a=1}^5 q_{\mathbf{1}_a}^2 \,,
\end{align}
which concludes our discussion of SO($N$) algebras.

\subsubsection*{Sp(2N) transitions}

In the case of symplectic groups\footnote{Note that there are several conventions for denoting the group. We follow the convention \cite{Slansky:1981yr} $\text{Sp}(2N) \sim C_N$.} we can restrict the discussion to Sp($2N$) with $N \in \{2,3,4\}$. For $N$ larger than that the first representation with non-trivial $C_{\mathbf{R}}$ needed for the transition is already too large.

For Sp(8) the relevant representations and group theory coefficients are given by
\begin{align*}
\begin{tabular}{| c | c || c | c | c | c |}
\hline
$\mathbf{R}$ & dim$(\mathbf{R})$ & $A_{\mathbf{R}}$ & $B_{\mathbf{R}}$ & $C_{\mathbf{R}}$ & $E_{\mathbf{R}}$ \\ \hline \hline
(1,0,0,0)&$8$     & 1 & 1      & 0 & 0 \\ \hline
(0,1,0,0)&$27$   & 6 &0 & 3 & 0 \\ \hline
\end{tabular}
\end{align*}
Solving the anomaly constraints, the change in the non-Abelian matter spectrum is uniquely fixed to
\begin{align}
\text{Sp(8):} \quad \Delta \mathcal{S} = - (\mathbf{27} \oplus \mathbf{1} \oplus \mathbf{1}) \,.
\end{align}
With additional U(1) factors and charge $q_{\mathbf{27}}$ for the non-Abelian matter field we find 
\begin{align}
9 \, q_{\mathbf{27}}^2 = q_{\mathbf{1}}^2 \,,
\end{align}
which fixes the relative charges with respect to the singlet.

For $G = \text{Sp}(6)$ we have
\begin{align*}
\begin{tabular}{| c | c || c | c | c | c |}
\hline
$\mathbf{R}$ & dim$(\mathbf{R})$ & $A_{\mathbf{R}}$ & $B_{\mathbf{R}}$ & $C_{\mathbf{R}}$ & $E_{\mathbf{R}}$ \\ \hline \hline
(1,0,0)&$6$     & 1 & 1      & 0 & 0 \\ \hline
(0,1,0)&$14$   & 4 & $-2$ & 3 & 0 \\ \hline
(0,0,1)&$14$   & 5 & $-7$ & 6 & 0 \\ \hline
\end{tabular}
\end{align*}
Again, the change in the non-Abelian matter spectrum is fixed by the consistency with anomaly cancellation,
\begin{align}
\text{Sp(6):} \quad \Delta \mathcal{S} = - (\mathbf{14} \oplus 2 \times \mathbf{6} \oplus 2 \times \mathbf{1} \oplus \mathbf{1}) \,.
\end{align}
Furthermore, we find 
\begin{align}
10 \, q_{\mathbf{14}}^2 = \sum_{a=1}^2 q_{\mathbf{1}_a}^2 \,,
\end{align}
for additional Abelian gauge group factors and charges parametrized by $q_{\mathbf{14}}$ and $\tfrac{1}{2} q_{\mathbf{6}_i}$. Note that the charges of the $\mathbf{6}$-plets drop out of the expression.

Finally, we discuss the case of gauge group $G = \text{Sp}(4)$, which on the level of the algebra is identical to\footnote{However, the Casimir coefficients $A,B,C,E$ are different due to the factor of 3 in $\lambda$ between SO and Sp groups.} $G = \text{SO}(5)$. Even though there are in principle six different representations that could be involved in the transition, the restriction $A_{\mathbf{R}} \leq 6$ together with the fact that the number of adjoint hypermultiplets remains unchanged allows only for
\begin{align*}
\begin{tabular}{| c | c || c | c | c | c |}
\hline
$\mathbf{R}$ & dim$(\mathbf{R})$ & $A_{\mathbf{R}}$ & $B_{\mathbf{R}}$ & $C_{\mathbf{R}}$ & $E_{\mathbf{R}}$ \\ \hline \hline
(1,0)&$4$   & 1 & 1      & 0 & 0 \\ \hline
(0,1)&$5$   & 2 & $-4$ & 3 & 0 \\ \hline
\end{tabular}
\end{align*}
The change in the non-Abelian matter spectrum hence is
\begin{align}
\text{Sp(4):} \quad \Delta \mathcal{S} = - (\mathbf{5} \oplus 4 \times \mathbf{4} \oplus 7 \times \mathbf{1} \oplus \mathbf{1}) \,,
\end{align}
and the Abelian charges are restricted to satisfy
\begin{align}
14 \, q_{\mathbf{5}}^2 + \sum_{i=1}^4 q_{\mathbf{4}_i}^2 = 2 \sum_{a=1}^7 q_{\mathbf{1}_a}^2 \,,
\end{align}
where we parametrized the charges of $\mathbf{5}$ and $\mathbf{4}_i$ as $q_{\mathbf{5}}$ and $\tfrac{1}{2} q_{\mathbf{4}_i}$, respectively.

\subsubsection*{G$_2$ transitions}

For $G = \text{G}_2$ the relevant group theory coefficients can be summarized as
\begin{align*}
\begin{tabular}{| c | c || c | c | c | c |}
\hline
$\mathbf{R}$&dim$(\mathbf{R})$ & $A_{\mathbf{R}}$ & $B_{\mathbf{R}}$ & $C_{\mathbf{R}}$ & $E_{\mathbf{R}}$ \\ \hline \hline
(1,0)&7 & 1 & 0 & $\tfrac{1}{4}$ & 0 \\ \hline
(0,1)&14 & 4 & 0 & $\tfrac{5}{2}$ & 0 \\ \hline 
(2,0)&27 & 9 & 0 & $\tfrac{27}{2}$ & 0 \\ \hline
\end{tabular}
\end{align*}
Note that G$_2$ does not have a third and fourth order Casimir invariant, which is why the coefficients $E_R$ and $B_R$ are zero. The adjoint representation is 14-dimensional and will not be involved in the transition. Therefore, we derive the constraints
\begin{align}
\begin{split}
\tfrac{4}{3} \sum_\mathbf{R} \Delta n[\mathbf{R}] C_{\mathbf{R}} &= \tfrac{1}{3} \, \Delta n[\mathbf{7}] + 18 \, \Delta n[\mathbf{27}] = -1 \,, \\
\tfrac{1}{3} \sum_{\mathbf{R}} \Delta n[\mathbf{R}] A_{\mathbf{R}} &= \tfrac{1}{3} \, \Delta n[\mathbf{7}] + 3 \, \Delta n[\mathbf{27}]  = -1 \,.
\end{split}
\end{align}
These are solved uniquely by
\begin{align}
\Delta n[\mathbf{7}] = -3 \,,
\end{align}
and we deduce the change in the matter spectrum to be
\begin{align}
\text{G}_2: \quad \Delta \mathcal{S} = - (3 \times \mathbf{7} \oplus 7 \times \mathbf{1} \oplus \mathbf{1}) \,.
\end{align}
The last singlets indicates the complex structure deformation and is uncharged, whereas the remaining states can be charged with respect to Abelian gauge algebras. For a single additional Abelian gauge algebra, equation \eqref{specAbelian} leads to the constraint
\begin{align}
5 \sum_{i=1}^3 q_{\mathbf{7}_i}^2 = \sum_{a=1}^7 q_{\mathbf{1}_a}^2 \,,
\end{align}
where the charges of the representation $\mathbf{7}_i$ is given by $q_{\mathbf{7}_i}$.

\subsubsection*{F$_4$ transitions}

For the algebra F$_4$ the relevant representations and their group theory coefficients are 
\begin{align*}
\begin{tabular}{| c|  c || c | c | c | c |}
\hline
$\mathbf{R}$&dim$(\mathbf{R})$ & $A_{\mathbf{R}}$ & $B_{\mathbf{R}}$ & $C_{\mathbf{R}}$ & $E_{\mathbf{R}}$ \\ \hline \hline
(0,0,0,1)&26 & 1 & 0 & $\tfrac{1}{12}$ & 0 \\ \hline
(1,0,0,0)&52 & 3 & 0 & $\tfrac{5}{12}$ & 0 \\ \hline 
\end{tabular}
\end{align*}
The 52-dimensional representation is the adjoint and thus its multiplicity does not change in the transition. The consistency equations for the anomalies yield
\begin{align}
\begin{split}
\Delta n[\mathbf{26}] = -1 \,.
\end{split}
\end{align}
Hence, the change in the matter spectrum is determined to be
\begin{align}
\text{F}_4: \quad \Delta \mathcal{S} = - (\mathbf{26} \oplus 2 \times \mathbf{1} \oplus \mathbf{1}) \,.
\end{align}
Defining $q_{\mathbf{26}}$ as the charge of the $\mathbf{26}$ representation, we find
\begin{align}
10 \, q_{\mathbf{26}}^2 = \sum_{a=1}^2 q_{\mathbf{1}_a}^2 \,,
\end{align}
which poses strict constraints on the involved singlet charges.

\subsubsection*{E$_6$ transitions}

The only representation of E$_6$ relevant for the transition is the lowest-dimensional, which is 27-dimensional. The group theory coefficients for $\mathbf{27}$ and the adjoint representation $\mathbf{72}$ are given by
\begin{align*}
\begin{tabular}{| c | c || c | c | c | c |}
\hline
$\mathbf{R}$& dim$(\mathbf{R})$ & $A_{\mathbf{R}}$ & $B_{\mathbf{R}}$ & $C_{\mathbf{R}}$ & $E_{\mathbf{R}}$ \\ \hline \hline
(1,0,0,0,0,0)&27 & 1 & 0 & $\tfrac{1}{12}$ & 0 \\ \hline
(0,0,0,0,0,1)&78 & 4 & 0 & $\tfrac{1}{2}$ & 0 \\ \hline
\end{tabular}
\end{align*}
The change in the non-Abelian spectrum is fixed uniquely and in total we find
\begin{align}
\text{E}_6: \quad \Delta \mathcal{S}: - (\mathbf{27} \oplus \mathbf{1} \oplus \mathbf{1}) \,.
\end{align}
For an additional U(1) factor, with the charge of the $\mathbf{27}$ given by $\tfrac{1}{3} q_{\mathbf{27}}$, we can further restrict 
\begin{align}
q_{\mathbf{27}}^2 = q_{\mathbf{1}}^2 \,,
\end{align}
which is realized in the examples discussed in \ref{e6exa}.

\subsubsection*{E$_7$ transitions}

The smallest irreducible representation of E$_7$ has dimension 56, which is already larger than 28. However, since the representation is pseudo-real, one can consider half-hypermultiplets which contribute only 28 degrees of freedom. These have to be necessarily uncharged. The group theory parameters for the adjoint and $\mathbf{56}$ representation are given by
\begin{align*}
\begin{tabular}{| c | c || c | c | c | c |}
\hline
$\mathbf{R}$&dim$(\mathbf{R})$ & $A_{\mathbf{R}}$ & $B_{\mathbf{R}}$ & $C_{\mathbf{R}}$ & $E_{\mathbf{R}}$ \\ \hline \hline
(0,0,0,0,0,1,0)&56 & 1 & 0 & $\tfrac{1}{24}$ & 0 \\ \hline
(1,0,0,0,0,0,0)&133 & 3 & 0 & $\tfrac{1}{6}$ & 0 \\ \hline
\end{tabular}
\end{align*}
Indeed, we find that a transition involving a half-hyper in the $\mathbf{56}$ representation is allowed,
\begin{align}
\text{E}_7: \quad \Delta \mathcal{S} = - (\tfrac{1}{2} \times \mathbf{56}_0 \oplus \mathbf{1}_0)
\end{align}
A specific realization is discussed in Section~\ref{e7exa}.

\subsubsection*{E$_8$ transitions}
For the algebra E$_8$ a transition of the form above is not possible, since all representations contain too many degrees of freedom.

\section{Tensor-matter transitions in toric hypersurfaces}
\label{sec:torichyper}

Elliptically fibered Calabi-Yau manifolds based on toric hypersurfaces allow for a direct way to construct tensor-matter transitions of the kind discussed above. In this section we discuss the construction of the transitions and investigate the change in the Abelian anomaly coefficients \eqref{Abeliananomcoef} in these models.

\subsection{Construction of toric hypersurfaces}
\label{subsec:toric}

In this section we summarize the construction of toric hypersurface models. For a more detailed discussion we refer to \cite{Klevers:2014bqa, Buchmuller:2017wpe} and Appendix~\ref{App:speccomp}.

In toric hypersurface constructions one embeds a torus fiber in one of the 16 toric ambient spaces given e.g.\ in \cite{Bouchard:2003bu}. Fibering this ambient space over a compact complex two-dimensional base $B$, one obtains an ambient toric variety $\mathcal{V}$ which can admit a torus-fibered Calabi-Yau manifold as a submanifold. The ambient variety $\mathcal{V}$ can be described by a 4d polytope $\Diamond$. If this polytope $\Diamond$ is reflexive with exactly one interior point and admits a projection onto one of the 16 2d polytopes, it admits a genus-one fibered 3-fold as a submanifold. This submanifold is specified as the vanishing of a polynomial that is a section of the canonical bundle of $\mathcal{V}$ that can be obtained in a combinatorial way from $\Diamond$ \cite{Batyrev:1994hm},
\begin{align}
p_{\Diamond} = \sum_i \hat{s}_i \, p_i (x_f) \,.
\end{align}
where the $\hat{s}_i$ are sections\footnote{In order to distinguish the sections parameterizing the base-dependence and the rational sections of the model, we use hats for the former.} of the base $B$ and $x_f$ denotes the toric coordinates of the ambient space of the torus fiber. This polynomial can be obtained from the polytope as described in Appendix~\ref{App:speccomp}.

The inclusion of a top, introduced in \cite{Candelas:1996su} and classified in \cite{Bouchard:2003bu}, tunes the complex structure in a way to ensure the presence of a resolved ADE singularity over a specified base divisor $\Z$. This corresponds to a non-Abelian gauge group factor in the F-theory model. Therefore, the resulting geometry is a completely resolved elliptically fibered Calabi-Yau manifold including the resolution divisors of the engineered gauge algebra factors. Mapping the model to a singular Weierstrass form induces a factorization of the sections $\hat{s}_i$,
\begin{align}
\hat{s}_i \rightarrow z_0^{n_i} d_i \,,
\label{secfac}
\end{align}
where the $d_i$ are again sections on the base, but do not depend on the coordinate $z_0$ anymore, and $\Z = \{ z_0 = 0 \}$. Moreover, for a consistent model, the base divisor classes defined by $\{ d_i = 0 \}$ have to be effective. 

Using the $\mathbbm{C}^*$-scalings for the fiber coordinates induced by the toric ambient space, one can encode the base-dependence in four base divisor classes. These are the anti-canonical class $\Kbi$, the base divisor $\Z$ carrying the non-Abelian gauge algebra, as well as two residual classes $\Ss$ and $\Sn$, which originate from the two section $\hat{s}_7$ and $\hat{s}_9$, cf.\ Appendix~\ref{App:speccomp}.

Moreover, the choice of the ambient space of the fiber also dictates the free part of the Mordell-Weil group, which corresponds to the Abelian part of the gauge algebra, see also the discussion in Section~\ref{subsec:toricrational}. These rational sections define generators corresponding to the Shioda maps $\sigma(s_i)$, see \eqref{Shioda}, which can be used to determine the Abelian charges of hypermultiplets in the six-dimensional model.

The charged matter spectrum can then be obtained by an investigation of co\-di\-men\-sion-two singularities in the base. From an enhancement of the fiber singularity over intersections of $\Z$ with another base divisor $D$, one obtains matter states transforming in certain representations, which correspond to the specific type of the enhancement, with respect to the non-Abelian gauge algebra localized on $\Z$. Their Abelian charges are determined by intersecting the matter curves with the Shioda maps $\sigma (s_i)$. Note that in the presence of non-Abelian gauge algebras $G$, the U(1) factors can mix with the center elements of $G$ \cite{Grimm:2015wda, Cvetic:2017epq}, which leads in general to fractional Abelian charges for the non-Abelian matter representations. Furthermore, if the base divisor $\Z$ has genus $g_{\Z}$, this leads to the appearance of $g_{\Z}$ hypermultiplets transforming in the adjoint representation \cite{Witten:1996qb}.

For codimension-two singularities that are not associated with an enhancement of a non-Abelian algebra factor, one finds charged singlets. Again, the charges are determined by the intersection with the Shioda maps. 

Since all the base-dependence is encoded in the divisor classes $\Kbi$, $\Z$, $\Ss$, and $\Sn$, the multiplicities of the charged matter states is given in terms of intersection numbers involving these classes. Hence, the full charged matter spectrum can be obtained base-independently.

Finally, one can compute the Euler number $\chi$ of the Calabi-Yau 3-fold $Y_3$ directly from the polytope data. First, we note that the number of K\"ahler deformations is generically given by
\begin{align}
\label{eq:Kahler}
h^{1,1} (Y_3) = \text{rank} (\overline{G}) + h^{1,1}(B) - 1 + \sum_i \delta_i \, n_{\text{SCP}_i}
\end{align}
where the last term accounts for the K\"ahler deformations in the presence of a non-flat fiber point, i.e.\ for models containing SCPs associated to points interior to some facet. This formula can be motivated from the combinatorial Batyrev formula to compute the Hodge numbers from the full polytope $\Diamond$ of the 4d ambient variety of the CY hypersurface as 
\begin{align}
h^{1,1}(Y_3) = \underbrace{ l (\Diamond) - 4 - \sum_\Gamma l^\circ (\Gamma) }_{h^{1,1}_{\text{toric}}}+ \underbrace{ \sum_\Theta l^\circ(\Theta)   l^\circ(\Theta^*)  }_{h^{1,1}_{\text{nt}}} \, ,
\end{align}
where $l (\Diamond)$ counts points in the polytope $\Diamond$, $\Gamma$ are edges, and $\Theta$ denotes codimension-two faces in $\Diamond$, whereas $\Theta^*$ are their dual faces in $\Diamond^*$ of dimension one. $l^\circ$ counts the points in the relative interior of $\Theta$ or $\Theta^*$. Hence the facets $\Theta_i$ can contribute non-toric K\"ahler deformations if they contain interior points and their multiplicity is given by the number of interior points of the dual face $\Theta^*_i$.

This motivates the inclusion of distinct SCPs labeled by $i$ in \eqref{eq:Kahler}, which contribute a multiplicity of K\"ahler deformations given by $\delta_i = l^\circ(\Theta_i)$ in the $i^\text{th}$ face. As we are mainly discussing E-string models, we have $\delta_i = 1$. However, in one example discussed in Section~\ref{e7exa}, we find an SCP of vanishing order larger than $(4,6,12)$, which leads to $\delta_i = 3$, justifying the above generalization.

The number of neutral singlets arising from the complex structure deformations is
\begin{align}
H_{\text{neut}} = h^{2,1} (Y_3) + 1= h^{1,1} (B) + \text{rank} (\overline{G}) + \sum_i \delta_i \, n_{\text{SCP}_i} + 2 - \tfrac{1}{2} \chi (Y_3) \,.
\end{align}
Consequently, the full matter spectrum can be obtained base-independently. As we create $n_{\text{SCP}_i}$ additional non-toric K\"ahler deformations at the cost of $m_i$ complex structure deformations for each of the $n_{\text{SCP}_i}$ additional SCPs, the difference in Hodge and Euler numbers are 
 \begin{align}
 (\Delta \chi, \Delta h^{1,1}, \Delta h^{2,1}) = ( 2(\delta_i + m_i), \delta_i, - m_i) \times n_{\text{SCP}_i}\, .
 \end{align}
Again, for E-string transitions, we have $\delta_i = 1$. As we have argued in the classification in Section~\ref{subsec:class}, $m_i \geq 1$. 

\subsection{SCPs in toric models}
\label{subsec:SCPtoric}

In this section we want to clarify the relation between 2d faces (facets) of the top $\Delta$ with interior points, and the appearance of non-flat fibers corresponding to $(4,6,12)$ singularities in the associated Weierstrass model. We begin the discussion with F-theory on K3. Since K3 is obtained from a polytope built from tops, a point $\hat{p}$ (with toric coordinate $x_{\hat{p}}$ and associated divisor $D_{\hat{p}}=\{x_{\hat{p}}=0\}$) in the facet of the top is a point in the facet of the entire polytope. The divisor $D_{\hat{p}}$ generically misses the anti-canonical hypersurface \eqref{pDiamond}, which manifests itself in the hypersurface becoming a non-zero constant, see also \cite{Braun:2013nqa}. This happens because the monomial $m_i$, which does not depend on $x_{\hat{p}}$, contains only toric coordinates that are in the Stanley-Reisner ideal with $x_{\hat{p}}$,
\begin{align}
\label{eq:HypersurfaceConst}
p_\Diamond = d_i m_i + \sum_{j\neq i}d_j \;x_{\hat{p}}\; m_j\stackrel!=0 \quad\xrightarrow{x_{\hat{p}}=0}\quad d_i\;\text{const} \neq 0\,.
\end{align} 
Now, in going to 6d on a Calabi-Yau 3-fold rather than a K3, the complex constant coefficient $d_i$ becomes a section of an (effective) line bundle corresponding to some divisor $D_i$. Along this divisor, the hypersurface equation is satisfied by setting in addition $x_{\hat{p}}$ \cite{Braun:2013nqa}. This leaves the entire toric ambient space of the fiber ($F_1$ to $F_{16}$), and hence the fiber becomes two-dimensional rather than one-dimensional over these points.

We thus note that at $d_i=0$, $x_{\hat{p}}$ factors out of all sections $\hat{s}_i$, $i=1,\ldots,10$. We can use \cite{An:2001aa} to obtain the Jacobian corresponding to \eqref{eq:HypersurfaceConst}. By using the corresponding maps of the $\hat{s_i}$ to $f$ and $g$ or by  noting that $f$ and $g$ are sections of $K_B^{-4}$ and $K_B^{-6}$, respectively, and studying those combinations of $\hat{s}_i$ that form sections of these bundles using \eqref{bidivisors2}, we find that each monomial contains at least $4$ or $6$ of the $\hat{s}_i$. Thus, $f$ vanishes to order 4, $g$ to order 6, and the discriminant $\Delta=4f^3+27g^2$ to order $12$ generically in $x_{\hat{p}}$ along $d_i=0$. Similarly, a top with $m$ facets with an interior point has $m$ distinct types of non-flat fibers, each having a vanishing order $\text{ord}(4,6,12)$ in codimension two.

The multiplicity of the SCPs is given by the intersection
\begin{align}
n_{\text{SCP}_i} = \Z \cdot D_i \,,
\end{align}
which, like the matter multiplicities, can be determined from the four base divisor classes $\Kbi$, $\Z$, $\Ss$, and $\Sn$. Those SCPs correspond to superconformal matter, due to the presence of tensionless strings. Accordingly, the strongly coupled subsectors should contribute to the degrees of freedom entering the gravitational anomaly in six dimensions. In particular, for the irreducible gravitational anomaly we expect contributions of the form
\begin{align}
H - V + 29 T + 29 \sum_i n_{\text{SCP}_i} = 273 \,,
\label{SCPanom}
\end{align}
which is indeed the case in all models we discuss.

Moreover, one can speculate that faces that contain multiple interior points correspond to non-flat fibers with vanishing order $\text{ord}(f,g,\Delta) > (4,6,12)$. In these cases equation \eqref{SCPanom} needs to be corrected due to the possibility of further gauge algebra and tensor factors in the resolution process of the higher order SCPs. Parametrizing the contribution to the gravitational anomaly of the strongly coupled subsector on $\text{SCP}_i$ by $\mathcal{H}_{i}$, we find
\begin{align}
H - V + 29 T + \sum_i \mathcal{H}_i n_{\text{SCP}_i} = 273 \,.
\end{align}
We claim that the coefficient $\mathcal{H}_i$ can be identified with the dimension of the Higgs branch of the corresponding $\text{SCP}_i$. This dimension can be computed on its tensor branch via
\begin{align}
\mathcal{H}_i = H_{\text{SCP}_i} - V_{\text{SCP}_i} + 29 T_{\text{SCP}_i} \,,
\label{Tensorbranchformula}
\end{align}
where $T_{\text{SCP}_i}$, $V_{\text{SCP}_i}$, and $H_{\text{SCP}_i}$ denote the number of tensor-, vector-, and hypermultiplets appearing after the resolution of the $i^\text{th}$ SCP, respectively. Since we restrict to E-string theories with $V_{\text{SCP}_i}=0=H_{\text{SCP}_i}$ and $T_{\text{SCP}_i}=1$ we obtain a dimension of the Higgs branch given by $\mathcal{H}_i = 29 T_{\text{SCP}_i}$ and recover the formula given in \eqref{SCPanom}. In Section~\ref{e7exa} we discuss a model with three internal points in one of its faces, leading to $\mathcal{H}_i = 63$. This  motivates the generalization of the anomaly constraints given above.

\subsection{Tensor-matter transitions in toric models}
\label{subsec:transtoric}

Having established the relation between faces with interior points and non-flat fibers, we can construct the tensor-matter transitions in toric hypersurface models in the following way.

Given a specific polytope $\Diamond$ which contains a top $\Delta$ that encodes the non-Abelian gauge algebra factor $G$, one performs a blow-up in the top, which results in an additional vertex of the polytope $\Diamond$. For appropriately chosen blow-ups, this leads to an interior point in one of the faces of the top. In other words, one performs a complex structure deformation that affects the fiber singularity over certain base intersections and leads to the appearance of non-flat fibers. 

For the following discussion we restrict to the vertices of the top $\Delta$ and separate them into the following classes
\begin{itemize}
\item{{\bf Invariant vertices:} These vertices are denoted by $f_i, g_i, h_i, \dots \in \Delta$ depending on their height (i.e.\ their Dynkin multiplicity). They correspond to the vertices associated to resolution divisors in the original top before tuning the complex structure.}
\item{{\bf Interior points:} In cases where the original top $\Delta$ already contains points interior to faces, we denote them by $\hat{f}_i,\hat{g}_i, \dots$, depending on their height. They do not correspond to resolution divisors and signal the presence of SCPs already in the original theory.}
\item{{\bf Blow-up vertex:} The blow-up we perform leads to the appearance of an additional vertex that we denote by $\mathfrak{f}_i$. The associated divisor is a resolution divisor of the ADE singularity in codimension one. This blow-up procedure is related to a complex structure deformation. The new top is denoted by $\hat{\Delta}$.}
\item{{\bf New interior points:} Due to the blow-up one of the original vertices becomes an interior point in one of the faces of $\hat{\Delta}$, which we will denote by $\mathbf{f_i}$. These points are associated to the additional SCPs appearing after the tuning of the complex structure.}
\end{itemize}
refore, the first part of the transition to a singular model is summarized by the following steps. One starts with a certain top $\Delta$ encoding the presence of a non-Abelian gauge algebra $G$. One then performs a blow-up leading to an additional vertex $\mathfrak{f}_i$ at height one for the new top $\hat{\Delta}$. The modified top now contains an additional interior point $\mathbf{f_i}$. Since interior points do not intersect the Calabi-Yau in codimension one, the divisor associated to $\mathbf{f_i}$ cannot be a resolution divisor of the ADE singularity anymore. Moreover, we demand that the codimension-one singularity over $\Z$ remains unchanged, i.e.\ that $\mathfrak{f}_i$ has the same intersections as $\mathbf{f_i}$ before in order to not change the gauge algebra. Also, the two-dimensional polygon at height zero is not altered, so the full gauge algebra remains unchanged. 

In the description of a singular Weierstrass model, this blow-up procedure can be formulated as a modified factorization of the sections $d_i$, see Section~\ref{subsec:toric},
\begin{align}
d_i \rightarrow z_0^{m_i} d_i \,.
\label{SCPfac}
\end{align}
The tuning of the coefficients necessary to induce this new factorization corresponds to a complex structure deformation. This transition is reminiscent of a conifold transition in the following sense: We start with a complex structure deformation, such the Calabi-Yau $Y_{3,\text{sing}}$ becomes singular with $k$ singular $(4,6,12)$ fibers in codimension two in the base. We  then perform a blow-up by adding a vertex to $\Diamond$ in such a way that a former vertex $\mathbf{f}$ of $\Delta$ becomes an interior point to a face of $\Delta$ , resulting in a smooth manifold $\hat{Y}_3$. The single point $\mathbf{f}$ does not intersect $\hat{Y}_3$ in codimension one in the base but exactly $k$ times in codimension two, leading to a non-flat fiber. The interior point $\mathbf{f}$ thus contributes $k$ non-toric K\"ahler deformations. 

\begin{figure}
\centering
\includegraphics[width=.95\textwidth]{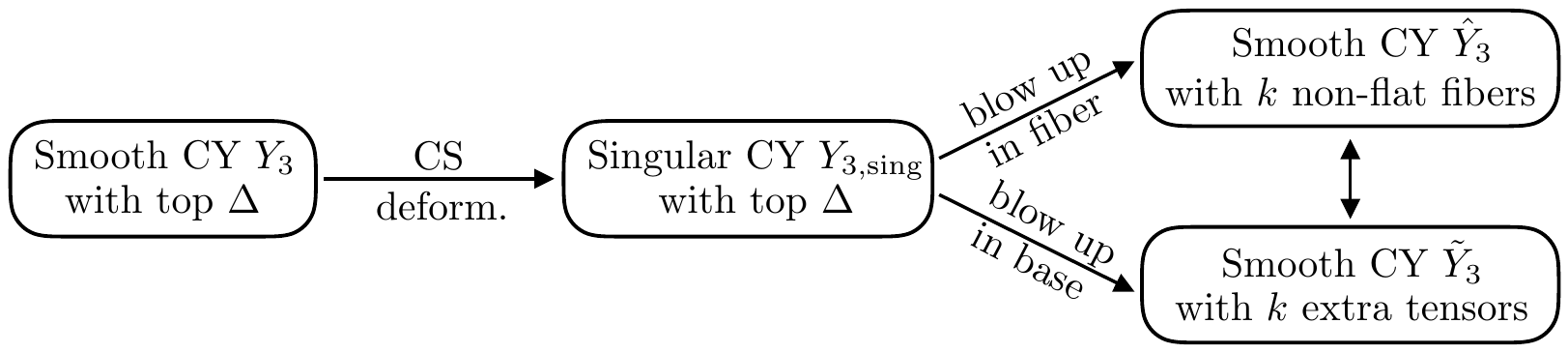}
\caption{Summary of the two transitions. After tuning the complex structure of $Y_3$ to obtain $Y_{3,\text{sing}}$, we can either add another vertex $\mathfrak{f}$ to the top such that we get $k$ SCP points in $\hat{Y}_3$, or we can blowup the base $k$ times, which leads to $k$ new tensor multiplets in $\tilde{Y}_3$. }
\label{fig:FlowchartBlowups}
\end{figure}

The second way to obtain another smooth 3-fold $\tilde{Y}_3$ is by blowing up the $k$ intersection points in the base $B$ directly by introducing $k$ additional vertices in $\Diamond$ in the base of the fibration, which corresponds to the construction outlined in Section~\ref{subsec:blow}. This induces $k$ toric divisors $E_a$  with $a \in \{1,\ldots,k \}$ resulting in a smooth fibration with no non-flat fibers and a well-defined 6d SUGRA limit. Finally, there exists also a transition from $\hat{Y}_3$ with $k$ non-flat fibers to $\tilde{Y}_3$  by the same blow-up procedure.
In this way, the $k$ non-toric K\"ahler deformations of the non-flat fibers get exchanged for toric ones in the base. Both ways are summarized in Figure~\ref{fig:FlowchartBlowups}

\subsection{Rational sections in toric hypersurfaces}
\label{subsec:toricrational}

Now that we discussed the realizations of tensor-matter transitions in toric hypersurface models, we can investigate the change in the rational sections and especially the Abelian anomaly coefficients of these models during the transition.

The homogeneous coordinates of the zero section $s_0$ and the rational sections $s_i$ are given by polynomial equations in the sections $\hat{s}_a$ parameterizing the base-dependence of the fiber ambient space\footnote{Since both the rational sections $s$ and the toric sections $\hat{s}$ appear here, we additionally distinguish them by using indices $i$ and $a$, respectively.}. We denote the coordinates by
\begin{align}
s_0: ~[P^0_1(\hat{s}_a): ... : P^0_k(\hat{s}_a)] \,, \qquad~~ s_i: ~ [P^i_1(\hat{s}_a): ... : P^i_k(\hat{s}_a)] \,,
\end{align}
where $k$ is the number of coordinates describing the toric ambient space of the fiber. In order to evaluate the quantities $\sigma^{\alpha}_{i0}$ and $\sigma^{\alpha}_{ij}$ appearing in the Abelian anomaly coefficients, we have to solve equations of the type
\begin{align}
\begin{split}
\sigma^{\alpha}_{i0}: \quad [P^0_1(\hat{s}_a): ... : P^0_k(\hat{s}_a)] \big|_{H_{\alpha}} &= [P^i_1(\hat{s}_a): ... : P^i_k(\hat{s}_a)] \big|_{H_{\alpha}} \,, \\
\sigma^{\alpha}_{ij}: \quad [P^i_1(\hat{s}_a): ... : P^i_k(\hat{s}_a)] \big|_{H_{\alpha}} &= [P^j_1(\hat{s}_a): ... : P^j_k(\hat{s}_a)]\big|_{H_{\alpha}} \,.
\end{split}
\end{align}
Both sides are restricted to the base divisor $H_{\alpha}$ in order to yield the corresponding coefficient with index $\alpha$. In general, these equations simplify to a single polynomial equation depending on the sections $\hat{s}_a$ \cite{Klevers:2014bqa},
\begin{align}
\hat{P}(\hat{s}_a) \big|_{H_{\alpha}} = 0 \,.
\end{align}
Including a top that corresponds to a non-Abelian gauge algebra $G$ induces the factorization discussed in \eqref{secfac} and can lead to a factorization of the polynomial $\hat{P}$ in terms of $z_0$ and the remaining base coordinates. Similarly, after complex structure deformations, the sections $d_a$ factorize further, see \eqref{SCPfac}, which leads to possible modifications in $\tilde{\sigma}^{\alpha}_{i0}$ and $\tilde{\sigma}^{\alpha}_{ij}$ as discussed at the end of Section~\ref{subsec:blow}. However, since the factorization only involves the coordinate $z_0$ which corresponds to the base divisor $\Z = b^{\alpha} H_{\alpha}$, we see that the additional terms have to be proportional to $b^{\alpha}$,
\begin{align}
(\tilde{\sigma}^{\alpha}_{i0} - \sigma^{\alpha}_{i0}) \propto b^{\alpha} \,, \qquad (\tilde{\sigma}^{\alpha}_{ij} - \sigma^{\alpha}_{ij}) \propto b^{\alpha} \,.
\end{align}
Similarly, the modification of the coefficients $c_{ij}$ only contributes via $b^{\alpha}$ and in the toric hypersurface models we find in the notation of \eqref{Abeliananomcoef} that
\begin{align}
\Delta b^{\alpha}_{ii} = \kappa_{ii} b^{\alpha} \,, \quad \Delta b^{\alpha}_{ij} = \kappa_{ij} b^{\alpha} \,,
\label{abelianprop}
\end{align}
The coefficients $\Delta b^a_{i0}$ and $\Delta b^a_{ij}$ can be constrained in a similar way by considering the base blow-up. However, as explained in Section~\ref{subsec:genconst}, they are not needed for our discussion.

Furthermore, it turns out that the factorization in all our toric examples does not change the coefficients $\sigma^{\alpha}_{i0}$ and $\sigma^{\alpha}_{ij}$. This can be seen by studying the factorization of the polynomials $\hat{P}(\hat{s}_a) \big|_{H_{\alpha}}$ with respect to the complex structure deformation leading to SCPs. Moreover, in the toric examples of the tensor-matter transitions discussed in this paper, the blow-up in the top does not modify the intersections of the Shioda map with respect to the curve over the non-Abelian divisor $\Z$. Hence, also the $c_{ij}$ generically remain unchanged in toric constructions. These considerations allow a further restriction of the Abelian charges of matter states involved in tensor-matter transitions. Even though for general models the change in the Abelian anomaly coefficients has to be checked on a case by case basis, it is satisfied in all our toric hypersurface examples.

\section{Examples}
\label{sec:Examples}

The following examples are F-theory realizations of six-dimensional supergravity theories coupled to a collection of non-Abelian gauge algebras as well as Abelian (discrete) symmetries. The non-Abelian gauge algebras are mostly constructed via tops over generic bases. For each example we discuss transitions towards a theory with SCPs that can be obtained by a toric conifold transition such that the resulting top admits a face with a point in its interior, as was discussed in Section~\ref{sec:torichyper}. 

The examples we pick have increasingly higher rank gauge algebras and individually demonstrate some unique features. The first example we construct in Section~\ref{su3exa} is an SU(3) theory coupled to a $\mathbbm{Z}_3$ discrete gauge theory, where the SCP conifold tunes points of vanishing order $\text{ord}(2,2,4)$ to SCPs with vanishing order $\text{ord}(4,6,12)$. The subsequent resolution of the SCPs involves matter charged in the fundamental representation of SU(3) as well as discretely charged and neutral singlets. Next, we discuss an SU(5)$\times$SU(2)$\times$U(1) example in Section~\ref{su5exa} and an SO(10)$\times$U(1)$\times\mathbbm{Z}_2$ example in Section~\ref{so10exa}, the latter of which features again discretely charged singlets. After that, we discuss in Section~\ref{e6exa} two different transitions within an E$_6 \times \text{U(1)}^2$ theory distinguished by different U(1) charged matter originating from non-homologous non-flat fiber points. Finally, we discuss E$_7$ theories in Section~\ref{e7exa}, which involve a neutral half-hypermultiplet in the representation $\mathbf{56}$ as well as non-flat fiber points with several points in one face.

\subsection[\texorpdfstring{SU(3)$\times\mathbbm{Z}_3$}{SU(3) x Z\_3} transitions]{\texorpdfstring{SU(3)$\boldsymbol{\times\mathbbm{Z}_3}$}{SU(3) x Z\_3} transitions}
\label{su3exa}

In this section we discuss an SCP transition within a toric theory containing an SU(3) gauge algebra since this is the minimal transition that we can engineer within toric geometry. For simplicity, we choose the cubic in $\mathbbm{P}^2$ as the ambient space of the generic fiber, which is a genus-one curve and admits a discrete $\mathbbm{Z}_3$ symmetry. The respective polytopes before and after the transition are shown in Figure~\ref{fig:SU3Top}. The base-independent spectrum is given in Table~\ref{tab:SU3}. The hypersurface equation obtained from the polytope leads to a factorization of the ten sections $\hat{s}_i$ of the cubic, cf.~\eqref{genericpolynomsections},
\begin{align}
\begin{split}
\hat{s}_1  &= f_0 d_1 \,, \enspace \hat{s}_2 = d_2 f_0 f_2 \,, \enspace \hat{s}_3 = d_3 f_0 f_2^2 \,, \enspace \hat{s}_4 = d_4 f_0 f_2^3 \,, \enspace \hat{s}_5 = d_5f_0 f_1  \,, \\ \hat{s}_6 &= d_6  \,, \enspace
\hat{s}_7  = d_7 f_2 \,, \enspace \hat{s}_8 = d_8 f_0  f_1^2\,, \enspace \hat{s}_9 = d_9 f_1 \,, \enspace \hat{s}_{10} = d_{10} f_0 f_1^3 \,,
\end{split}
\end{align}
where we choose $f_0$ as the affine node of SU(3).
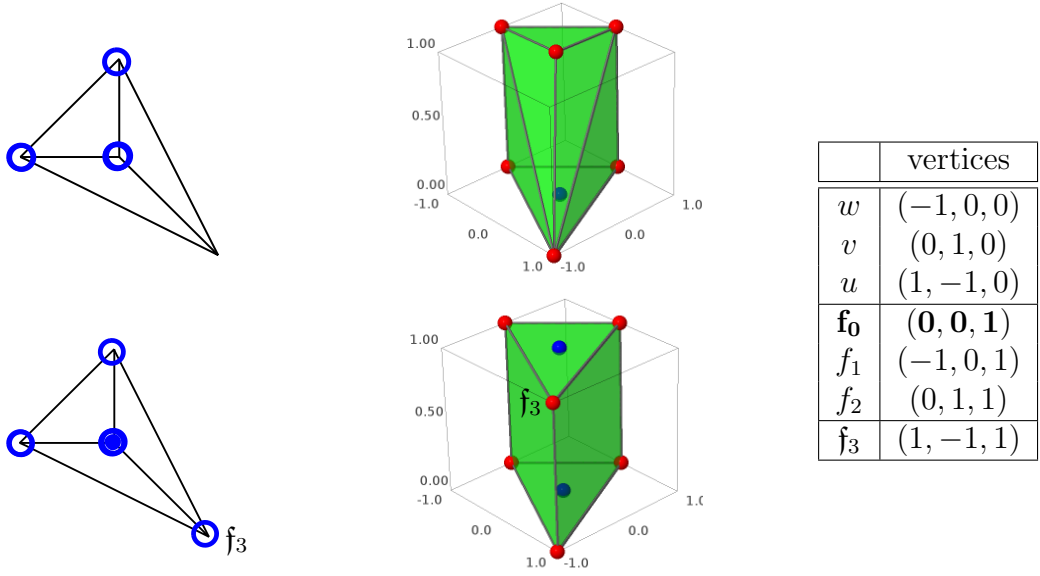
\begin{figure}
\begin{center}
\begin{picture}(400,200)
	\put(0,150){\includegraphics[width=0.18\textwidth]{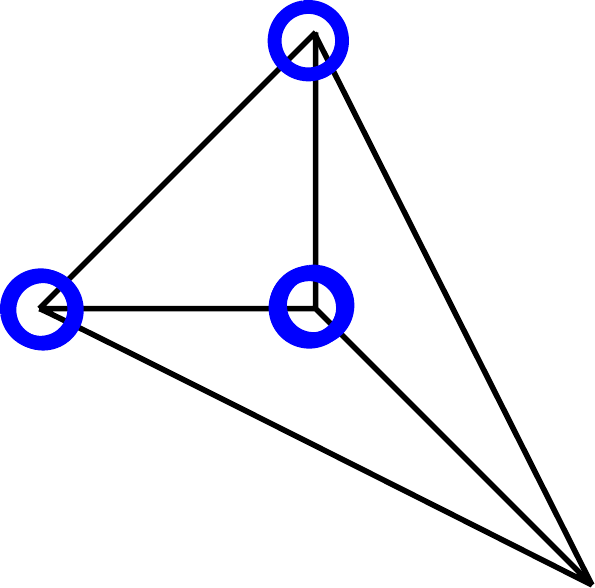}} 
	\put(82,40){$\mathfrak{f}_3$}
	\put(150,140){\includegraphics[width=0.25\textwidth]{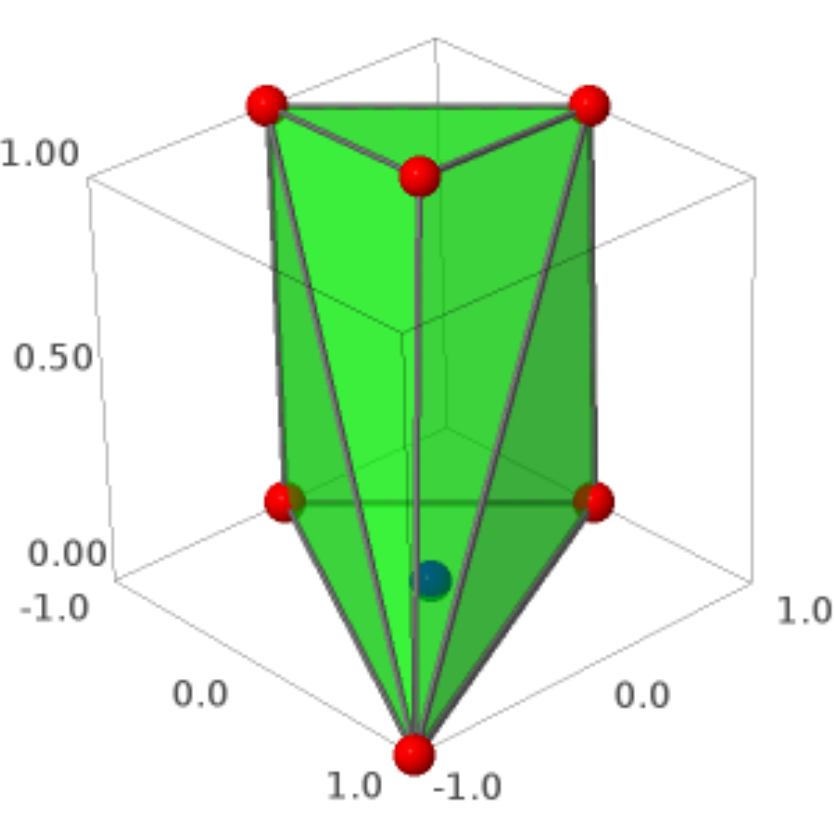}}
	\put(0,40){\includegraphics[width=0.18\textwidth]{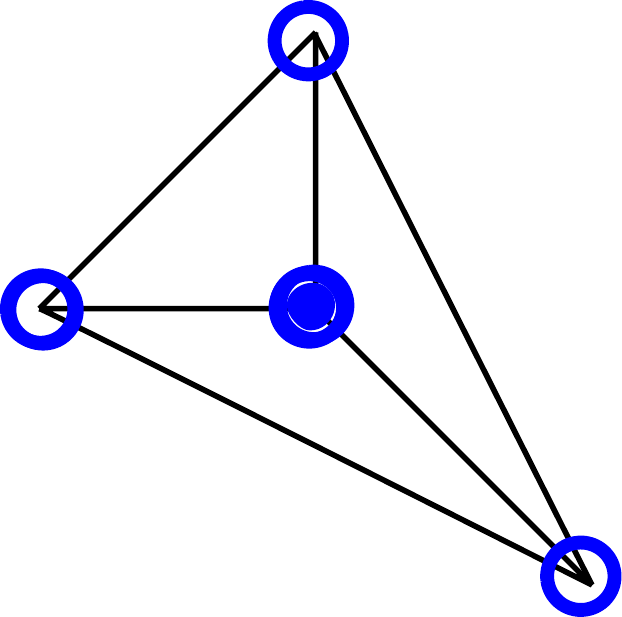}}
	\put(150,30){\includegraphics[width=0.25\textwidth]{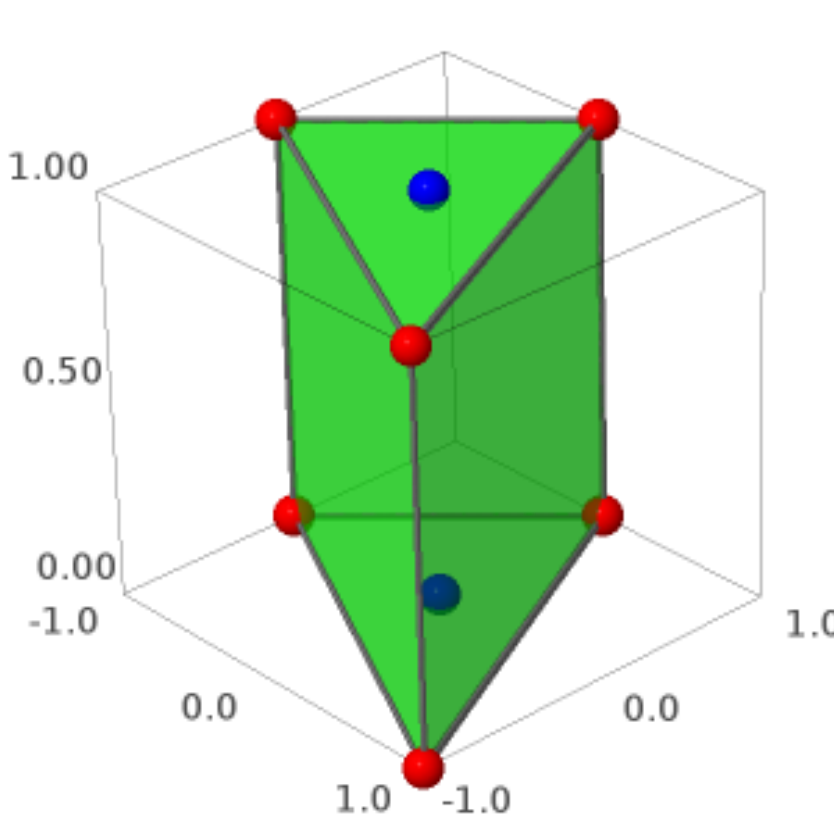}}
	\put(192,92){$\mathfrak{f}_3$}
	\put(300,130){
		\begin{tabular}{|c|c|}
		\hline
		 & vertices \\ \hline \hline
		$w$	& $(-1,0,0)$ \\
		$v$ & $(0,1,0)$ \\
		$u$& $(1,-1,0)$ \\ \hline
		$\mathbf{f_0} $ & $\mathbf{(0,0,1)}$ \\  
		$f_1$ &  $(-1,0,1)$ \\  
		$f_2$ & $(0,1,1)$ \\ \hline
		$\mathfrak{f}_3$ &$(1, -1, 1)$\\
		\hline
		\end{tabular}}
\end{picture}
\vspace{-1.2cm}
\end{center}
\caption{\label{fig:SU3Top} The SU(3) top polytopes over $F_1$ before (upper figure) and after (lower figure) the transition. Simplified depictions of the polytope are shown on the left, where vertices at height one are drawn as blue circles and points internal to facets are drawn as filled blue circles. The toric blow-up by $\mathfrak{f}_3$ leaves the vertex $\mathbf{f_0}$ in a face.}
\end{figure}
The discrete charges can be computed by the discrete Shioda map,
\begin{align}
\sigma(s^{(3)}) = [u] + [f_1] + [f_2] \, , \text{ with } \sigma(s^{(3)})\cdot([f_0],[f_1],[f_2]) = (3,0,0)\, .
\end{align}
To find the matter loci, we use the Jacobian of the cubic as well as the induced factorization after blow down of all curves $(f_0, f_1, f_2) \rightarrow (z_0, 1,1)$ and check the discriminant.
\begin{table}
\centering
\small
\begin{tabular}{| c | c | c | c |} \hline
locus& ord$(f,g,\Delta)$ & Multiplicity & $\mathbf{R}$ \\ \hline
$z_0 = 0 $& $(0,0,3)$ & $ 1+ \tfrac{1}{2} \mathcal{Z} (\mathcal{Z}- \Kbi )) $ & $(\mathbf{1},\mathbf{8})_{0}$ \\   \hline
$d_1=z_0=0$&$ (0,0,4)$& $ (3 \Kbi - \Ss - \Sn - \Z) \Z$ & $\mathbf{3}_{0}$ \\ \hline
$\begin{array}{l} -d_4 d_6^3 + d_3 d_6^2 d_7   \\-d_2 d_6 d_7^2 + d_1 d_7^3\\ = z_0 = 0\end{array}$ & $(0,0,4)$ & $(2 \Ss - \Sn - \Z + 3 \Kbi) \Z $& $ \mathbf{3}_{1}$ \\ \hline
$\begin{array}{l} -d_{10} d_6^3 + d_6^2 d_8 d_9 \\- d_5 d_6 d_9^2 + d_1 d_9^3\\ = z_0 = 0\end{array}$ & $(0,0,4)$ & $(2 \Sn - \Ss - \Z + 3 \Kbi) \Z $& $ \mathbf{3}_{2}$ \\ \hline
$d_6 = z_0 = 0$ & $(2,2,4) $& $ \Kbi \mathcal{Z} $&  $(-)$ \\ \hline 
$V(I_{(1)})$ p.25 in \cite{Klevers:2014bqa}  & $(0,0,2)$ &$ \begin{array}{c} 1 - 28 (\Kbi)^2 - 3 \Ss\Sn - 3 \Kbi (\Ss + \Sn)\\ + 3 (\Ss^2 + \Sn^2) + 17 \Kbi \Z - 5 \Z^2 \end{array}$& $\mathbf{1}_{1}$ \\ \hline
$h^{2,1}(X)+1$ & -&$\begin{array}{l}  11 (\Kbi)^2 - 3 \Kbi (\Ss + \Sn + 3 \Z)\\ + 3 (\Ss^2 - \Ss \Sn + \Sn^2 + \Z^2)+14  \end{array}$ & $\mathbf{1}_{0}$ \\ \hline 
\end{tabular}
\caption{Base-independent matter spectrum for the upper polytope in Figure~\ref{fig:SU3Top}.}
\label{tab:SU3}
\end{table}
The discrete charges of SU(3) matter originate from the intersections of $\sigma(s^{(3)})$ with the irreducible components of the smooth fiber in codimension two. We point out that the locus $d_6 = z_0$ corresponds to a $(2,2,4)$ locus that does not contribute any charged matter. Finally, we compute the neutral singlets via the base-independent Euler number,
\begin{align}
\chi = -6 (4 (\Kbi)^2 + \Sn^2 - \Sn \Ss + \Ss^2 + \Z^2 - \Kbi (\Sn + \Ss + 3 \Z))\,.
\end{align}
The transition towards the second top can be performed by adding $\mathfrak{f}_3 = (1,-1,1)$ to the top, which results in $\mathbf{f_0}$ becoming an interior point of a face. In terms of the singular geometry, this amounts to a tuning $(d_7, d_9) \rightarrow (d_7 z_0 , d_9 z_0)$. We can read off the change of the matter spectrum from Table~\ref{tab:SU3} by imposing this factorization. The various non-toric $\mathbf{3}$-plet loci become toric, inducing a change in their multiplicity. Moreover, these loci contribute additional powers of $d_6$, such that the $(2,2,4)$ locus enhances to an SCP with vanishing order $(4,6,12)$ with multiplicity
\begin{align}
n_{\text{SCP}}=\Kbi \Z \, .
\end{align}
Due to the two tunings in $d_7$ and $d_9$, we loose two complex structure degrees of freedom per SCP. This statement is confirmed by the computation of the difference in Euler and Hodge numbers,
\begin{align}
(\Delta \chi, \Delta h^{1,1}, \Delta h^{2,1}) = (6, 1, - 2) \times n_{\text{SCP}} \,.
\end{align}
The difference in $h^{1,1}$ comes from non-toric K\"ahler deformations associated to the additional non-flat fiber points. Hence, we obtain the total change of the matter spectrum
\begin{align}
\Delta \mathcal{S} = - (3 \times \mathbf{3}_1 \oplus 3 \times \mathbf{3}_2 \oplus 9 \times \mathbf{1}_1 \oplus 2 \times \mathbf{1}_0) \times n_{\text{SCP}} \, .
\end{align}
The change in the matter spectrum matches perfectly the general considerations in Section~\ref{subsec:class}. Moreover, we find that mainly discretely charged matter is lost in this example. As also noted in \cite{Anderson:2018heq}, it would be desirable to link the absence of those singlets to discrete anomaly cancellation.

\subsection[\texorpdfstring{SU(5)$\times$SU(2)$\times$U(1)}{SU(5)xSU(2)xU(1)} transitions]{\texorpdfstring{SU(5)$\boldsymbol{\times}$SU(2)$\boldsymbol{\times}$U(1)}{SU(5)xSU(2)xU(1)} transitions}
\label{su5exa}

Next, we discuss a toric example of an SCP transition within an SU(5) gauge algebra. We consider again one of the toric resolved models that have been classified in \cite{Bouchard:2003bu}. We start with an SU(5) top over the polygon $F_6$ with an SU(2)$\times$U(1) gauge symmetry as has been considered already in \cite{Braun:2013nqa,Borchmann:2013hta}. The polytope before and after the transition is given in Figure~\ref{fig:SU5Top}. 
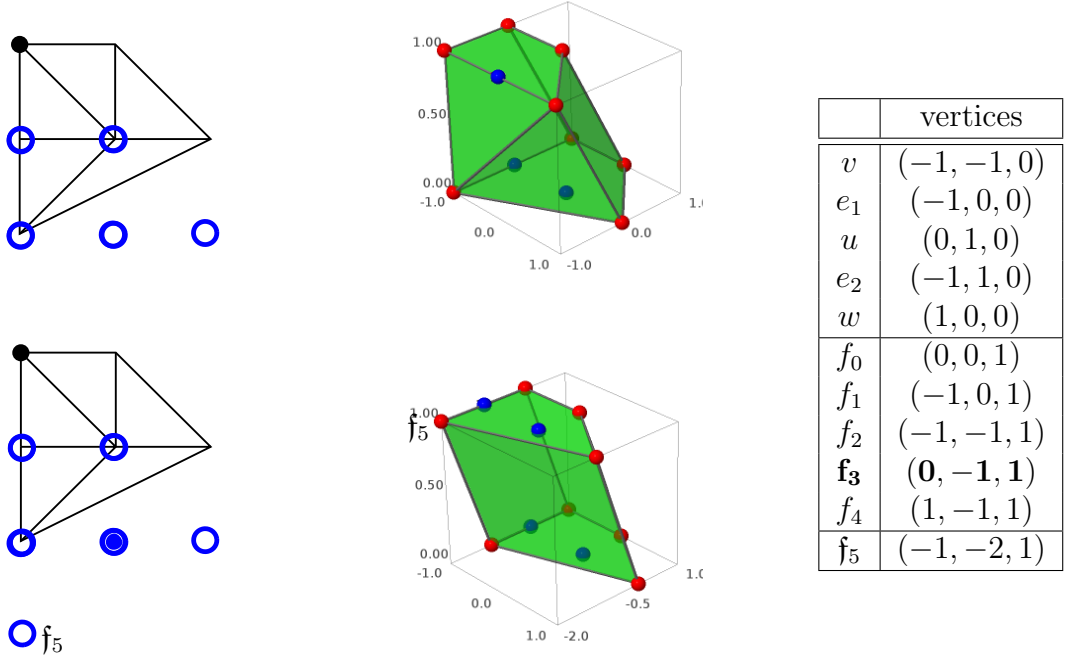
\begin{figure}
\begin{center}
\begin{picture}(400,200)
	\put(0,150){\includegraphics[width=0.18\textwidth]{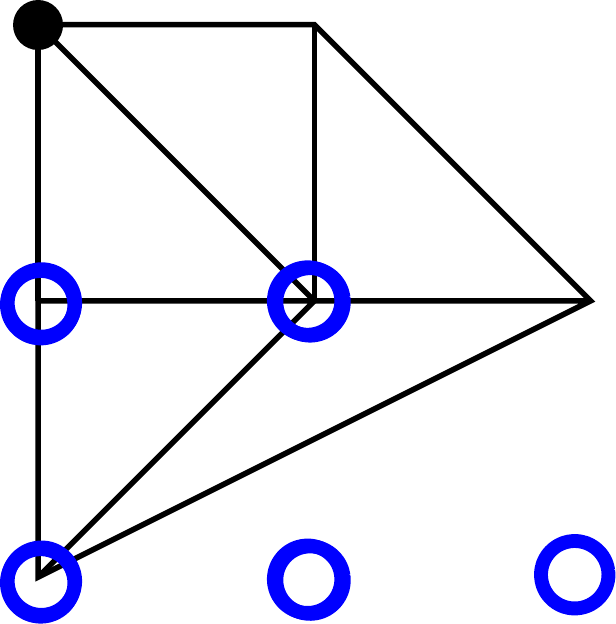}} 
	\put(150,140){\includegraphics[width=0.25\textwidth]{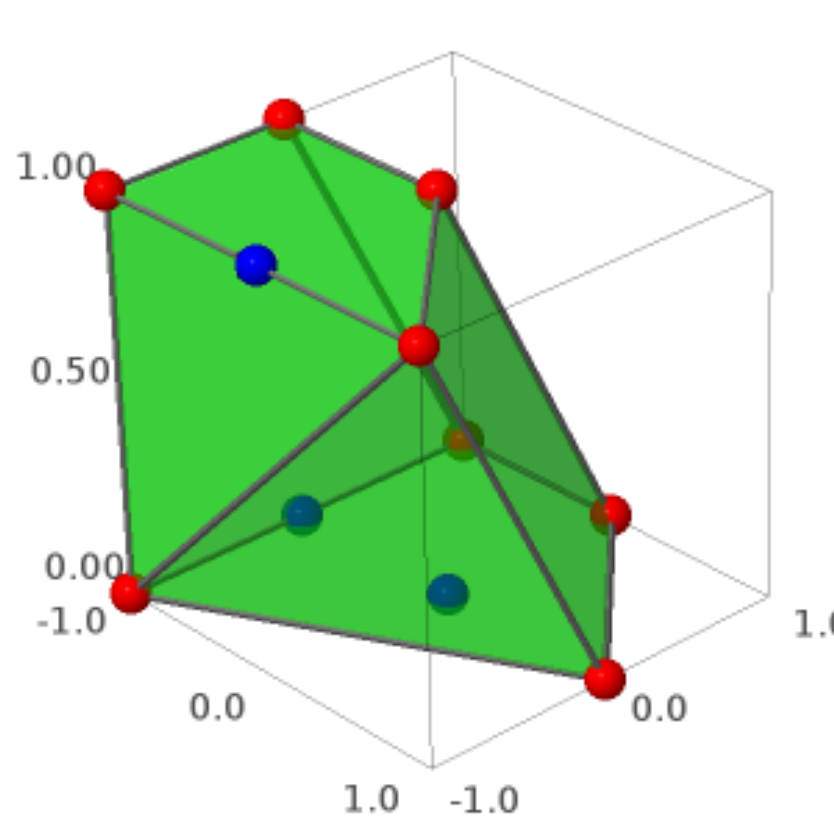}} 
	\put(0,0){\includegraphics[width=0.18\textwidth]{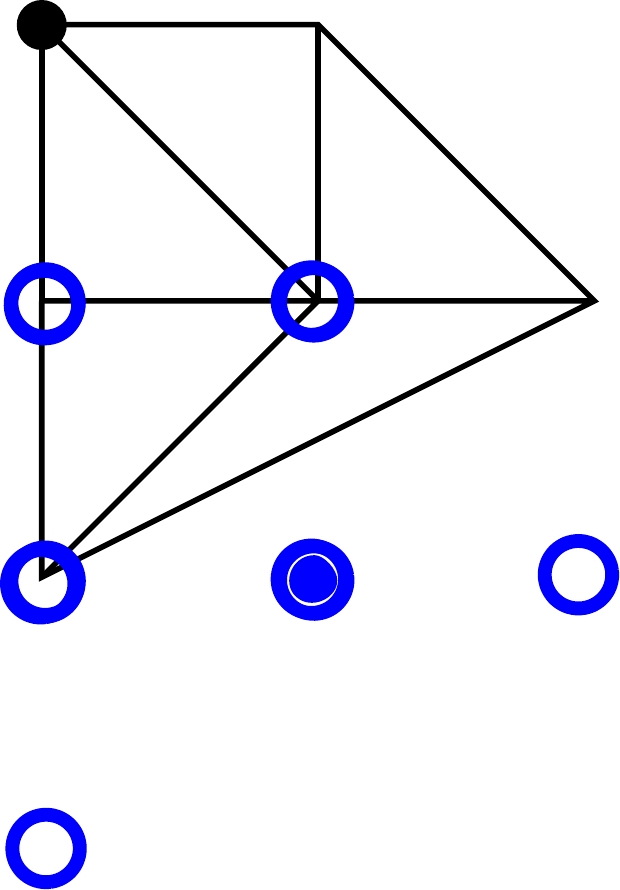}} 
	\put(150,0){\includegraphics[width=0.25\textwidth]{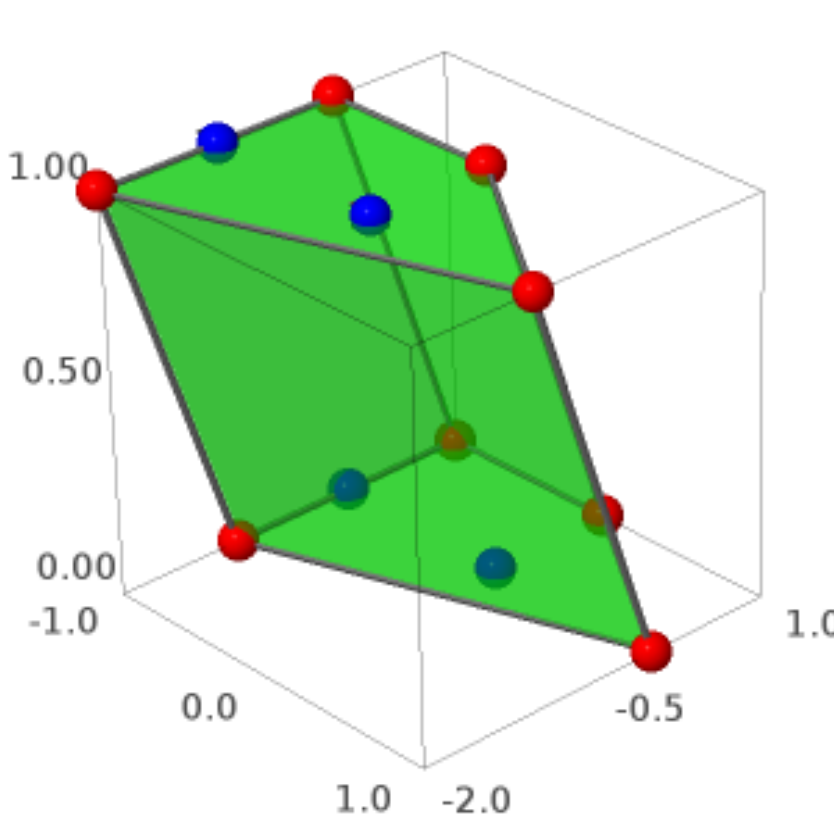}} 
	\put(13,0){$\mathfrak{f}_5$}
	\put(150,80){$\mathfrak{f}_5$}
	\put(300,115){ \begin{tabular}{|c|c|}
	\hline
 		 & vertices \\ \hline \hline
		$v$ & $(-1,-1,0)$ \\
		$e_1$ & $(-1,0,0)$ \\
		$u$& $(0,1,0)$ \\
		$e_2$ & $(-1,1,0)$ \\  
		$w$ &  $(1,0,0)$ \\ \hline
		$f_0$ & $(0,0,1)$ \\
 		$f_1$ & $(-1,0,1)$ \\
		$f_2$ & $(-1,-1,1)$ \\ 
		$\mathbf{ f _3}$ &$ \mathbf{(0,-1,1)}$ \\ 
		$f _4$ & $(1,-1,1)$ \\  \hline
		$\mathfrak{f}_5$ & $(-1,-2,1)$\\
		\hline
		\end{tabular}}
\end{picture}
\end{center}
\caption{\label{fig:SU5Top} The SU(5) top polytopes over $F_6$ before (upper figure) and after (lower figure) the transition. Simplified depictions of the polytope are shown on the left with vertices at height one as blue circles. The toric blow-up by $\mathfrak{f}_5$ leaves one vertex $\mathbf{f_3}$ in a face.
}
\end{figure}
The factorization is given by
\begin{align} 
\begin{split}
\hat{s}_1 &= d_1 f_0^3 f_2^4 f_3^2 f_4 \,, \enspace \hat{s}_2 = d_2 f_0^2 f_2^3 f_3^2 f_4 \,, \enspace \hat{s}_3 = d_3 f_0 f_2^2 f_3^2 f_4 \,, \enspace \hat{s}_4 = d_4 f_2 f_3^2 f_4 \,, \\
\hat{s}_5 &= d_5 f_0 f_2 \,, \enspace \hat{s}_6 = d_6  \,, \enspace \hat{s}_7 = d_7 f_3 f_4 f_5\,, \enspace \hat{s}_8 = d_8 f_0 f_4 f_5^2 \,,
\end{split}
\label{eq:SU5factorization}
\end{align}
where $f_0$ denotes the affine node which we use in order to obtain the shift of the sections $d_i$ by the SU(5) divisor $\Z$. Due to the intersection of the rational sections with the SU(5) divisors, the Shioda map reads
\begin{align}
\sigma (s_1)=  [u]-[e_2]+\tfrac{1}{2} [e_1] - \Kbi -\Ss -\tfrac{1}{5} \left(4 [f_1]+3 [f_2]+2[f_3]+[f_4] \right) \, .
\end{align}
Hence we have fractional U(1) charges for the SU(5) representations. The neutral hypermultiplets can be computed from the complex structure deformations which we obtain from the base-independent expression for the Euler number of the full 3-fold,
\begin{align}
\begin{split}
\chi =& -2 \big(12 (\Kbi)^2 + 4 \Ss^2 - 3 \Ss \Sn + 2 \Sn^2 + 5 \Ss \Z \\
&+ \Sn \Z + 8 \Z^2 - 2 \Kbi (2 (\Ss + \Sn) + 9 \Z)) \,.
\end{split}
\end{align} 
The full spectrum is summarized in Table \ref{Tab:SU5matter}. It is consistent with all 6d anomaly conditions. The second model given in Figure~\ref{fig:SU5Top} can be obtained by adding the vertex $\mathfrak{f}_5 = (-1,-2,1)$ to the polytope such that the vertex $\mathbf{f_3} = (0,-1,1)$ becomes an interior point of a face and a $(4,6,12)$ singularity at $d_6 = z_0 = 0$ is created with multiplicity
\begin{align}
n_{\text{SCP}} = \Kbi \Z \, .
\end{align}
In terms of the factorized singular model, this change is achieved by tuning $d_5 \rightarrow d_5 z_0$ followed by a resolution. In this way, one sees that the non-toric locus of the matter representation $(\mathbf{1},\mathbf{5})_{1/5}$ changes in the following way:
\begin{align}
(d_4 d_5^3 - d_3 d_5^2 d_6 + d_2 d_5 d_6^2 - d_1 d_6^3) \rightarrow (d_4 d_5^3 z_0^3 - d_3 d_5^2 d_6 z_0^2 + d_2 d_5 d_6^2 z_0 - d_1 d_6^3) \,.
\end{align}
Hence, over $z_0 = 0$ the above ideal becomes reducible into two toric loci $d_1 = 0$ and $d_6=0$. The $(\mathbf{1},\mathbf{5})_{1/5}$-plets can be found at $d_1=0$. The additional three powers of $d_6=0$ enhance the $(2,3,7)$ locus of the $(\mathbf{1},\mathbf{10})_{-3/5}$-plet states to $(4,6,12)$ points.
\begin{table}
\centering
\small
\begin{tabular}{| c | c | c | c |} \hline
locus& ord$(f,g,\Delta)$ & multiplicity & $\mathbf{R}$ \\ \hline
$z_0 = 0 $& $(0,0,5)$ & $1+ \tfrac{1}{2} \mathcal{Z} (\mathcal{Z}- \Kbi) $ & $(\mathbf{1},\mathbf{24})_{0}$ \\  \hline
$d_4=z_0=0$&$ (0,0,6)$& $ (2 \mathcal{S}_7-\mathcal{S}_9) \mathcal{Z} $& $(\mathbf{1},\mathbf{5})_{-4/5}$ \\ \hline
$ d_6 = z_0 = 0$ & $(2,3,7) $& $ \Kbi \mathcal{Z} $&  $(\mathbf{1},\mathbf{10})_{-3/5}$ \\ \hline 
$d_7 = z_0 = 0$ &   $(0,0,6)$ & $ \mathcal{S}_7 \mathcal{Z} $&  $(\mathbf{1},\mathbf{5})_{6/5}$ \\ \hline
$\begin{array}{l} d_4 d_5^3 - d_3 d_5^2 d_6  \\ + d_2 d_5 d_6^2 - d_1 d_6^3\\ = z_0 = 0\end{array}$ & $(0,0,6)$ & $(6 \Kbi -\mathcal{S}_7-\mathcal{S}_9-3 \mathcal{Z}) \mathcal{Z} $& $(\mathbf{1},\mathbf{5})_{1/5}$ \\ \hline
$d_8 = z_0 =0$ & $(0,0,8)  $ &$(\Kbi-\mathcal{S}_7+\mathcal{S}_9-\mathcal{Z}) \mathcal{Z} $& $(\mathbf{2},\mathbf{5})_{3/10}$ \\ \hline \hline
$d_8 = d_7 =0$ & $(0,0,3)$ &  $ \mathcal{S}_7 (\Kbi-\mathcal{S}_7+\mathcal{S}_9-\mathcal{Z}) $& $(\mathbf{2},\mathbf{1})_{-3/2}$ \\ \hline
$d_4 = d_7 = 0$ & $(0,0,2)$ & $\mathcal{S}_7 (2 \mathcal{S}_7-\mathcal{S}_9) $& $(\mathbf{1},\mathbf{1})_{2}$ \\ \hline
$V(I_{(2)})$ p.48 in \cite{Klevers:2014bqa} & $(0,0,3)$ & $\begin{array}{c} (6 \Kbi +\mathcal{S}_7-2 \mathcal{S}_9-3 \mathcal{Z}) \\ \times (\Kbi-\mathcal{S}_7+\mathcal{S}_9-\mathcal{Z})\end{array} $& $(\mathbf{2},\mathbf{1})_{1/2}$ \\ \hline
$V(I_{(4)})$ p.48 in \cite{Klevers:2014bqa}  & $(0,0,2)$ & $ \begin{array}{c} 6 (\Kbi)^2+ \Kbi (13 \mathcal{S}_7-5 \mathcal{S}_9-3 \mathcal{Z})\\-(3 \mathcal{S}_7-\mathcal{S}_9) (\mathcal{S}_7+\mathcal{S}_9+3 \mathcal{Z})\end{array} $& $(\mathbf{1},\mathbf{1})_{1}$ \\ \hline
$d_8 = 0$ & $(0,0,2)$ & $ \begin{array}{c} 1+ \tfrac{1}{2} (\mathcal{S}_9-\mathcal{Z}-\mathcal{S}_7) \times \\ (\Kbi -\mathcal{S}_7+\mathcal{S}_9-\mathcal{Z}) \end{array}$& $(\mathbf{3},\mathbf{1})_{0}$ \\ \hline 
$h^{2,1}(X)+1$ & - & $\begin{array}{c} 18 + 11 (\Kbi)^2 + 4 \Ss^2 - 3 \Ss \Sn \\ + 2 \Sn^2 + 5 \Ss \Z + \Sn \Z +  8 \Z^2 \\ - 2 \Kbi (2 (\Ss + \Sn) + 9 \Z) \end{array}$ & $\mathbf{1}_{0}$ \\ \hline 
\end{tabular}
\caption{Base-independent matter spectrum for the upper polytope in Figure~\ref{fig:SU5Top}.}
\label{Tab:SU5matter}
\end{table}
Also the total topological numbers change as
\begin{align}
(\Delta \chi, \Delta h^{1,1}, \Delta h^{2,1}) = (4, 1, -1) \times n_{\text{SCP}} \,,
\end{align}
which we can use to compute the total change of the neutral spectrum. Using resultant techniques we thus obtain a total change in the matter spectrum,
\begin{align}
\Delta \mathcal{S} = - \big( (\mathbf{1},\mathbf{10})_{-3/5} \oplus 3 \times (\mathbf{1},\mathbf{5})_{1/5} \oplus 3 \times (\mathbf{1},\mathbf{1})_1 \oplus (\mathbf{1},\mathbf{1})_0 \big) \times n_{\text{SCP}} \,,
\end{align}
which is consistent with the constraints of global tensor-matter transitions with gauge algebra SU(5) discussed in Section~\ref{subsec:warmup}. Since the transition satisfies the additional assumption \eqref{assumption}, the U(1) charges match the constraints presented in Section~\ref{subsec:warmup} as well.

\subsection[\texorpdfstring{SO(10)$\times$U(1)$\times \mathbbm{Z}_2$}{SO(10)xU(1)xZ\_2} transitions]{\texorpdfstring{SO(10)$\boldsymbol{\times}$U(1)$\boldsymbol{\times\mathbbm{Z}_2}$}{SO(10)xU(1)xZ\_2} transitions}
\label{so10exa}

In this section we consider an SCP transition within a toric model with gauge algebra SO(10)$\times$U(1)$\times \mathbbm{Z}_2$. These models can be obtained from SO(10) tops over the polytope $F_2$, i.e.\ SO(10) resolved models with a generic torus fibration described as a hypersurface in the ambient space given by the Hirzebruch surface $\mathbbm{F}_0$. This model has been constructed in \cite{Buchmuller:2017wpe}, which we refer to for further details. The vertices of the top are summarized in Figure~\ref{fig:SO10Top} and the matter spectrum is presented in Table~\ref{tab:SO10}.
\begin{figure}
\begin{center}
\begin{picture}(400,230)
	\put(0,170){\includegraphics[width=0.25\textwidth]{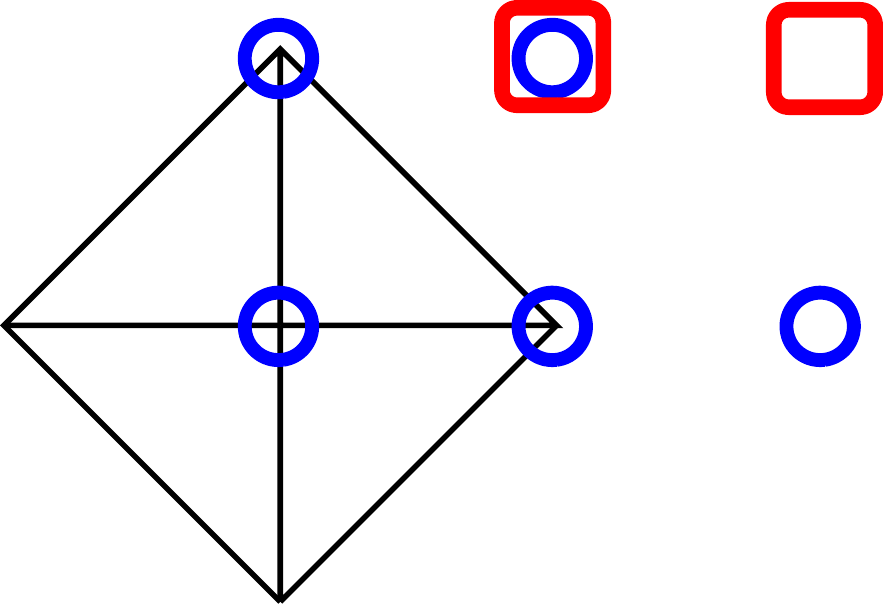}} 
	\put(150,150){\includegraphics[width=0.3\textwidth]{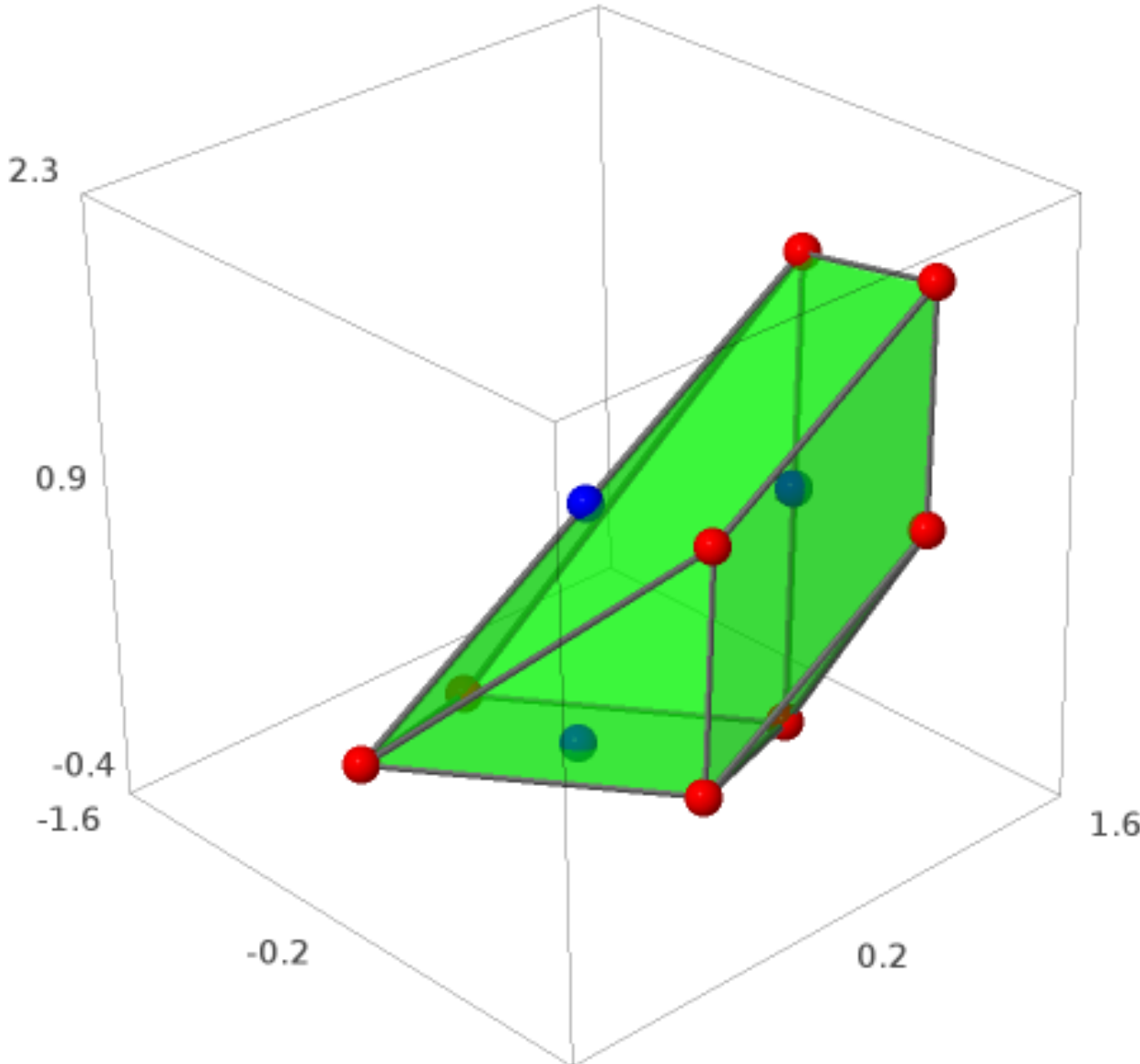}} 
	\put(0,20){\includegraphics[width=0.25\textwidth]{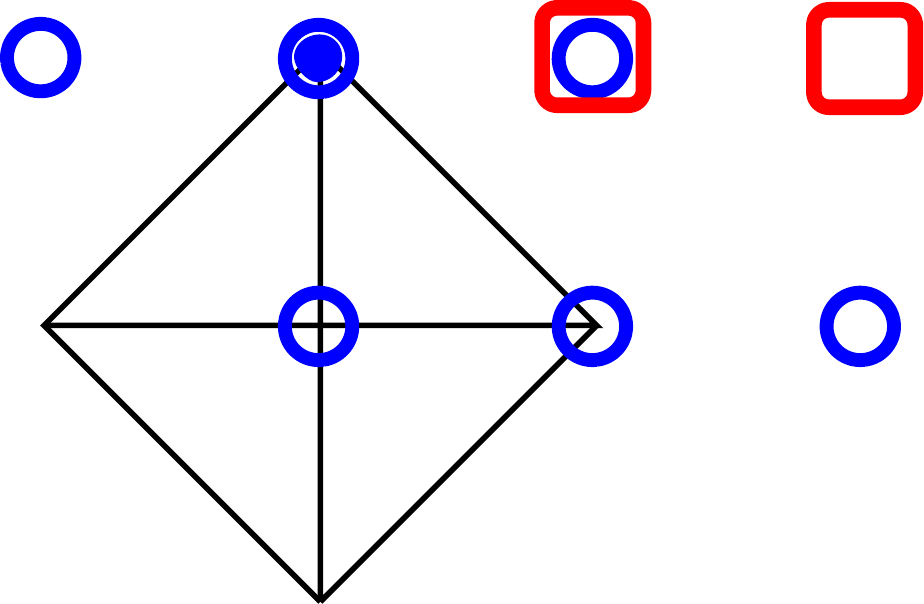}} 
	\put(150,0){\includegraphics[width=0.3\textwidth]{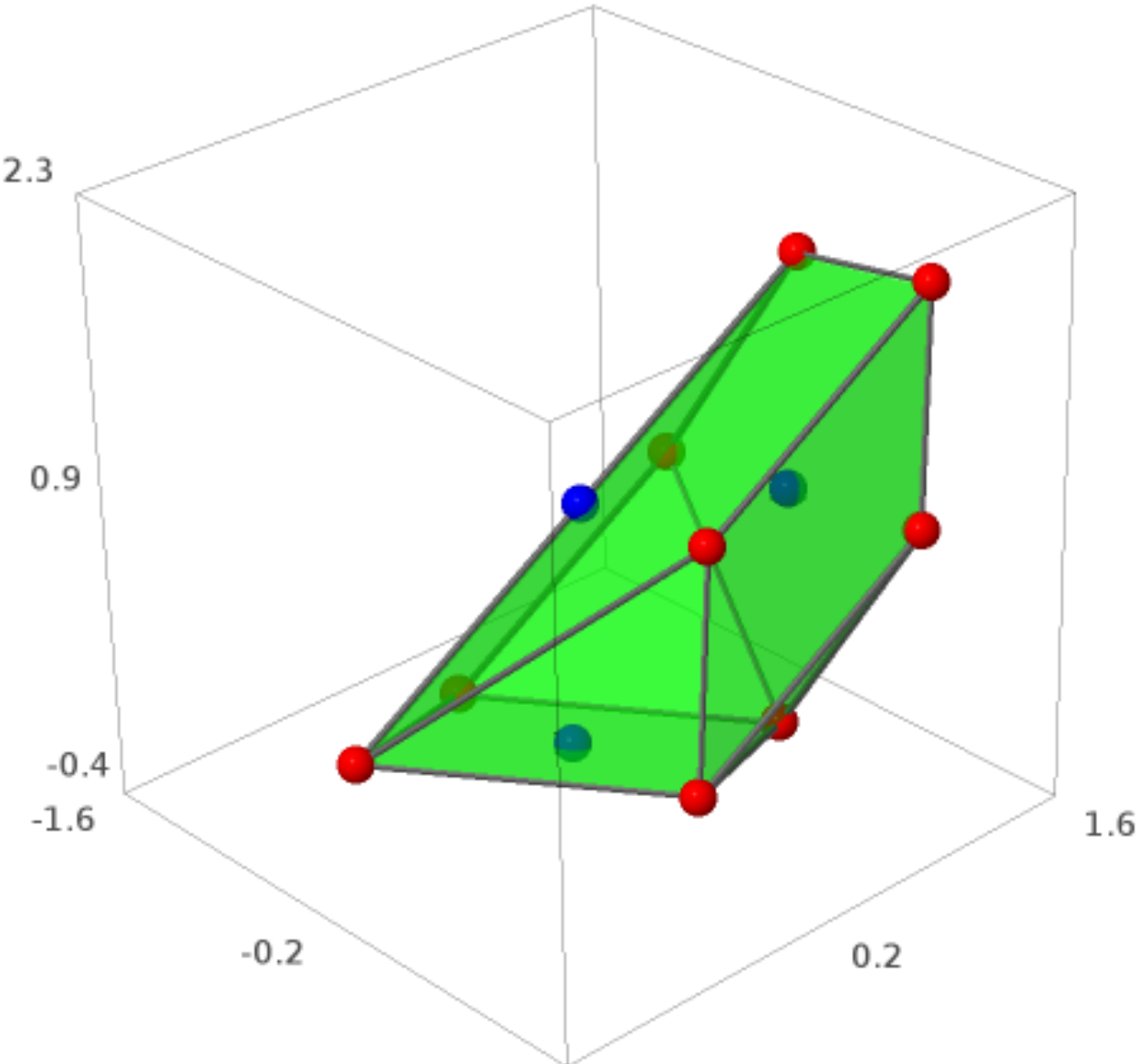}} 
	\put(8,72){$\mathfrak{f}_4$}
	\put(217,74){$\mathfrak{f}_4$}
	\put(310,140){ \begin{tabular}{|c|c|}
	    \hline
 		 & vertices \\ \hline \hline
		$t$ & $(1,0,0)$ \\
		$x$ & $(-1,0,0)$ \\
		$s$ & $(0,-1,0)$ \\
		$y$ & $(0,1,0)$ \\ \hline
		$f_0$ & $(0,0,1)$ \\
 		$f_1$ & $(0,1,1)$ \\
		$\mathbf{f_2}$ & $\mathbf{(1,0,1)}$ \\  
		$f_3$ & $(1,1,1)$ \\ \hline
		$g_1$ & $(0,1,2)$ \\  
		$g_2$ & $(1,1,2)$ \\   \hline
	$\mathfrak{f}_4$ & $(-1,1,1)$\\
	\hline
	\end{tabular}}
\end{picture}
\end{center}
\caption{The SO(10) tops over $F_2$ before (upper figure) and after (lower figure) the transition. Simplified depictions of the polytope are given on the left with vertices at height one as blue circles and vertices at height two as red squares. The toric blow-up by $\mathfrak{f}_4$ leaves the former vertex $\mathbf{f}_2$ as an interior point of a face.}
\label{fig:SO10Top}
\end{figure}
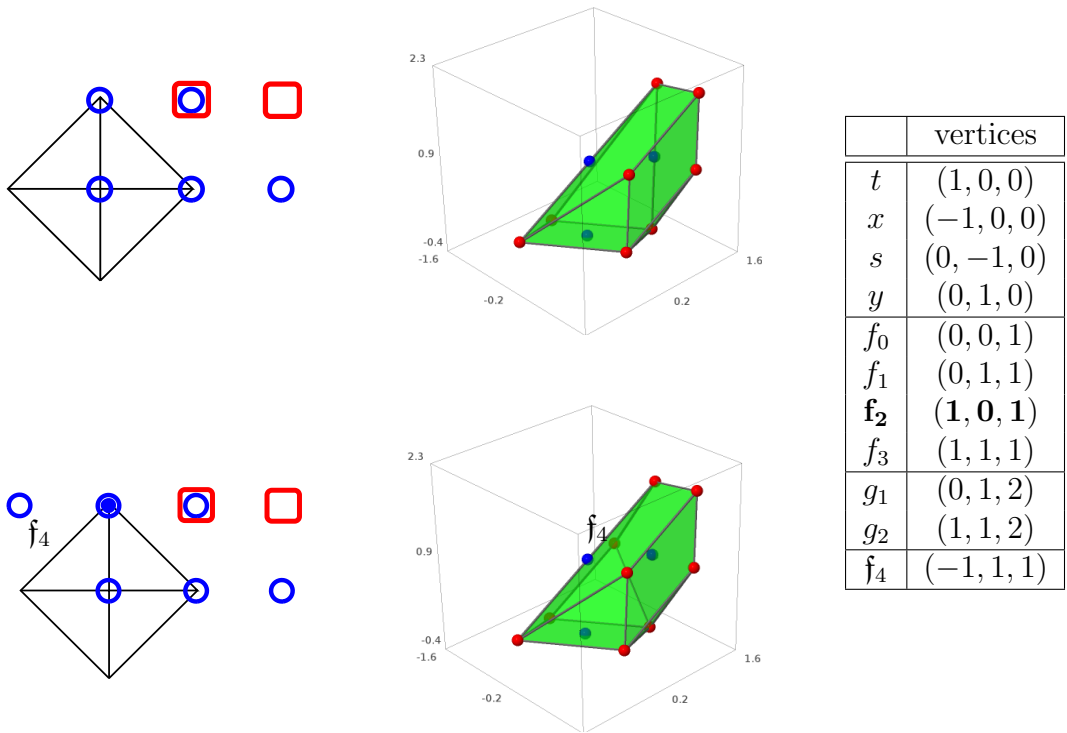
For convenience, we repeat the base-independent Euler number here,
\begin{align}
\begin{split}
 \chi =& -24 (\Kbi)^2 + 8 \Kbi \Ss - 4 \Ss^2 + 8 \Kbi \Sn \\ &- 4 \Sn^2+ 
  24 \Kbi \Z   + 4 \Ss \Z - 2 \Sn \Z - 14 \Z^2 \, .
  \end{split}
\end{align}
The SCP transition is induced by a complex structure deformation, which corresponds to the additional factorization $d_{10} \rightarrow d_{10} z_0$ in the singular model.
In this way, the non-toric $\mathbf{10}_{1/2,0}$ locus becomes toric over $z_0=0$ with reduced multiplicity. In the resolved model, we add the vertex $\mathfrak{f}_4 = (-1,1,1)$ which results in the former vertex $\mathbf{f_2} =(1,0,1)$ becoming an interior point in a face, corresponding to the $z_0 = d_7 = 0$ locus. 
Thus the $\mathbf{16}_{1/4,0}$ locus becomes a $(4,6,12)$ singularity with multiplicity
\begin{align}
n_{\text{SCP}} = (\Ss-\Z) \Z \, .
\end{align}
Moreover, we compute the change in the Euler and Hodge numbers,
\begin{align}
 (\Delta \chi, \Delta h^{1,1}, \Delta h^{2,1}) = (4, 1, -1) \times n_{\text{SCP}} \, ,
\end{align}
In addition, the number of charged SO(10) singlets gets reduced, such that we obtain a total change in the spectrum of
\begin{align}
\Delta \mathcal{S} = - (\mathbf{16}_{1/4,0} \, \oplus \mathbf{10}_{1/2,0} \, \oplus \mathbf{1}_{1,1} \, \oplus \mathbf{1}_{0,1} \, \oplus \mathbf{1}_{0,0}) \times n_{\text{SCP}} \, .
\end{align}
Again, $\Delta \mathcal{S}$ satisfies all constraints of Section~\ref{sec:transanom} for transitions with gauge algebra SO(10)$\times$U(1), since also the additional assumption \eqref{assumption} is respected by the transition.
\begin{table}
\centering
\small
\begin{tabular}{| c | c | c | c |}\hline
locus& ord$(f,g,\Delta)$ & multiplicity & $\mathbf{R}$ \\ \hline 
$z_0 = d_5 = 0$ & $(3,4,8)$ & $(2 \Kbi - \Ss) \Z$ & $\mathbf{16}_{1/4,1}$ \\ \hline
$z_0 = d_2 = 0 $& $(2,3,8)$ & $ (2 \Kbi - \Sn - \Z) \Z $ & $\mathbf{10}_{1/2,1}$    \\ \hline 
$\begin{array}{c} d_{10} d_5 - d_8 d_7 = \\ z_0 = 0 \end{array} $  & $(2,3,8)$ & $(\Kbi + \Sn - \Z) \Z $ & $\mathbf{10}_{1/2,0}$ \\ \hline 
$z_0 = d_7 = 0 $ & $(3,4,8)$ & $(\Ss - \Z) \Z$ & $\mathbf{16}_{1/4,0}$ \\ \hline
$V(I_{(3)})$ p.30 in \cite{Klevers:2014bqa}  & $(0,0,2)$ & $\begin{array}{c} 6 (\Kbi)^2 - 2 \Ss^2 + 2 \Sn^2 + 3 \Z^2 + \Sn \Z \\+ \Kbi (4 \Ss - 4 \Sn - 12 \Z) + 2 \Ss \Z \end{array}$ & $\mathbf{1}_{1,0}$ \\ \hline 
$V(I_{(2)})$ p.30 in \cite{Klevers:2014bqa}  & $(0,0,2)$ & $\begin{array}{c} 6 (\Kbi)^2 + 2 \Ss^2 - 2 \Sn^2 + 3 \Z^2 - \Sn \Z \\ + \Kbi (- 4 \Ss + 4 \Sn - 5 \Z) - 2 \Ss \Z \end{array}$ & $\mathbf{1}_{1,1}$ \\ \hline
$V(I_{(1)})$ p.30 in \cite{Klevers:2014bqa}  & $(0,0,2)$ & $\begin{array}{c} 6 (\Kbi)^2 - 2 \Ss^2 - 2 \Sn^2 + 3 \Z^2 - \Sn \Z \\ + \Kbi (4 \Ss + 4 \Sn - 13 \Z) + 2 \Ss \Z \end{array}$ & $\mathbf{1}_{0,1}$ \\ \hline
 & $h^{2,1}+1$ & $\begin{array}{c} 18 + 11 (\Kbi)^2 + 2 \Ss^2 + 2 \Sn^2 + 7 \Z^2 + \Sn \Z \\ - 4 \Kbi (\Ss + \Sn + 3 \Z) - 2 \Ss \Z \end{array}$ & $\mathbf{1}_{0,0}$ \\ \hline
\end{tabular}
\caption{Base-independent matter spectrum for the upper polytope in Figure~\ref{fig:SO10Top}.}
\label{tab:SO10}
\end{table}
Interestingly, also one discretely charged singlet participates in the transition, making sure that all discretely charged hypermultiplet degrees of freedom sum up to an even number, which might be explained by a 6d version of discrete anomaly cancellation, similar to Section~\ref{su3exa}. Note that the above transition can be unhiggsed to a transition in a model with gauge algebra SO(10)$\times$U(1)$^2$, where both U(1) factors satisfy the strict constraints for the transition imposed in \eqref{specAbelian}.

\subsection[\texorpdfstring{E$_6 \times$U(1)$^2$}{E\_6xU(1)\^{}2} transitions]{\texorpdfstring{E$\boldsymbol{_6 \times}$U(1)$\boldsymbol{^2}$}{E\_6xU(1)\^{}2} transitions}
\label{e6exa}

We consider three different E$_6$ tops over $F_5$, which are depicted in Figure~\ref{fig:E6Top}. The generic gauge algebra of this model is E$_6 \times$U(1)$^2$. It can be constructed from the following factorization enforced by building a top over a dP$_2$ ambient space,
\begin{align}
\begin{split}
\hat{s}_1 &= d_1 f_1 f_2^2 f_4 g_2 \,, \enspace \hat{s}_2 =d_2 f_1^2 f_2^2 f_3 f_4 g_1^2 g_2^2 g_3 h_1^2 \,, \enspace \hat{s}_3 = d_3 f_1^2 f_2 f_3 g_1^2 g_2 h_1 \,, \\
\hat{s}_5 &= d_5 f_1 f_2^2 f_3 f_4^2 g_1 g_2^2 g_3^2 h_1^2 \,, \enspace \hat{s}_6 =  d_6 f_1 f_2 f_3 f_4 g_1 g_2 g_3 h_1  \,, \enspace \hat{s}_7 =d_7 f_1 f_3 g_1 \,, \\
\hat{s}_8 &= d_8 f_2 f_3 f_4^2 g_2 g_3^2 h_1\,, \enspace \hat{s}_9 =d_9 f_3 f_4 g_3 \,.  
\end{split}
\end{align}
\begin{figure}
\begin{center}
\begin{picture}(400,420)
	\put(0,280){\includegraphics[width=0.15\textwidth]{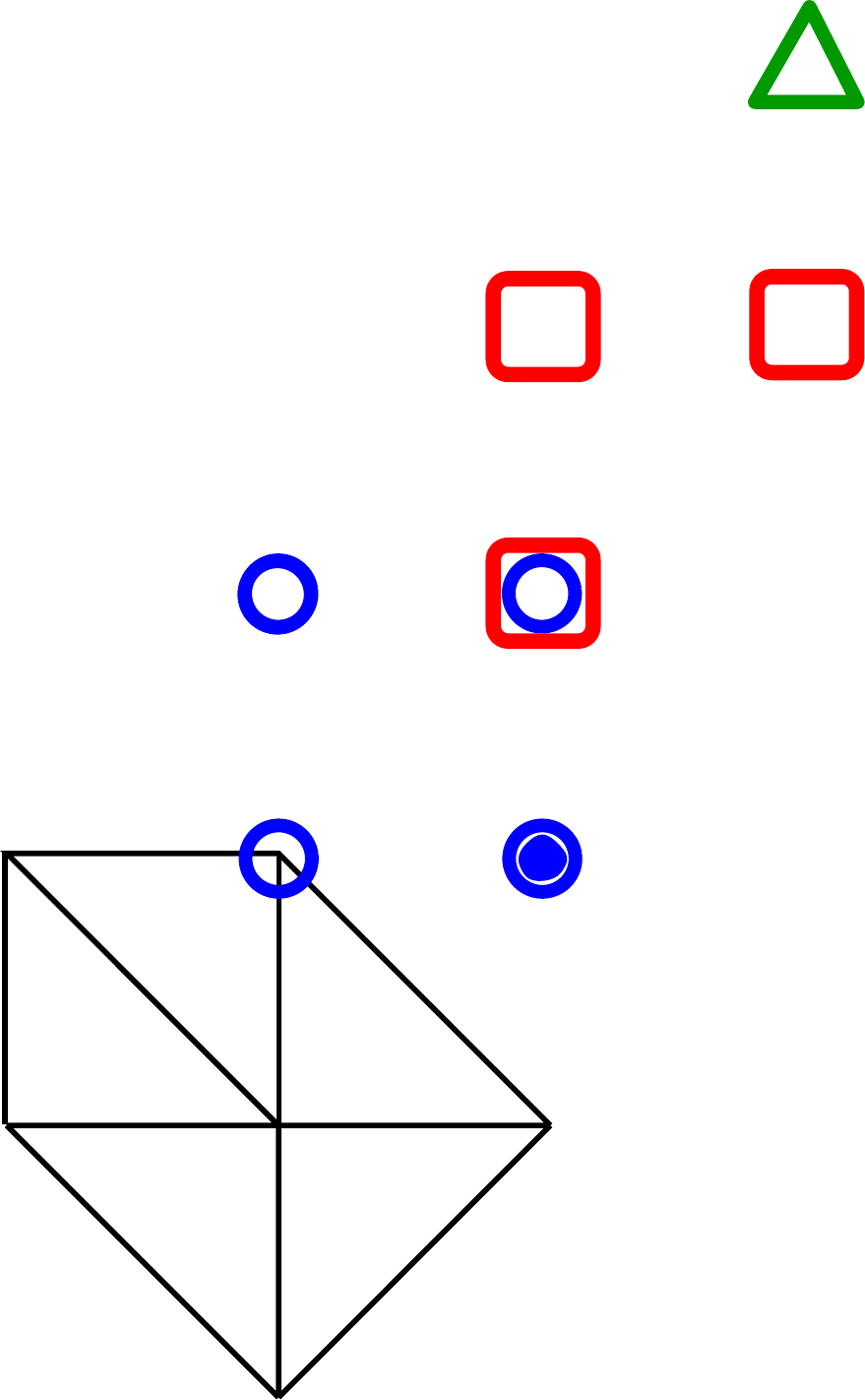}} 
	\put(140,280){\includegraphics[width=0.3\textwidth]{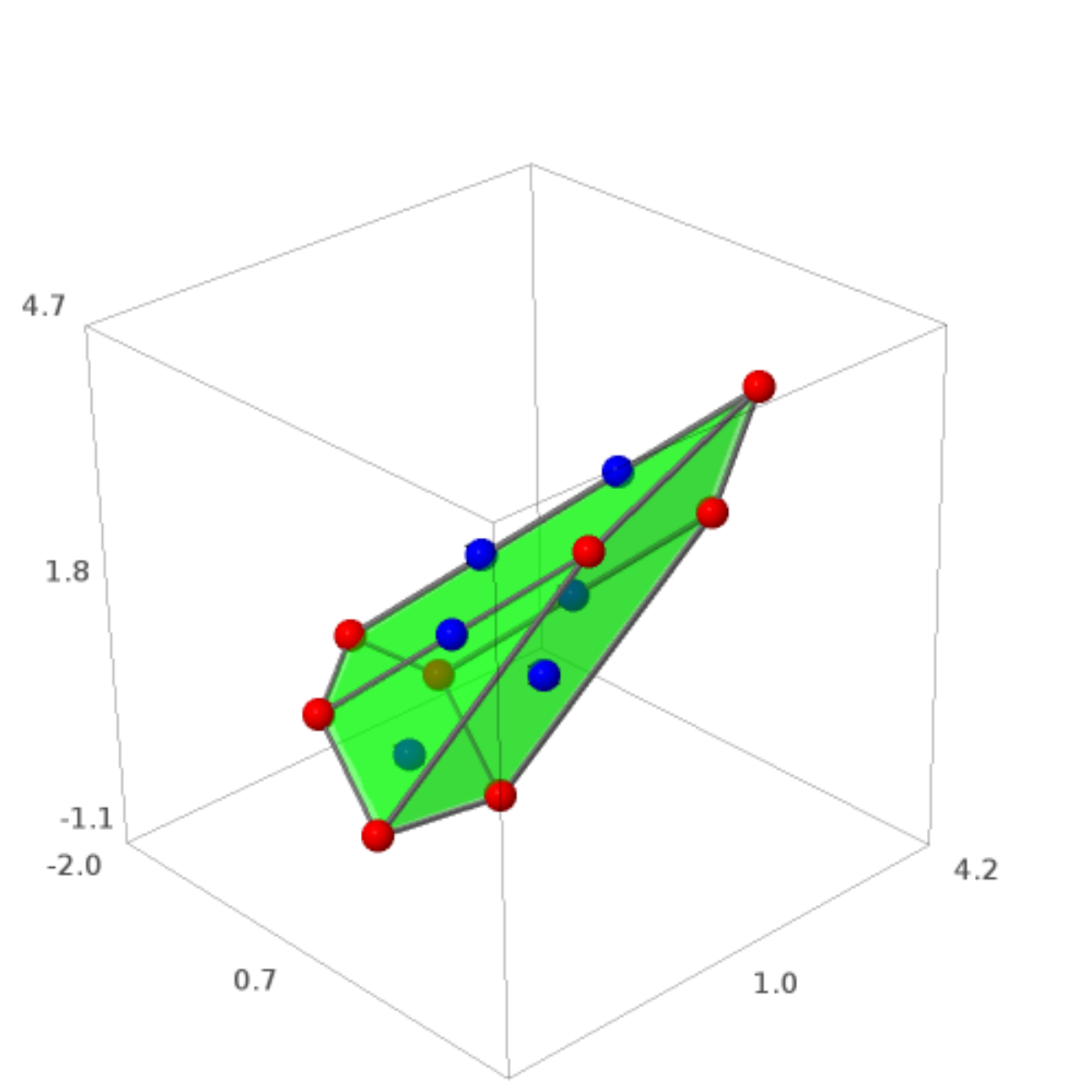}} 
	\put(0,140){\includegraphics[width=0.15\textwidth]{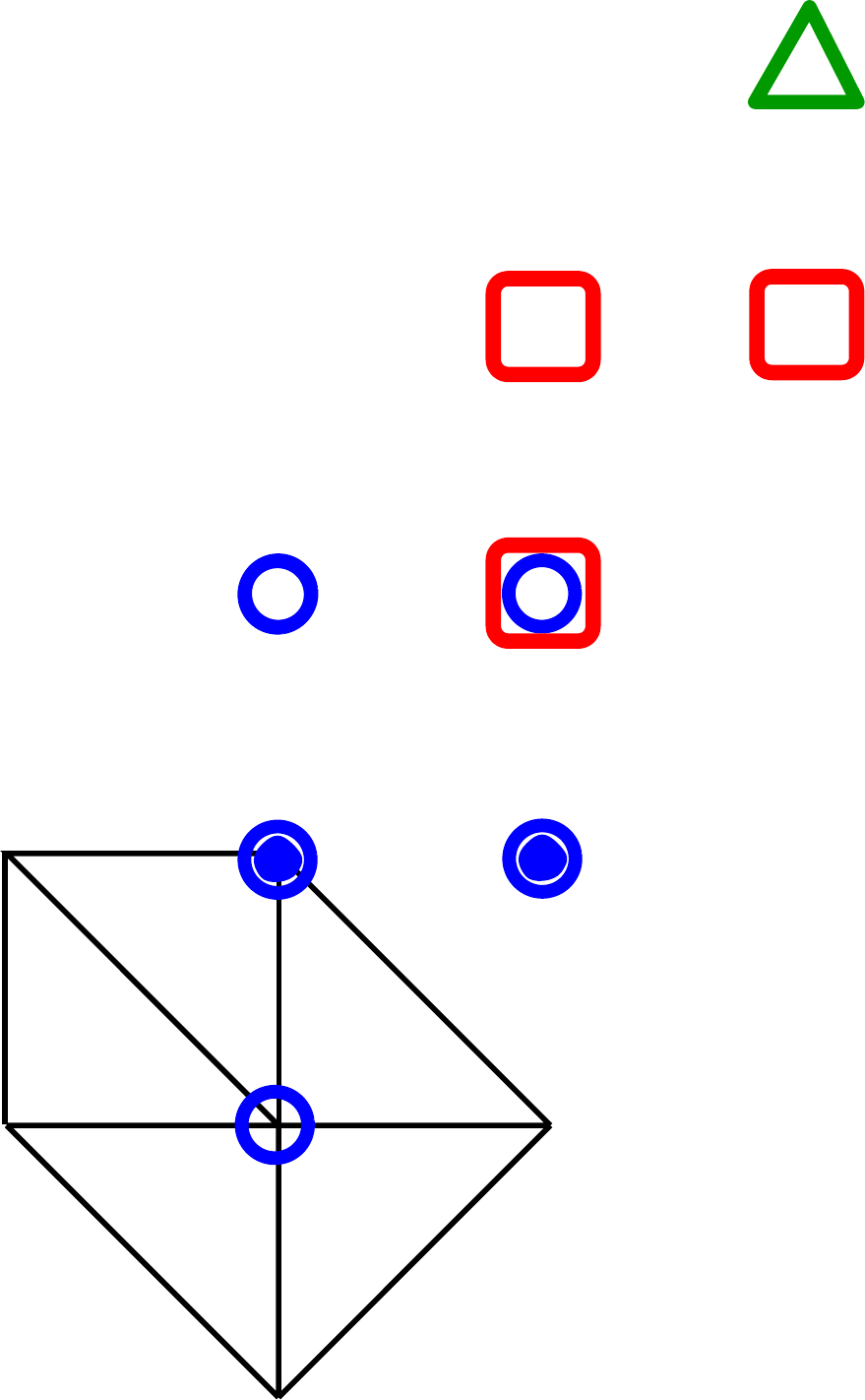}} 
	\put(140,140){\includegraphics[width=0.3\textwidth]{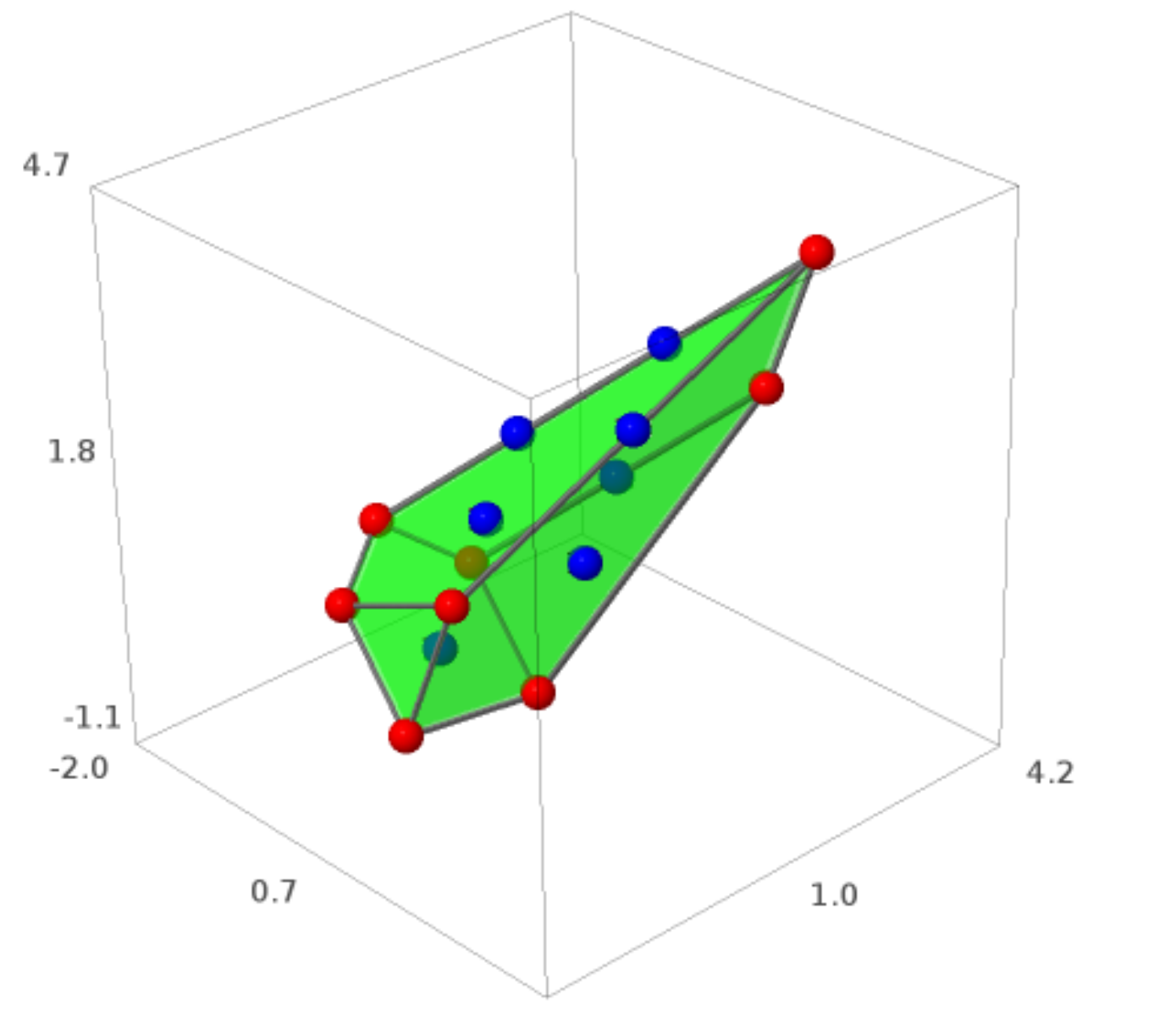}} 
	\put(0,0){\includegraphics[width=0.15\textwidth]{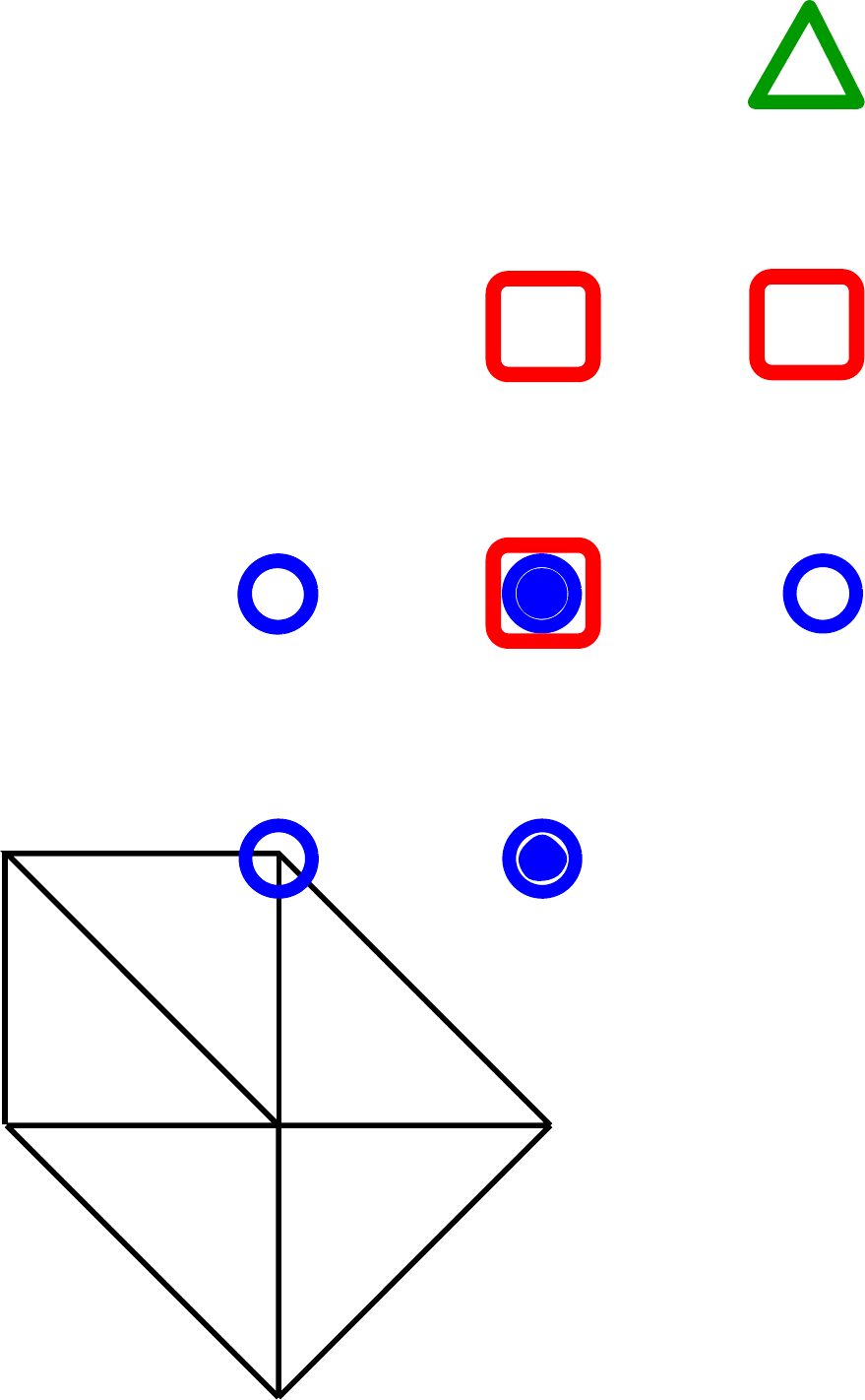}} 
	\put(140,0){\includegraphics[width=0.3\textwidth]{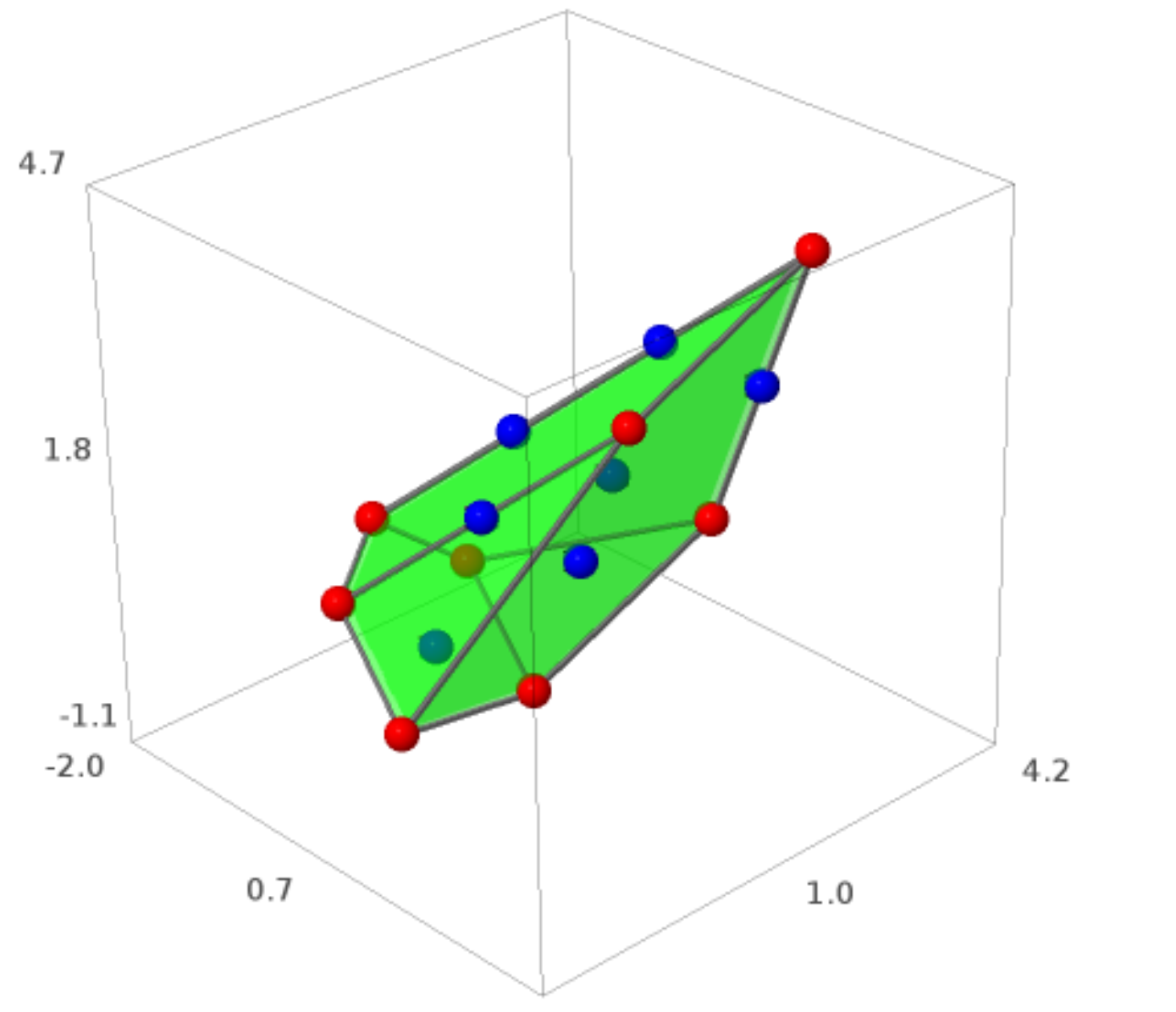}}
	\put(25,152){$\mathfrak{f}_0$}
	\put(182,180){$\mathfrak{f}_0$}
	\put(65,52){$\mathfrak{f}_5$}
	\put(223,50){$\mathfrak{f}_5$}
	\put(300,200){  \begin{tabular}{|c|c|}
	    \hline
		 & vertices \\ \hline \hline
		$u$ & $(-1,1,0)$ \\
		$w$ & $(1,0,0)$ \\
		$v$ & $(0,-1,0)$ \\
		$e_1$ & $(0,1,0)$ \\ 
		$e_2$ & $(-1,0,0)$ \\ \hline
		$\mathbf{f_1}$ & $\mathbf{(0,1,1)}$ \\
  		$f_2$ & $(0,2,1)$ \\
   		$\hat{f}_3$ & $(1,1,1)$ \\ 
    		$\mathbf{f_4}$ & $\mathbf{(1,2,1)}$ \\ \hline
    		$g_1$ & $(1,2,2)$ \\
       		$g_2$ & $(1,3,2)$ \\ 
  		$g_3$ & $(2,3,2)$ \\ \hline
    		$h_1$ & $(2,4,3)$ \\ \hline \hline
    		$\mathfrak{f}_0$ & $(0,0,1)$ \\
    		$\mathfrak{f}_5$ & $(2,2,1)$\\
    		\hline
	\end{tabular}} 
\end{picture}
\end{center}
\caption{\label{fig:E6Top} The E$_6$ tops over $F_5$ before (upper figure) and after (middle and lower figure) the transition. Simplified depictions are given on the left with vertices at height one, two, and three depicted as blue circles, red squares, and green triangles, respectively. The two different blow-ups $\mathfrak{f}_0$ ($\mathfrak{f}_5$) leave the former vertices $\mathbf{f_1}$ ($\mathbf{f_4}$) as a point interior to a face.}
\end{figure}
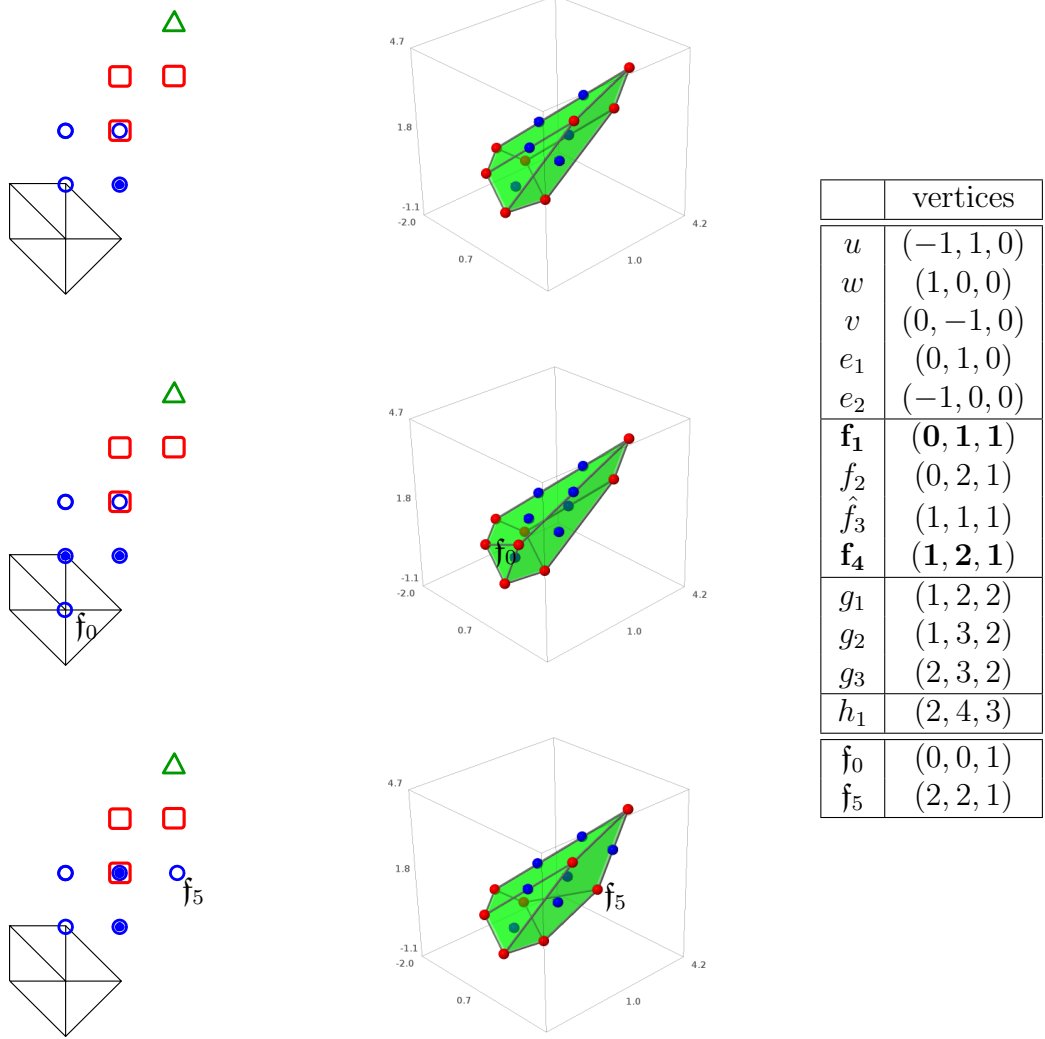
Both U(1) generators intersect E$_6$ and their Shioda maps read
\begin{align}
\begin{split} 
\sigma (s_1) &= [e_1]-[e_2]- \Kbi +\tfrac13(4 [f_4]+ 5 [g_3]+6 [h_1]+ 4 [g_2]+ 2 [f_2]+3 [g_1])  \, ,\\ 
\sigma (s_2) &= [u]-[e_2] - \Kbi -\Sn+\Z+ \tfrac13(2 [f_4]+ 4 [g_3]+6 [h_1]+ 5 [g_2]+ 4 [f_2] +3[g_1]) \, .
\end{split}
\end{align}
We see that the U(1) generators mix with the $\mathbbm{Z}_3$ center of the E$_6$ gauge algebra and one consequently obtains fractional Abelian charges for E$_6$ matter multiplets. Moreover, we find that the starting polytope in Figure~\ref{fig:E6Top} already admits SCPs over $z_0=d_1$ due to the face with interior point $\hat{f}_3$, and comes with multiplicity
\begin{align}
n_{\text{SCP}_1} = (3 \Kbi-\Ss-\Sn)\Z \, .
\end{align}
The full spectrum is summarized in Table~\ref{tab:E6}.
\begin{table}
\centering
\small
\begin{tabular}{| c | c | c | c |} \hline
locus &  ord$(f,g,\Delta)$  & multiplicity &  $\mathbf{R}$ \\ \hline \hline
\!$d_1 = z_0 = 0$\!\! & $(4,6,12)$ & $(3 \Kbi-\Ss-\Sn)\Z$  & SCP$_1$ \\ \hline
\!$d_7 = z_0 = 0$\!\! & $(4,5,9)$ & $(\Ss-\Z) \Z$  & $\mathbf{27}_{1/3,-1/3}$ \\ \hline
\!$d_9 = z_0 = 0$\!\! & $(4,5,9)$ & $\Sn \Z $ & $\mathbf{27}_{1/3,2/3}$ \\ \hline
\!$z_0 = 0$\!\! & $(3,4,8)$ & $1+ \tfrac{1}{2} \Z(\Z-\Kbi)$ & $\mathbf{78}_{0,0}$ \\ \hline \hline 
\!$d_3 = d_7 = 0$\!\! & $(0,0,2)$ & $(\Kbi+\Ss-\Sn-2 \Z) (\Ss-\Z)$ & $\mathbf{1}_{1,-1}$ \\ \hline
\!$d_8 = d_9 = 0$\!\! & $(0,0,2)$ & $\Sn (\Kbi-\Ss+\Sn)$ & $\mathbf{1}_{1,2}$ \\ \hline
\!$d_7 = d_9 = 0$\!\! & $(0,0,2)$ & $\Sn (\Ss-\Z)$ & $\mathbf{1}_{0,2}$ \\ \hline
\!$V(I_{(4)})$ p.44 in \cite{Klevers:2014bqa}\!\! & $(0,0,2)$  & $\begin{array}{c} 6 (\Kbi)^2-3 \Sn \Z+\Z^2+\Ss^2-2 \Sn^2\\+\Kbi (-5 \Ss+4 \Sn-5 \Z)+\Ss (\Sn+2 \Z)\end{array}$ & $\mathbf{1}_{-1,-1}$ \\ \hline
\!$V(I_{(2)})$ p.48 in \cite{Klevers:2014bqa}\!\! & $(0,0,2)$ & $\begin{array}{c} 6 (\Kbi)^2+\Sn^2+\Kbi (4 \Ss-5 \Sn-14 \Z)\\+3 \Sn \Z+4 \Z^2+\Ss (\Sn+2 \Z) -2 \Ss^2\end{array}$ & $\mathbf{1}_{1,0}$ \\ \hline
\!$V(I_{(6)})$ p.48 in \cite{Klevers:2014bqa}\!\! & $(0,0,2)$ & $\begin{array}{c} -2 (\Sn \Z -3 (\Kbi)^2+\Ss^2+\Sn^2-\Ss \Z \\ -2 \Z^2+\Kbi (-2 \Ss-2 \Sn+7 \Z)) \end{array}$ & $\mathbf{1}_{0,1}$ \\ \hline
\!$h^{2,1}(X)+1$\!\! & $-$ & $\begin{array}{c} 20 + 11 (\Kbi)^2 + 2 \Ss^2 + 2 \Sn^2\\ - 4 \Kbi (\Ss + \Sn) - 17 \Kbi \Z + 
 4 \Sn \Z\\ + 9 \Z^2 - \Ss (\Sn + \Z) \end{array}$ & $\mathbf{1}_{0,0}$ \\ \hline 
\end{tabular}
\caption{Base-independent matter spectrum for the upper polytope in Figure~\ref{fig:E6Top}.}
\label{tab:E6}
\end{table}
The base-independent Euler number for the E$_6$ top is given by
\begin{align}
\begin{split}
\chi =& -24 (\Kbi)^2 + 8 \Kbi (\Ss + \Sn + 5 \Z) \\&- 2 (2 \Ss^2 - \Ss \Sn + 2 \Sn^2 + 5 \Sn\ \Z + 10 \Z^2) \, .
\end{split}
\end{align}
In the following, we consider two transitions to theories with two different SCPs with their tops both given in Figure~\ref{fig:E6Top}. The first transition is induced by the factorization $d_3 \rightarrow z_0 d_3$ which enhances the vanishing order at $z_0 = d_7 = 0$ to $(4,6,12)$ and thus creates an SCP with multiplicity
\begin{align}
n_{\text{SCP$_2$}} = (\Ss-\Z)\Z\,. 
\end{align} 
In the toric description this is realized by adding the vertex $\mathfrak{f}_5=(2,2,1)$ to the top which results in $\mathbf{f_4} =(1,2,1)$ becoming an interior point. The induced change in the topological quantities is given by
\begin{align} 
(\Delta \chi, \Delta h^{1,1}, \Delta h^{2,1}) = (4, 1, - 1) \times n_{\text{SCP}_2} \,.
\end{align}
Similarly, we can compute the change in the total matter spectrum, which is given by
\begin{align}
\Delta \mathcal{S} = - (\mathbf{27}_{1/3,-1/3} \oplus \mathbf{1}_{1,-1} \oplus  \mathbf{1}_{0,0}) \times n_{\text{SCP$_2$}} \,,
\end{align}
which satisfies all the general constraints discussed in Section~\ref{subsec:class}. Moreover, the assumption \eqref{assumption} is again valid and the Abelian charges are also consistent with the restrictions imposed in Section~\ref{subsec:class}.

The second transition in this model is performed by the tuning $d_8 \rightarrow d_8 z_0$ and resolved by adding the vertex $\mathfrak{f}_0 = (0,0,1)$, resulting in the vertex $\mathbf{f_1} = (0,1,1)$ becoming a point in a face of the E$_6$ top, where now the fiber at $z_0 = d_9 = 0$ becomes non-flat of vanishing order $(4,6,12)$ with multiplicity
\begin{align}
n_{\text{SCP$_3$}}=\Sn \Z \, .
\end{align}The change in the topological quantities in this case reads
\begin{align} 
(\Delta \chi, \Delta h^{1,1}, \Delta h^{2,1}) = (-4, 1, -1) \times n_{\text{SCP}_3} \,,
\end{align}
with which we can determine the change in the matter spectrum,
\begin{align}
\Delta \mathcal{S} = (\mathbf{27}_{1/3,2/3} \oplus \mathbf{1}_{-1,-2} \oplus  \mathbf{1}_{0,0}) \times n_{\text{SCP$_3$}} \, .
\end{align} 
Again all consistency constraints derived in Section~\ref{subsec:class}, including the ones for the Abelian charges, are fulfilled. Finally, we note that the transitions commute and that it is possible to construct a top that has only SCPs and no $\mathbf{27}$-plets. In such a case it is fascinating to see that the non-flat fibers corresponding to SCP$_1$, SCP$_2$ and SCP$_3$ are not homologous. Hence, from the F-theory intuition we would assume that they have different quantum numbers that are not visible on the tensor branch, where we simply get a collection of $\mathbbm{P}^1$'s. However, as argued, SCP$_2$ and SCP$_3$ originate from different tensor-matter transitions where hypermultiplets with different U(1) charges disappear in the SCP transition. 

\subsection[\texorpdfstring{E$_7 \times$SU(2)$\times$U(1)}{E\_7xSU(2)xU(1)} transitions]{\texorpdfstring{E$\boldsymbol{_7 \times}$SU(2)$\boldsymbol{\times}$U(1)}{E\_7xSU(2)xU(1)} transitions}
\label{e7exa}

In this section we consider a transition between two E$_7$ tops over polygon $F_6$, which is the same one we used in the example of Section~\ref{su5exa}. The two different tops are depicted in Figure~\ref{fig:E7Top}. We discuss this model in some detail, since this model additionally exhibits a different kind of superconformal matter.  
\begin{figure}
\begin{center}
\begin{picture}(400,220)
	\put(0,180){\includegraphics[width=0.3\textwidth]{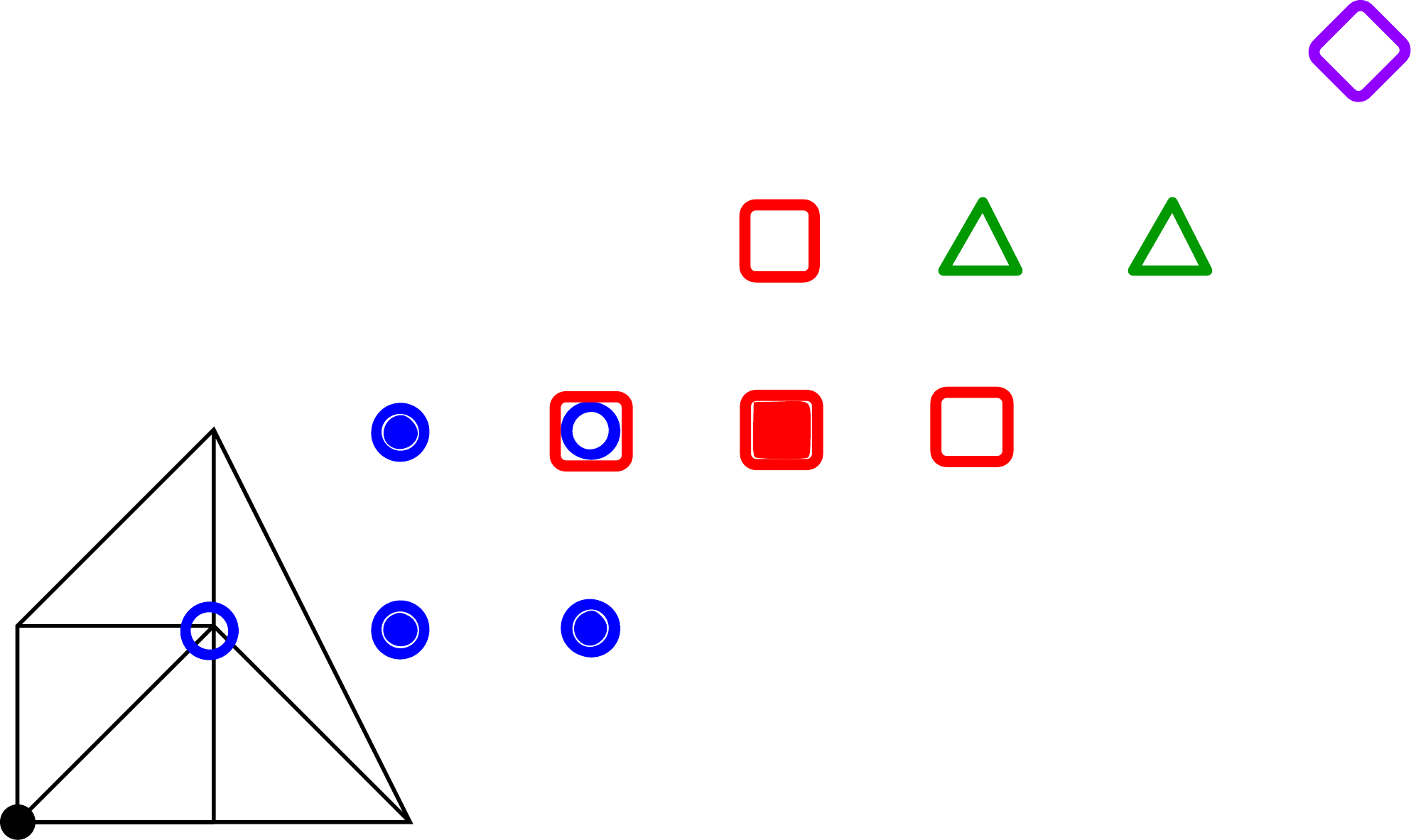}} 
	\put(170,150){\includegraphics[width=0.27\textwidth]{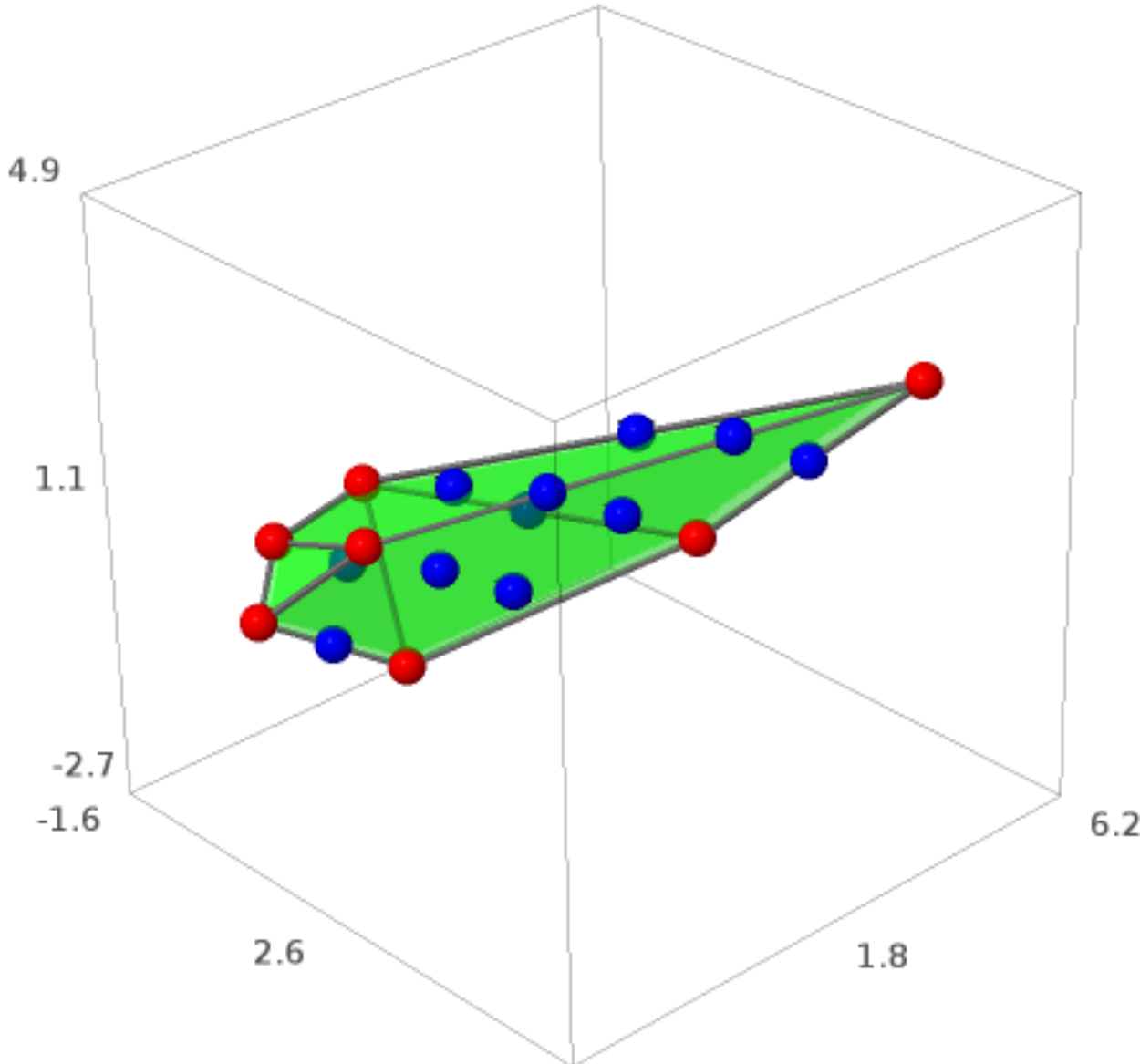}} 
	\put(0,30){\includegraphics[width=0.3\textwidth]{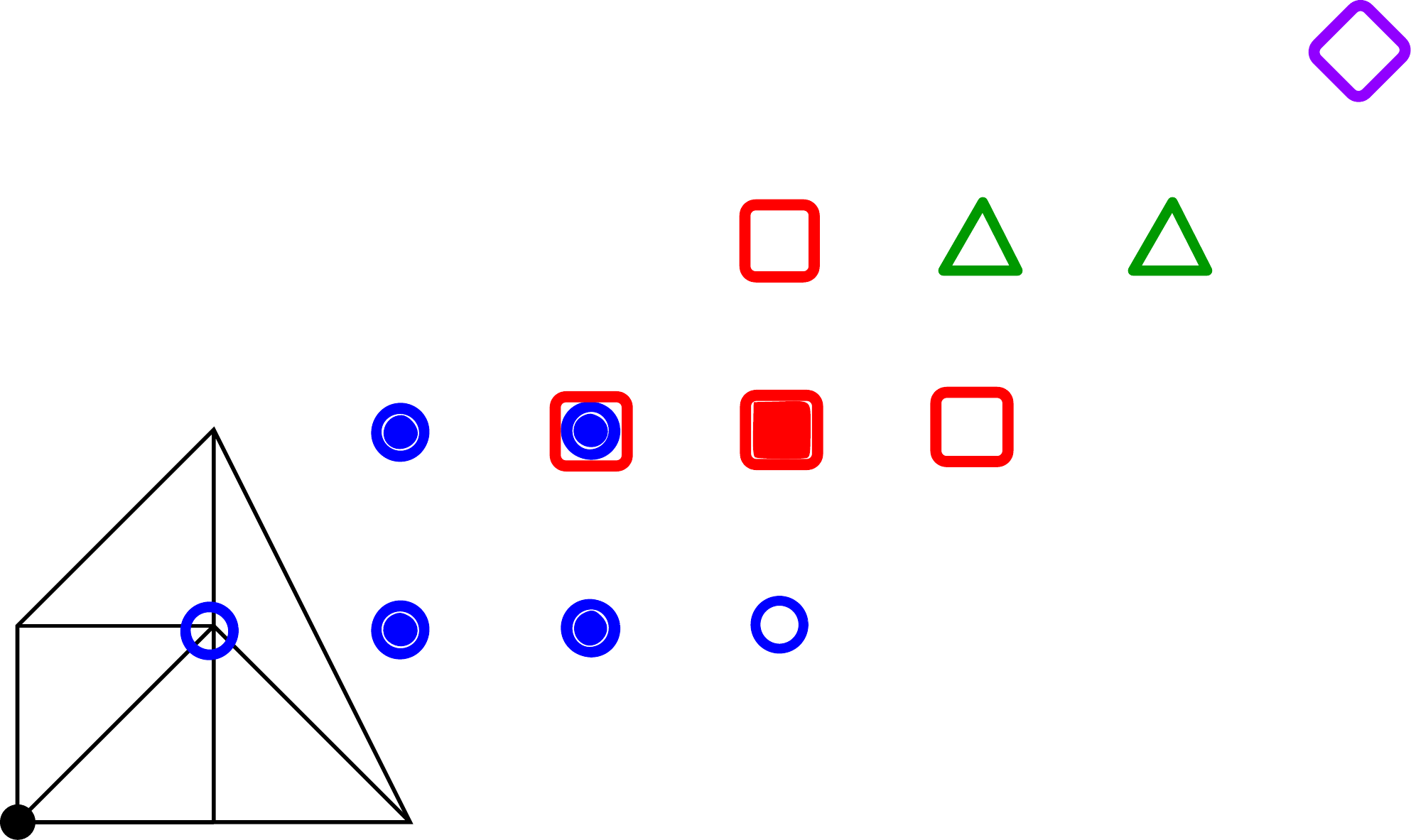}} 
	\put(170,10){\includegraphics[width=0.27\textwidth]{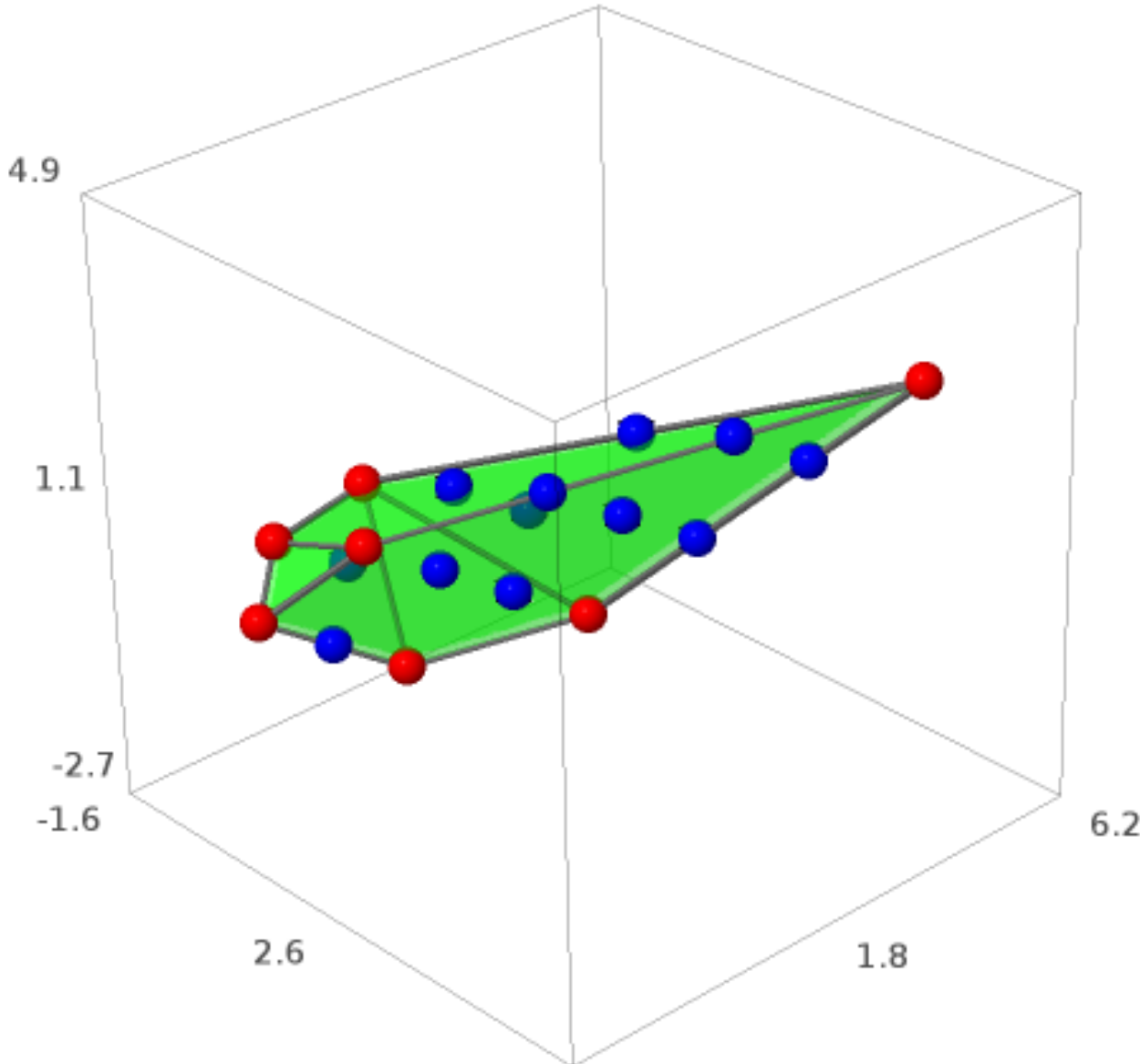}} 
	\put(78,45){$\mathfrak{f}_5$}
	\put(230,45){$\mathfrak{f}_5$}
	\put(330,130){\begin{tabular}{|c|c|}
	    \hline
		 & vertices \\ \hline \hline
		$e_2$& $(-1,-1,0)$ \\
		$u$ & $(-1,0,0)$ \\
		$e_1$& $(0,-1,0)$ \\
		$v$ & $(1,-1,0)$ \\ 
		$w$ & $(0,1,0)$ \\ \hline
		$f_0$ &  $(0,0,1)$ \\
  		$\hat{f}_1$ & $(1,0,1)$ \\
	   	$\hat{f}_2$ & $(1,1,1)$ \\ 
     		$\hat{f}_3$ & $(2,0,1)$ \\   
      		$\mathbf{f_4}$ & $\mathbf{(2, 1, 1)}$\\ \hline 
    		$g_1$ & $(2,1,2)$ \\
       		$\hat{g_2}$ & $(3,1,2)$ \\ 
        		$g_3$ & $(3,2,2)$ \\ 
              	$g_4$ & $(4,1,2)$ \\ \hline
    		$h_1$ & $(4,2,3)$ \\ 
      		$h_2$ & $(5,2,3)$ \\ \hline 
        		$k_1$ & $(6,3,4)$ \\ \hline \hline
     		$\mathfrak{f}_5$   & $(3,0,1)$\\
     		\hline
	\end{tabular}}  
\end{picture}
\end{center}
\caption{\label{fig:E7Top} The E$_7$ tops over polygon $F_6$ before (upper figure) and after (lower figure) the transition. Simplified depictions are given on the left with vertices at height one, two, three and four as blue circles, red squares, green triangles and violet diamonds, respectively. The toric blow-up with vertex $\mathfrak{f}_5 = (3,0,1)$ moves $\mathbf{f_4 = (2,1,1)}$ into a facet.}
\end{figure}
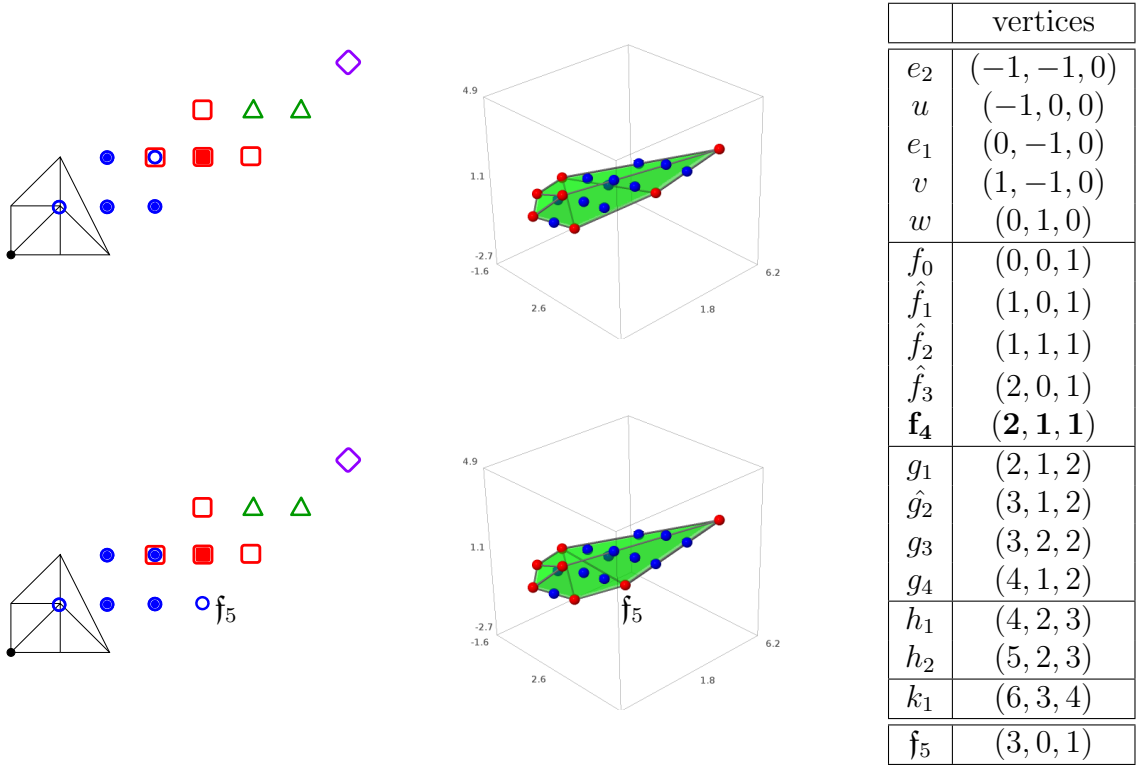 
It has a gauge algebra of E$_7 \times$SU(2)$\times$U(1) and can be obtained from the following factorization: 
\begin{align}
\begin{split}
\hat{s}_1 &=  d_1 f_0^5 f_1^3 f_2^2 f_3 g_1^4 g_2^2 g_3 h_1^3 h_2 k_1^2 \,, \hspace{11.5mm} \hat{s}_2 = d_2 f_0^3 f_1^2 f_2 f_3 g_1^2 g_2 h_1  \,, \\
\hat{s}_3 &= d_3 f_0^2 f_1^2 f_2 f_3^2 f_4 g_1^2 g_2^2 g_3 g_4^2 h_1^2 h_2^2 k_1^2 \,, \quad \hat{s}_4 = d_4 f_1 f_3^2 f_4 g_2 g_4^2 h_2 \,, \\
\hat{s}_5 &= d_5 f_0^3 f_1^2 f_2^2 f_3 f_4 g_1^3 g_2^2 g_3^2 g_4 h_1^3 h_2^2 k_1^3 \,, \quad \hat{s}_6 = d_6 f_0 f_1 f_2 f_3 f_4 g_1 g_2 g_3 g_4 h_1 h_2 k_1 \,, \\
\hat{s}_7 &= d_7 f_1 f_2 f_3^2 f_4^2 g_1 g_2^2 g_3^2 g_4^3 h_1^2 h_2^3 k_1^3 \,, \hspace{8.5mm} \hat{s}_8 = d_8 f_2 f_4 g_3 \,.
\end{split}
\end{align}
The U(1) generator only intersect the SU(2) but not the additional E$_7$, and can be written as  
 \begin{align}
\sigma (s_1) = [u] - [e_2] -\Kbi-\Ss+ \tfrac{1}{2} [e_1] \, .
\end{align}
Similar to the E$_6$ example, the starting polytope already has two faces with interior points. The first facet corresponds to SCP$_1$ over $d_4 = z_0 =0$ with a single interior point $\hat{f_2}=(1,1,1)$ and multiplicity
\begin{align}
n_{\text{SCP}_1} = (2 \Ss-\Sn) \Z \, .
\end{align}
The second facet, corresponding to the locus $d_8 = z_0 = 0$ has three interior points $\hat{f_1}$, $\hat{f_3}$, and $\hat{g_2}$. As opposed to the first locus, which leads to a $(4,6,12)$ singularity for SCP$_1$, the second is a $(4,6,14)$ point and hence is of a different type compared to the singularities discussed in the main part. It comes with a multiplicity
\begin{align}
n_{\text{SCP}_2} = (\Kbi-\Ss+\Sn) \Z  \, .
\end{align}
The matter loci and representations are summarized in Table~\ref{tab:E7}.
\begin{table}
\centering
\small
\begin{tabular}{| c | c | c | c |} \hline
locus & $\text{ord}(f,g,\Delta)$ & \text{multiplicity} & $\mathbf{R}$ \\ \hline \hline
$d_4=z_0=0$ & $(4,6,12)$ & $(2 \Ss-\Sn) \Z$ & SCP$_1$ \\ \hline
$d_8=z_0=0$ & $(4,6,14)$ & $(\Kbi-\Ss+\Sn) \Z$ & SCP$_2$ \\ \hline
$d_2=z_0=0$ & $(4,5,10)$ & $(2 \Kbi-\Sn-3 \Z) \Z$ & $(\mathbf{56},\mathbf{1})_{0}$ \\ \hline
$z_0=0$ & $(4,5,9)$ & $1 + \tfrac{1}{2} \Z(\Z-\Kbi)$ & $(\mathbf{133},\mathbf{1})_{0}$ \\ \hline \hline 
$d_8= d_7=0$ & $(0,0,3)$ & $\Ss (\Kbi-\Ss+\Sn)$ & $(\mathbf{1},\mathbf{2})_{-3/2}$ \\ \hline
$d_7 = d_4=0$ & $(0,0,2)$ & $\Ss (2 \Ss-\Sn)$ & $(\mathbf{1},\mathbf{1})_{2}$ \\ \hline
$V(I_{(2)})$ p.48 in \cite{Klevers:2014bqa}  & $(0,0,3)$ & $\begin{array}{c}(\Kbi-\Ss+\Sn)\\\times (6 \Kbi+\Ss-2 (\Sn+4 \Z))\end{array}$ & $(\mathbf{1},\mathbf{2})_{1/2}$ \\ \hline
$V(I_{(4)})$ p.48 in \cite{Klevers:2014bqa}  & $(0,0,2)$ & $\begin{array}{c}6 (\Kbi)^2-3 \Ss^2+\Sn (\Sn+4 \Z))\\+\Kbi (13 \Ss-5 \Sn-8 \Z))\\-2 \Ss (\Sn+8 \Z)) \end{array}$ & $(\mathbf{1},\mathbf{1})_{1}$ \\ \hline
$d_8=0$ & $(0,0,2)$ & $1+ \tfrac{1}{2} (-\Ss+\Sn) (\Kbi-\Ss+\Sn)$ & $(\mathbf{1},\mathbf{3})_{0}$ \\ \hline
$h^{2,1}(X)$ & - & $\begin{array}{c} 21 + 11 (\Kbi)^2 + 4 \Ss^2 - 3 \Ss \Sn\\ + 2 \Sn^2  + 5 \Ss \Z + 6 \Sn \Z\\ + 21 \Z^2 - 4 \Kbi (\Ss + \Sn + 8 \Z) \end{array}$ & $(\mathbf{1},\mathbf{1})_{0}$ \\ \hline 
\end{tabular}
\caption{\label{tab:E7} Base independent spectrum of the upper E$_7$ top in Figure~\ref{fig:E7Top}.}
\end{table}
The base independent Euler number, which we used in order to compute the complex structures of the upper E$_7$ top is given by
\begin{align}
\begin{split}
\chi = -2 \big[&12 (\Kbi)^2 + 4 \Ss^2 - 3 \Ss \Sn + 2 \Sn^2 - 4 \Kbi (\Ss + \Sn) \\
& - 35 \Kbi \Z + 6 \Ss \Z + 4 \Sn \Z + 21 \Z^2 \big] \,.
\end{split}
\end{align}
In order to compute the correct amount of complex structure deformations, we have to take into account that for each superconformal matter point SCP$_2$ we obtain three non-toric K\"ahler deformations instead of just one. Hence, the formula for the complex structures reads
\begin{align}
\begin{split}
h^{2,1}(X) &= h^{1,1}(X) - \tfrac{1}{2} \chi(X) \,, \\
&= \text{rank}(G) + h^{1,1}(B)+1 + n_{\text{SCP$_1$}} + 3 n_{\text{SCP$_2$}} -\tfrac{1}{2}\chi(X) \, ,
\end{split}
\end{align}
as argued in Section~\ref{sec:torichyper}.

We now consider a transition to the second theory given in Figure~\ref{fig:E7Top} obtained by the factorization $d_1 \rightarrow z_0 d_1$, which enhances the singularity at $d_2 = z_0 =0$ to vanishing order  $\text{ord}(f,g,\Delta) = (4,6,12)$ with multiplicity
\begin{align}
n_{\text{SCP$_3$}}= ( 2 \Kbi-\Sn-3 \Z) \Z \,.
 \end{align}
In the toric description this is realized by adding the vertex $\mathfrak{f}_5 = (3,0,1)$ to the top, resulting in $\mathbf{f_4} = (2,1,1)$ becoming an interior point of a face. The induced change in the topological quantities is given by
\begin{align}
(\Delta \chi, \Delta h^{1,1}, \Delta h^{2,1}) = (4, 1, - 1) \times n_{\text{SCP}_3},
\end{align} 
and the total change of the spectrum reads
\begin{align}
\Delta \mathcal{S} = -( \tfrac{1}{2} \times (\mathbf{56},\mathbf{1})_{0} \oplus (\mathbf{1},\mathbf{1})_{0}) \times n_{\text{SCP$_3$}} \, , 
\end{align}
consistent with the general constraints in Section~\ref{subsec:class}. Since the  $\mathbf{56}$-plet is a half-hypermultiplet, it must be uncharged.

\subsubsection*{Gravitational anomalies and the SCP$\boldsymbol{_2}$ tensor branch} 

In this subsection we comment on the anomalies of the $(4,6,14)$ points of the E$_7$ model discussed above. These points with a higher vanishing order of the discriminant are not simply E-string theories like the ones we have considered throughout this work. Indeed, this can be already inferred by investigating the gravitational anomaly, which reads
\begin{align}
H - V - 29 T + 29 n_{\text{SCP}_1} + 63 n_{\text{SCP}_2} = 273 \,.
\end{align}
Consequently, we expect the effective degrees of freedom $\mathcal{H}_2$ appearing at the tensor branch to be
\begin{align}
\label{eq:hbranch}
\mathcal{H}_2 = H_{\text{SCP}_2} - V_{\text{SCP}_2} + 29 T_{\text{SCP}_2} = 63 \,,
\end{align}
which is the dimension of the Higgs branch \cite{Mekareeya:2017jgc} of the superconformal matter, see also \eqref{Tensorbranchformula}. That this is indeed the case can be checked by going to the tensor branch of one SCP$_2$ and analyzing the resulting spectrum. Indeed, resolving the $(4,6,14)$ singularity requires two resolutions in the base resulting in the following diagram with the left and right flavor groups being gauged
\begin{align}
[\text{E}_7]  \, [\text{SU}(2)] \quad \rightarrow  \quad [\text{E}_7] \,~  -1 ~ \, \stackrel{\text{\footnotesize SU(2)$'$}}{-2~}  ~ \, [\text{SU}(2)] \, .
\end{align}
Indeed there is an additional SU(2) over the the $-2$ curve. The full spectrum reads
\begin{align}
T_{\text{SCP}_2}= 2\,,\quad V_{\text{SCP}_2}=  3\,,\quad H_{\text{SCP}_2} = 2 \times (\mathbf{1}, \mathbf{2}) \oplus (\mathbf{2}, \mathbf{2})\,,
\end{align}
which satisfies all anomalies, in particular \eqref{eq:hbranch}. The blow-up procedure has been summarized in Figure~\ref{fig:E7SU2BU}. 
\begin{figure}
\begin{center}
\begin{tikzpicture}[scale=1.5]
	\draw[very thick] (-4,0) to[out = 30, in = 200] (-1,0);
	\draw[very thick] (-3,1.5) to[out = -60, in = 110] (-2,-1.5);
	\draw[very thick, red, fill=red] (-2.43,0.05) circle (0.04);
	\node at (-4,-0.2) {$[\text{E}_7]$};
	\node at (-2.2, -1.7) {$[\text{SU(2)}]$};
	\node at (-2.1,0.3) {SCP$_2$};
	\draw[very thick] (1,1.5) to[out = -60, in = 60] (1,-1.5);
	\draw[very thick] (1,0) to[out = 10, in = 210] (3,1);
	\draw[very thick] (2,1) to[out = -20, in = 160] (4,0);
	\draw[very thick] (4,1.5) to[out = 220, in = 100] (4,-1.5);
	\draw[very thick, fill = black] (1.43,0.11) circle (0.04);
	\draw[very thick, fill = black] (3.58, 0.17) circle (0.04);
	\draw[very thick, fill = black] (2.58, 0.73) circle (0.04);
	\node at (0.7,-1.5) {$[\text{E}_7]$};
	\node at (3.5,-1.5) {$[\text{SU(2)}]$};
	\node at (2,0.2) {$-1$};
	\node at (3,0.2) {$-2$};
	\node at (4.5,0) {$[\text{SU(2)}']$};
	\draw[thick, dashed] (3.58,0.17) -- (4.2,0.8);
	\node at (4.55,0.9) {$(\mathbf{2}, \mathbf{2})$};
	\draw[thick, dashed] (3.2,0.4) -- (3.1,1.3);
	\draw[thick, dashed] (2.2,0.92) -- (2.9,1.3);
	\node at (3,1.5) {$(\mathbf{1},\mathbf{2})$};
\end{tikzpicture}
\end{center}
\caption{The collision of an E$_7$ and SU(2) leading to an SCP$_2$ subsector is given on the left, its resolution on the tensor branch on the right. The full flat resolution requires s a $-1$ and a $-2$ curve. The latter hosts and additional SU(2)$'$ gauge algebra and charged matter multiplets.}
\label{fig:E7SU2BU}
\end{figure}
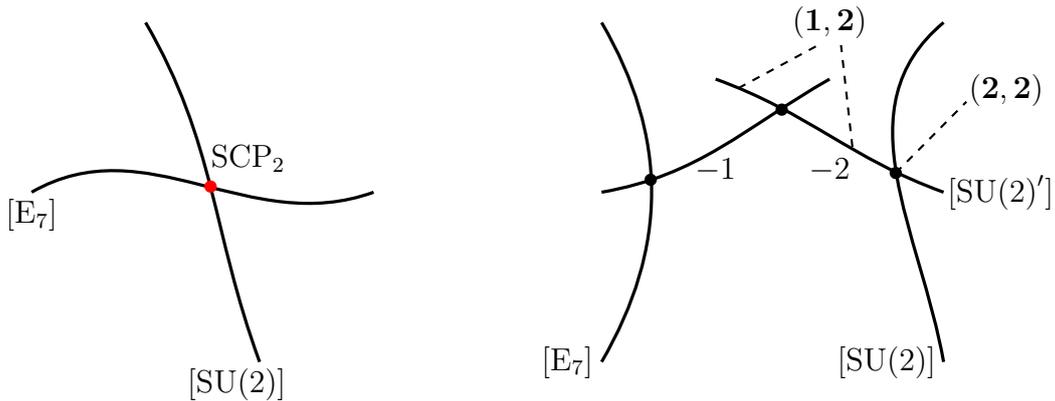
Such a theory is known as $(\text{E}_7, \text{SO}(7))$  minimal conformal matter describing a half-M5 brane on top of the E$_7$ singularity  \cite{DelZotto:2014hpa,Heckman:2015bfa,Mekareeya:2017jgc,Merkx:2017jey}. One might speculate whether the top encodes some non-trivial information of the above theory. This seems to be the case, since for each of the above points, three non-toric fiber components intersect the Calabi-Yau, which exactly accounts for the two tensor multiplets and the additional SU(2) Cartan generator. The investigation of more general non-flat fiber points is, however, left for future work.

\section{Conclusions and Outlook}
\label{sec:conclusions}

In this work we discussed tensor-matter transitions in six-dimensional theories, originating from F-theory compactifications on a genus one-fibered Calabi-Yau 3-fold, which leave the full six-dimensional gauge algebra $\overline{G}$ invariant. These transitions pass through theories containing strongly coupled subsectors described by superconformal matter and corresponding SCPs in the compactification manifold. The blow-ups in the base that resolve these SCPs are associated with the appearance of an additional tensor multiplet in the theory. For the stringent anomaly constraints to be satisfied before and after the transition, also other parts of the spectrum have to change.

The geometric data encoding the anomaly coefficients is also modified by the additional exceptional divisors in the base which accounts for the additional tensor multiplets entering in the Green-Schwarz mechanism. We determine the change in the non-Abelian matter spectrum, taking into account the specific form of the blow-up and consistency with anomaly cancellation, leading to equations \eqref{specirred}, \eqref{changenonA} and \eqref{changemixed}. For the type of tensor-matter transitions considered here, these constraints fix the change in the hypermultiplet sector which transforms non-trivially under the non-Abelian gauge algebra uniquely.

Under the additional assumption that the form of the Shioda maps of the U(1) generators stays invariant, which seems to be satisfied in toric hypersurface constructions, we further obtain restrictions \eqref{specAbelian} on the Abelian charges of the hypermultiplets involved in the tensor-matter transition.

Even though these constraints do not fix the singlet spectrum uniquely, they are often strong enough to determine the singlet charges in terms of the charges of the non-Abelian representations. Therefore, they facilitate the evaluation of the nature of the full set of 29 degrees of freedom vanishing in the hypermultiplet sector during the transition, especially in explicit constructions. This allows for a full classification of tensor-matter transitions with an arbitrary semi-simple Lie algebra $G$ localized on a smooth divisor $\Z$, plus additional Abelian gauge groups. We perform this classification for all Lie algebra representations $\mathbf{R}$ that can be involved in the transition.

We illustrate our results in five different examples constructed from models based on toric hypersurfaces. These models elucidate several aspects of the general discussion of the transition and confirm the general formulae. Moreover, the Abelian charges in all of these models satisfy the additional assumption imposed for the Abelian anomaly coefficients and the general discussion for the restriction of U(1) charges can be applied. 

The tensor-matter transitions are a non-perturbative way of connecting 6d SUGRA vacua and thus this classification aims to understand its full landscape. Our results are consistent with the higher symmetric and other exotic matter representations involved in similar transitions considered in \cite{Anderson:2015cqy, Klevers:2017aku}. However, those transitions exhibit singular divisors which we have not considered in this work, which is why these exotic representations are absent in our classification. Having classified all transitions along smooth curves, it would be desirable to incorporate singular ones as well and to obtain their possible representations. Moreover, in order to fully restrict the U(1) charges of singlets and exotic matter in such transitions, we have to understand the full scope of possible induced changes of the Shioda map in the future. 

In addition, we have illustrated that superconformal points are realized as non-flat fibers in the smooth geometry, which have a toric interpretation of the underlying top. In the various examples we made two intriguing observations: First, we find that non-flat fibers are homologous if they are associated to the same tensor-matter transitions and non-homologous if they originate from inequivalent transitions, where other U(1) charged hypermultiplets are involved. As non-homologous splits of the F-theory fiber generically leads to states with different quantum numbers, we might expect a similar identification for SCPs that originate from different non-flat fibers. Second, we have also given an example for non-flat fibers of even higher vanishing order and its toric realization. We identify those fibers as (E$_7$, SO(7)) minimal superconformal matter in Section~\ref{e7exa}, as seen from the gravitational anomaly and the tensor branch. Those examples point towards the possibility of understanding superconformal matter and their symmetries, possibly coupled to gravity, directly from their realization as non-flat fibers.
   
\subsection*{Acknowledgments}
We thank Lara Anderson, James Gray, and Fabio Apruzzi for valuable discussions. The work of P.K.O.\ is supported by an individual DFG grant OE 657/1-1. M.D.'s work is part of the D-ITP consortium, a program of the Netherlands Organisation for Scientific Research (NWO) that is funded by the Dutch Ministry of Education, Culture and Science (OCW). P.K.O.\ thanks DESY for hospitality during the completion of this work. The work of F.R.\ is supported by the EPSRC network grant EP/N007158/1.

\newpage
\appendix
\numberwithin{equation}{section}

\section{Details of the spectrum computation}
\label{App:speccomp}

In this Appendix we review more details of the general spectrum computation in models constructed from tops, using the methods applied e.g.\ in \cite{Klevers:2014bqa, Buchmuller:2017wpe}.

For a construction of tensor-matter transitions in models based on toric hypersurfaces, we start with one of the 16 toric ambient spaces for a genus-one fiber. We then engineer the non-Abelian gauge algebra via the top construction.

\subsection*{The geometry of the top}
Tops were introduced in \cite{Candelas:1996su} and classified in \cite{Bouchard:2003bu}. A 3d top $\Delta$ can be thought of as the ambient space polytope of a half K3 that is torus-fibered. It is given by the vertices $v_i = (v_i^1,v_i^2,v_i^3)$ with $v_i^3 \geq 0$ and has one interior point in the $v^3 =0$ facet. This facet at height zero corresponds to the 2d polytope $F_0$ of the generic fiber. Analogously to the Batyrev construction \cite{Batyrev:1994hm}, there is a dual object $\Delta^*$ obtained as
\begin{align}
\Delta^*: \quad \{ m \in \mathbbm{Z}^3 : \langle m,v_i \rangle \geq -1\; \forall \, v_i \in \Delta \} \,.
\label{topdual}
\end{align} 
This has the form of an half-infinite prism that is unbounded in the $m^3$ direction. All vertices at height $v^3 > 0$ in $\Delta$ correspond to divisors $D_{v_i}$ that project to the same point in the non-compact base and are resolution divisors of some ADE gauge algebra. The height $v_i^3$ of $v_i$ encodes the Dynkin multiplicity of its corresponding root. With this input data one can derive the hypersurface equation via
\begin{align}
p_{\Delta} =\sum_{m_j \in \Diamond^*} d_j \prod_{v_i \in \Delta} x_i^{\langle m_j, v_i \rangle + 1 } = \sum_{m_j \in \Delta^*} d_j \left(\prod_{v_s \in F_0} x_s^{ \langle m_j,v_s\rangle+1}\right) \left(\prod_{v_t \in \Delta , v^3_t>0} x_t^{ \langle m_j,v_t\rangle+1}\right) \,.
\label{pDiamond}
\end{align}
The first product in the sum includes the vertices $v_s$ of the 2d reflexive polytope $F_0$ at height zero. It thus encodes the form of the generic fiber. The second product in the sum contains the contribution from the ADE resolution divisors.

Hence, the second product in \eqref{pDiamond} can be viewed as a specialization of the generic fiber, given by the $x_s$ coordinates of the height zero 2d polytope. One starting point for that generic fiber might be the cubic, embedded in the 2d polytope $F_1=\mathbbm{P}^2$. This polytope is generated by the three vertices
$u: (-1,1),~w: (1,0),~v: (0,-1)$ transforming in the following line bundle classes
\begin{align}
 [u] \sim H + \Ss - \Kbi\, , \qquad [v] \sim H + \Sn - \Ss\, , \qquad [w] \sim H \, .
\end{align}
The generic cubic is then given by
\begin{align}
 p= \hat{s}_1 u^3 + \hat{s}_2 u^2 v + \hat{s}_3 u v^2 + \hat{s}_4 v^3 + \hat{s}_5 u^2 w + \hat{s}_6 u v w + 
 \hat{s}_7 v^2 w + \hat{s}_8 u w^2 + \hat{s}_9 v w^2 + \hat{s}_{10} w^3 \, ,
 \label{genericpolynomsections}
\end{align}
where the ten sections $\hat{s}_i$ transform in the following way
\begin{align}\label{bidivisors2}
 \begin{array}{clr@{\;}l}
& [\hat{s}_1] \sim 3 \Kbi - \mathcal{S}_7 - \mathcal{S}_9 \,,&
 [\hat{s}_2] &\sim 2 \Kbi - \mathcal{S}_9  \,, \\
& [\hat{s}_3] \sim \Kbi + \mathcal{S}_7 - \mathcal{S}_9 \,,& 
[\hat{s}_4] &\sim 2 \mathcal{S}_7 - \mathcal{S}_9 \,, \\
& [\hat{s}_5] \sim 2 \Kbi - \mathcal{S}_7   \,,  &
[\hat{s}_6] &\sim \Kbi    \,, \\ 
& [\hat{s}_7] \sim  \mathcal{S}_7    \,,  & 
[\hat{s}_8] &\sim \Kbi - \mathcal{S}_7 + \mathcal{S}_9 \,, \\ 
& [\hat{s}_9] \sim  \mathcal{S}_9   \,, &
[\hat{s}_{10}] &\sim 2 \mathcal{S}_9 - \mathcal{S}_7  \, . 
 \end{array}
 \end{align}
In most of the examples the fiber can be thought of as a restricted cubic with fewer coefficients but the same general base dependence as given in \eqref{bidivisors2}. The base-dependent coefficients of the polynomial $\hat{s}_j$ factorize in a specific form after inclusion of the top in order to ensure the presence of the corresponding ADE singularity. As can be seen from \eqref{pDiamond}, this factorization is
\begin{align}
\hat{s}_i \rightarrow d_i \left(\prod_{v_t \in \Delta , v^3 >0} x_t^{ \langle m_j,v_t\rangle+1}\right) \,.
\label{eq:factorization}
\end{align}
We provide the factorization of the sections $\hat{s}_i$ in all examples.

Upon including a bottom, one ends up with a reflexive polytope $\Diamond$ and a compact geometry where all $d_i$ transform as effective divisors in line bundles of the base. Turning this statement around, all completions to a compact geometry can be parametrized by the four base classes $\Z$, $K_B^{-1}$, $\Ss$ and $\Sn$ with the constraint that the $d_i$ are effective.

To find the loci of the charged matter, it is sufficient to work in the singular Weierstrass model and blow down all exceptional divisors of the top except for the affine node, which is fixed by its intersection with the zero-section. This coordinate we call $f_0$ and it projects onto the base as
\begin{align}
f_0 \xrightarrow{\pi} z_0 \,.
\end{align}
In this way, the discriminant of the generic fiber factorizes as
\begin{align}
 \Delta= z_0^m \left( P + \mathcal{O}(z_0) Q +\ldots \right) \, ,
\end{align}
with $m$ being fixed by the gauge group of the top. Hence we find the desired ADE gauge group. The genus $g$ of the $z_0=0$ curve leads to $g$ adjoint hypermultiplets \cite{Witten:1996qb}. Furthermore, the vanishing of the polynomial $P$ together with $z_0 =0$ signals an enhanced singularity. Finding all irreducible components of $P$ amounts to finding all matter loci charged under the ADE group localized at $z_0 = 0$. The loci of representations of the generic fiber can be obtained similarly. We start with the results of \cite{Klevers:2014bqa} for these loci and adapt them to account for the presence of the top.

The multiplicities of the matter multiplets are obtained by intersecting the respective divisor classes of their underlying ideals. 
We note that the multiplicities of certain non-toric ideals of the generic fiber include often simpler matter ideals as irreducible components. We denote the multiplicity with which these ideals occur by $r$. If $r\neq0$, the multiplicities of these simpler ideals have to be subtracted from the more complicated ideal under consideration. The mulitplicity $r$ can be computed from the resultant, following \cite{Klevers:2014bqa}.  

After having identified all codimension-two loci, we substitute their location into the equation for the resolved fiber to observe a split of the form
\begin{align}
 \mathcal{E} \rightarrow \mathcal{C}_{m,1} + \mathcal{C}_{m,2} \, .
\end{align}
The two irreducible curves can be used to compute the weights and U(1) charges of the associated matter representations by intersecting them with the ADE resolution divisors and Shioda maps respectively.

As the top encodes the geometry of a non-compact torus-fibered half K3, we can use this structure and impose a compact two-fold base to describe a wide range of torus fibered global 3-folds $Y_3$.  The structure of the top is then enough to compute Euler and Hodge numbers of these classes of 3-folds. First, we note that the number of $(1,1)$-forms on $Y_3$ is given by
\begin{align}
\label{eq:KahlerApp}
h^{1,1}(Y_3) = \text{rank}(G)+  h^{1,1}(B) -1 + \sum_{i} \delta_i \cdot n_{\text{SCP}_i}\, .
\end{align}
This receives contributions from ADE divisors of the fiber, from divisors that generate the base homology, and from non-flat fibers coming from the superconformal points with dimensionality $\delta_i$ as argued in the Section~\ref{subsec:toric}.
The number neutral singlets can be obtained from the Euler number via
\begin{align}
\label{eq:CS}
H_{\text{neu}}= h^{2,1}(Y_3)+1 = h^{1,1}(B) + \text{rank}(G) +\sum_i \delta_i \cdot n_{\text{SCP}_i} + 2 - \tfrac{1}{2} \chi(Y_3) \, .
\end{align}
Hence we can compute the complex structure moduli simply from the Euler number, which can be obtained base independently using the methods summarized in~\cite{Buchmuller:2017wpe}.

\bibliographystyle{JHEP}
\bibliography{biblio}

\end{document}